\documentclass[twocolumn,amsmath,nofootinbib]{revtex4} 

\usepackage{url}
\usepackage{amssymb}
\usepackage{amsfonts}
\usepackage{amsbsy}
\usepackage{graphicx}
\usepackage{hyperref}
\usepackage{array} 
\usepackage{multirow}

\newcolumntype{x}[1]{%
>{\centering\hspace{0pt}}p{#1}}%
\providecommand{\openone}{\leavevmode\hbox{\small1\kern-3.8pt\normalsize1}}

\def\etal{{\frenchspacing\it et al.}}
\def\ie{{\frenchspacing\it i.e.}}
\def\eg{{\frenchspacing\it e.g.}}
\def\etc{{\frenchspacing\it etc.}}

\def\spose#1{\hbox to 0pt{#1\hss}}
\def\simlt{\mathrel{\spose{\lower 3pt\hbox{$\mathchar"218$}}
   \raise 2.0pt\hbox{$\mathchar"13C$}}}
\def\simgt{\mathrel{\spose{\lower 3pt\hbox{$\mathchar"218$}}
     \raise 2.0pt\hbox{$\mathchar"13E$}}}
 \def\simpropto{\mathrel{\spose{\lower 3pt\hbox{$\mathchar"218$}}
     \raise 2.0pt\hbox{$\propto$}}}

\def\beq#1{\begin{equation}\label{#1}}
\def\eeq{\end{equation}}
\def\beqa#1{\begin{eqnarray}\label{#1}}
\def\eeqa{\end{eqnarray}}
\def\eq#1{equation~(\ref{#1})}	
\def\Eq#1{Equation~(\ref{#1})}
\def\eqn#1{~(\ref{#1})}

\def\fig#1{Figure~\ref{#1}}
\def\Fig#1{Figure~\ref{#1}}

\def\Tab#1{Table~\ref{#1}}

\def\Sec#1{Section~\ref{#1}}
\def\App#1{Appendix~\ref{#1}}

\def\ed{\end{document}}

\def\a{{\bf a}}

\def\h{{\bf h}}
\def\k{{\bf k}}
\def\kappab{{\mathbf\kappa}}
\def\lambdavec{{\boldsymbol\lambda}}
\def\e{{\bf e}}
\def\p{{\bf p}}
\def\q{{\bf q}}
\def\r{{\bf r}}
\def\sigmab{{\mathbf\sigma}}
\def\vv{{\bf v}}
\def\vmax{{v_{\rm max}}}
\def\x{{\bf x}}

\def\I{{\bf I}}
\def\A{{\bf A}}
\def\B{{\bf B}}
\def\C{{\bf C}}
\def\D{{\bf D}}

\def\Eint{\mathring{E}}
\def\Emin{E_{\rm min}}
\def\Emax{E_{\rm max}}
\def\F{{\bf F}}

\def\H{{\bf H}}
\def\K{{\bf K}}

\def\P{{\bf P}}
\def\PP{{\mathbf\Pi}}
\def\Phimatrix{{\mathbf\Phi}}

\def\S{{\bf S}}
\def\Slin{S^{\rm lin}}
\def\T{{\bf T}}
\def\tdyn{\tau_{\rm dyn}}
\def\tind{\tau_{\rm ind}}
\def\U{{\bf U}}

\def\tdyn{\tau_{\rm dyn}}

\def\diag{\hbox{diag}\,}
\def\pr{\hbox{pr}\,}
\def\tr{\hbox{tr}\,}
\def\proj{\operatornamewithlimits{\pr}}
\def\trace{\operatornamewithlimits{\tr}}
\def\tensormultiplication{\operatornamewithlimits{\tensormult}}

\def\tensormult{\otimes}

\def\ket#1{|#1\rangle}
\def\psiket{\ket{\psi}}

\def\up{\ket{\!\!\uparrow}}
\def\down{\ket{\!\!\downarrow}}

\def\bra#1{\langle #1|}
\def\psibra{\bra{\psi}}

\def\upbra{\bra{\uparrow\!\!}}
\def\downbra{\bra{\downarrow\!\!}}

 \def\ms{\mskip-3mu}
  \def\ns{\mskip-5mu}

\def\rn{}
\def\nn#1 #2{#2. #1}				
\def\nnn#1 #2 #3{#2. #3. #1}			
\def\nnnn#1 #2 #3 #4{#2. #3. #4 #1}		
\def\nnnnn#1 #2 #3 #4 #5{#2. #3. #4 #5. #1}	
\def\dualand{ and\hbox{ }}				
\def\multiand{, and\hbox{ }}				
\def\rf#1;#2;#3;#4;#5 {{\frenchspacing\par\rn#1, #3 {\bf #4}, #5 (#2). \par}}
\def\rfe#1;#2;#3;#4 {{\frenchspacing\par\rn#1, #3, #4 (#2). \par}}
\def\rg#1;#2;#3;#4;#5;#6 {{\frenchspacing\par\rn#1, #3 {\bf #4}, #5 (#2). \par}}
\def\rfbook#1;#2;#3;#4;#5 {{\frenchspacing\par\rn#1, {\it #3} (#5, #4, #2).\par}}
\def\rfprep#1;#2;#3 {{\par\frenchspacing\rn#1, #3 (#2).\par}}
\def\rfproc#1;#2;#3;#4;#5;#6 {{\frenchspacing\par\rn#1 #2, in {\it #3}, ed. #4 (#5: #6)\par}}
\def\rfprocp#1;#2;#3;#4;#5;#6;#7 {{\frenchspacing\par\rn#1 #2, in {\it #3}, ed. #4 (#5: #6), p#7\par}}

\begin{document}
\pdfoptionalwaysusepdfpagebox=5


\title{Consciousness as a State of Matter}

\author{Max Tegmark}

\address{Dept.~of Physics \& MIT Kavli Institute, Massachusetts Institute of Technology, Cambridge, MA 02139}

\date{\today}
\date{Accepted for publication in {\it Chaos, Solitons \& Fractals} March 17, 2015}

\vspace{10mm}

\begin{abstract}
We examine the hypothesis that consciousness can be understood as a state of matter, ``perceptronium", with distinctive information processing abilities. We explore four basic principles that may distinguish conscious matter from other physical systems such as solids, liquids and gases: the information, integration,  independence and dynamics principles.
If such principles can identify conscious entities, then they can help solve 
the {\it  quantum factorization problem}: why do conscious observers like us perceive the particular Hilbert space factorization corresponding to classical space (rather than Fourier space, say), and more generally, why do we perceive the world around us as a dynamic hierarchy of objects that are strongly integrated and relatively independent?
Tensor factorization of matrices is found to play a central role, and our technical results include a theorem about Hamiltonian separability (defined using Hilbert-Schmidt superoperators) being maximized in the energy eigenbasis.
Our approach generalizes Giulio Tononi's integrated information framework for neural-network-based consciousness to arbitrary quantum systems, and we find interesting links to error-correcting codes, condensed matter criticality, and the Quantum Darwinism program,
as well as an interesting connection between the emergence of consciousness and the emergence of time.
\end{abstract}

\maketitle

\section{Introduction}
\label{IntroSec}

\subsection{Consciousness in physics}

A commonly held view is that consciousness is irrelevant to physics and should therefore not be discussed in physics papers. One oft-stated reason is a perceived lack of rigor in past attempts to link consciousness to physics. Another argument is that physics has been managed just fine for hundreds of years by avoiding this subject, and should therefore keep doing so. Yet the fact that {\it most} physics problems can be solved without reference to consciousness does not guarantee that this applies to {\it all} physics problems.
Indeed, it is striking that many of the most hotly debated issues in physics today involve the notions of observations and observers, and we cannot dismiss the possibility that part of the reason why these issues have resisted resolution for so long is our reluctance as physicists to discuss consciousness and attempt to 
rigorously define what constitutes an observer.

For example, does the non-observability of spacetime regions beyond horizons imply that they in some sense do not exist independently of the regions that we can observe? This question lies at the heart of the controversies surrounding the holographic principle, black hole complementarity and firewalls, and depends crucially on the role of observers 
\cite{Almheiri12,Banks14}. What is the solution to the quantum measurement problem? This again hinges crucially on the role of observation: does the wavefunction undergo a non-unitary collapse when an observation is made, are there Everettian parallel universes, or does it make no sense to talk about an an observer-independent reality, as argued by QBism advocates \cite{SaundersBook}?
Is our persistent failure to unify general relativity with quantum mechanics linked to the different roles of observers in the two theories?
After all, the idealized observer in general relativity has no mass, no spatial extent and no effect on what is observed, whereas the quantum observer notoriously does appear to affect the quantum state of the observed system. Finally, out of all of the possible factorizations of Hilbert space, why is the particular factorization corresponding to classical space so special? Why do we observers perceive ourselves are fairly local in real space as opposed to Fourier space, say, which according to the formalism of quantum field theory corresponds to an equally valid Hilbert space factorization? This ``quantum factorization problem'' appears intimately related to the nature of an observer.

The only issue there is consensus on is that there is no consensus about how to define an observer and its role.
One might hope that a detailed observer definition will prove unnecessary because some simple properties such as the ability to record information might suffice; however, we will see that at least two more properties of observers may be necessary to solve the quantum factorization problem, and that a closer examination of consciousness may be required to identify these properties.

Another commonly held view is that consciousness is unrelated to quantum mechanics because the brain is a wet, warm system where decoherence destroys
quantum superpositions of neuron firing much faster than we can think, preventing our brain from acting as a quantum computer \cite{brain}.
In this paper, I argue that consciousness and quantum mechanics are nonetheless related, but in a different way: 
it is not so much that quantum mechanics is relevant to the brain, as the other way around.
Specifically, consciousness is relevant to solving an open problem at the very heart of quantum mechanics: the quantum factorization problem.

\subsection{Consciousness in philosophy}

Why are you conscious right now? Specifically, why are you having a subjective experience of reading these words, seeing colors and hearing sounds, while the inanimate objects around you are presumably not having any subjective experience at all? 
Different people mean different things by ``consciousness", including awareness of environment or self. I am asking the more basic question of why you experience anything at all, which is the essence of what philosopher David Chalmers has termed ``the hard problem" of consciousness  and which has preoccupied philosophers throughout the ages (see \cite{Chalmers95} and references therein). 
A traditional answer to this problem is dualism --- that living entities differ from inanimate ones because they contain some non-physical element such as an ``anima" or ``soul". Support for dualism among scientists has gradually dwindled with the realization that we are made of quarks and electrons, which as far as we can tell move according to simple physical laws. If your particles really move according to the laws of physics, then your purported soul is having no effect on your particles, so your conscious mind and its ability to control your movements would have nothing to do with a soul.
If your particles were instead found not to obey the known laws of physics because they were being pushed around by your soul, then we could treat the soul as just another physical entity able to exert forces on particles, and study what physical laws it obeys, just as physicists have studied new forces fields and particles in the past. 

The key assumption in this paper is
that consciousness is a property of certain physical systems, with no ``secret sauce'' or non-physical elements.\footnote{More specifically, 
we pursue an extreme Occam's razor approach and explore whether all aspects of reality can be derived from quantum mechanics with a density matrix evolving unitarily according to a Hamiltonian. It this approach should turn out to be successful, then all observed aspects of reality must emerge from the mathematical formalism alone: for example, the Born rule for subjective randomness associated with observation would emerge from the underlying deterministic density matrix evolution through Everett's approach, and both a semiclassical world and consciousness should somehow emerge as well, perhaps though processes generalizing decoherence. Even if quantum gravity phenomena cannot be captured with this simple quantum formalism, it is far from clear that gravitational, relativistic or non-unitary effects are central to understanding consciousness or how conscious observers perceive their immediate surroundings.
There is of course no {\it a priori} guarantee that this approach will work; this paper is motivated by the view that 
an Occam's razor approach is useful if it succeeds and very interesting if it fails, by giving hints as to what alternative  assumptions or ingredients are needed.
}, 
This transforms Chalmers' hard problem. 
Instead of starting with the hard problem of why an arrangement of particles can feel conscious, we will start with the hard fact that some arrangement of particles (such as your brain) {\it do} feel conscious while others (such as your pillow) do not, and ask what properties of the particle arrangement make the difference.

This paper is {\it not} a comprehensive theory of conciousness. Rather, it is an investigation into the physical properties that conscious systems must have.
If we understood what these physical properties were, then we could in principle answer all of the above-mentioned open physics questions by studying the equations of physics: we could identify all conscious entities in any physical system, and calculate what they would perceive.  
However, this approach is typically not pursued by physicists, with the argument that we do not understand consciousness well enough. 

\subsection{Consciousness in neuroscience}

Arguably, recent progress in neuroscience has fundamentally changed this situation, so that we physicists can no longer blame neuroscientists for our own lack of progress. 
I have long contended that consciousness is the way information feels when being processed in certain complex ways \cite{mmm,toe2}, \ie, that it corresponds to certain complex patterns in spacetime that obey the same laws of physics as  other complex systems. 
In the seminal paper {\it ``Consciousness as Integrated Information: a Provisional Manifesto''} \cite{TononiManifesto}, 
Giulio Tononi made this idea more specific and useful, making a compelling argument that for an information processing system to be conscious, it needs to have two separate traits:
\begin{enumerate}
\item {\bf Information:} It has to have a large repertoire of accessible states, \ie, the ability to store a large amount of information.
\item {\bf Integration:} This information must be integrated into a unified whole, \ie, it must be impossible to decompose the system into nearly independent parts, because otherwise these parts would subjectively feel like two separate conscious entities. 
\end{enumerate}
Tononi's work has  generated a flurry of activity in the neuroscience community, spanning the spectrum from theory to experiment (see \cite{Dehaene11,TononiBook,Casali13,IIT3,Dehaene14} for recent reviews), making it timely to investigate its implications for physics as well. This is the goal of the present paper --- a goal whose pursuit may ultimately provide additional tools for the neuroscience community as well.

Despite its successes, Tononi's Integrated Information Theory (IIT)\footnote{Since it's inception \cite{TononiManifesto},
IIT has been further developed \cite{IIT3}. In particular,  IIT 3.0 considers both the past and the future of a mechanism in a particular state (it's so-called cause-effect repertoire) and replaces the Kullback-Leibler measure with a proper metric.} 
leaves many questions unanswered. If it is to extend our consciousness-detection ability to animals, computers and arbitrary physical systems, then we need to ground its principles in fundamental physics. IIT takes information, measured in bits, as a starting point. But when we view a brain or computer through our physicistsÕ eyes, as myriad moving particles, then what physical properties of the system should be interpreted as logical bits of information? I interpret as a ``bit" both the position of certain electrons in my computerÕs RAM memory (determining whether the micro-capacitor is charged) and the position of certain sodium ions in your brain (determining whether a neuron is firing), but on the basis of what principle? Surely there should be some way of identifying consciousness from the particle motions alone, or from the quantum state evolution, even without this information interpretation? If so, what aspects of the behavior of particles corresponds to conscious integrated information? We will explore different 
measures of integration below.
Neuroscientists have successfully mapped out which brain activation patterns correspond to certain types of conscious experiences, and named these patterns Òneural correlates of consciousnessÓ.  How can we generalize this and look for physical correlates of consciousness, defined as the patterns of moving particles that are conscious? What particle arrangements are conscious?

\subsection{Consciousness as a state of matter}

Generations of physicists and chemists have studied what happens when you group together vast numbers of atoms, finding that their collective behavior depends on the pattern in which they are arranged: the key difference between a solid, a liquid and a gas lies not in the types of atoms, but in their arrangement. In this paper, I conjecture that consciousness can be understood as yet another state of matter.
Just as there are many types of liquids, there are many types of consciousness. However, this should not preclude us from identifying, quantifying, modeling and ultimately understanding the characteristic properties that all liquid forms of matter (or all conscious forms of matter) share.

\noindent
\begin{table}
\begin{center}
\begin{tabular}{|l|c|c|c|c|c|c|}
\hline
			&Many		&	&	&\\
State of		&long-lived&Information	&Easily&Complex?\\
matter		&states?&integrated?&writable?&dynamics?\\
\hline
Gas			&N	&N	&N	&Y\\
Liquid		&N	&N	&N	&Y\\
Solid			&Y	&N	&N	&N\\
Memory		&Y	&N	&Y	&N\\
Computer		&Y	&?	&Y	&Y\\
Consciousness	&Y	&Y	&Y	&Y\\
\hline
\end{tabular}
\end{center}
\caption{Substances that store or process information can be viewed as novel states of matter and 
investigated with traditional physics tools.
\label{StatesTable}
}
\end{table}

To classify the traditionally studied states of matter, we need to measure only a small number of physical parameters: 
viscosity, compressibility, electrical conductivity and (optionally) diffusivity.
We call a substance a solid if its viscosity is effectively infinite (producing structural stiffness), and call it a fluid otherwise.
We call a fluid a liquid if its compressibility and diffusivity are small and otherwise call it either a gas or a plasma, depending on its electrical conductivity.

What are the corresponding physical parameters that can help us identify conscious matter, and what are the key physical features that characterize it?   If such parameters can be identified, understood and measured, this will help us identify (or at least rule out)  consciousness ``from the outside'', without access to subjective introspection. This could be important for reaching consensus on many currently controversial topics, ranging from the future of artificial intelligence to determining when an animal, fetus or unresponsive patient can feel pain.
If would also be important for fundamental theoretical physics, by allowing us to identify conscious observers in our universe by using 
the equations of physics and thereby answer thorny observation-related questions such as those mentioned in the introductory paragraph.


\subsection{Memory}
As a first warmup step toward consciousness, let us first consider a state of matter that we would 
characterize as {\it memory}\footnote{Neuroscience research has demonstrated that long-term memory is not 
necessary for consciousness. However, even extremely memory-impaired conscious humans such as Clive Wearing \cite{Wilson95}
are able to retain information for several seconds; in this paper, I will assume merely that information needs to be remembered long enough to be subjectively
experienced --- perhaps 0.1 seconds for a human, and much less for entities processing information more rapidly.} 
--- what physical features does it have?
For a substance to be useful for storing information, it clearly needs to have a large repertoire of possible long-lived states or attractors (see Table~\ref{StatesTable}). 
Physically, this means that its potential energy function has a large number of well-separated minima. The information storage capacity (in bits) is simply the base-2 logarithm of the number of minima.
This equals the entropy (in bits) of the degenerate ground state if all minima are equally deep.
For example, solids have many long-lived states, whereas liquids and gases do not: if you engrave someone's name on a gold ring, the information will still be  there years later, but if you engrave it in the surface of a pond, it will be lost within a second as the water surface changes its shape. Another desirable trait of a memory substance, distinguishing it from generic solids,  is that it is not only easy to read from (as a gold ring), but also easy to write to: altering the state of your hard drive or your synapses requires less energy than engraving gold. 

\subsection{Computronium}
As a second warmup step, what properties should we ascribe to what Margolus and Toffoli have termed  {\it ``computronium''} \cite{computronium}, the most general substance that can process information as a computer? Rather than just remain immobile as a gold ring, it must exhibit complex {\it dynamics} so that its future state depends in some complicated (and hopefully controllable/programmable) way on the present state. 
Its atom arrangement must be less ordered than a rigid solid where nothing interesting changes, but more ordered than a liquid or gas. 
At the microscopic level, computronium need not  be particularly complicated, because computer scientists have long known that as long as a device can perform certain elementary logic operations, it is {\it universal}: it can be programmed to perform the same computation as any other computer with enough time and memory.
Computer vendors often parametrize computing power in FLOPS, floating-point operations per second for 64-bit numbers; more generically, we can parametrize computronium capable of universal computation 
by ``FLIPS'': the number of elementary logical operations such as bit flips that it can perform per second. 
It has been shown by Lloyd \cite{Lloyd99} that a system with average energy $E$ can perform a maximum of $4E/h$ elementary logical operations per second, where $h$ is Planck's constant.
The performance of today's best computers is about 38 orders of magnitude lower than this, because they use huge numbers of particles to store each bit and because most of their energy is tied up in a computationally passive form, as rest mass.
%


\subsection{Perceptronium}

What about {\it ``perceptronium''}, the most general substance that feels subjectively self-aware? 
If Tononi is right, then it should not merely be able to store and process information like computronium does,
but it should also satisfy the principle that its information is integrated, forming a unified and indivisible whole. 

Let us also conjecture another principle that conscious systems must satisfy: that of {\it autonomy}, \ie, 
that information can be processed with relative freedom from external influence.
Autonomy is thus the combination of two separate properties: {\it dynamics} and {\it independence}.
Here dynamics means time dependence (hence information processing capacity) and
independence means that the dynamics is dominated by forces from within rather than outside the system.
Just like integration, autonomy is postulated to be a necessary but not sufficient condition for a system to be conscious: 
for example, clocks and diesel generators tend to exhibit high autonomy, but  lack substantial information storage capacity.

\def\mytab{\hglue5mm}
\begin{table}
\begin{center}
\begin{tabular}{|>{\raggedright}p{2.0cm}|p{6.4cm}|}
\hline
Principle			&Definition\\
\hline
Information		&A conscious system has substantial\\
\mytab principle		&\mytab  information storage capacity.\\
Dynamics			&A conscious system has substantial\\
\mytab principle		&\mytab information processing capacity.\\
Independence		&A conscious system has substantial\\
\mytab principle		&\mytab independence from the rest of the world.\\
Integration			&A conscious system cannot consist of\\
\mytab principle		&\mytab nearly independent parts.\\
\hline
Autonomy			&A conscious system has substantial\\
\mytab principle		&\mytab dynamics and independence.\\
\hline
Utility			&An evolved conscious system records mainly\\
\mytab principle		&\mytab information that is useful for it.\\
\hline
\end{tabular}
\end{center}
\caption{Four conjectured necessary conditions for consciousness that we  explore in this paper.
The fifth principle simply combines the second and third. The sixth is not a necessary condition, but may explain the evolutionary origin of the others.
\label{PrincipleTable}
}
\end{table}

\subsection{Consciousness and the quantum factorization problem}

Table~\ref{PrincipleTable} summarizes the four candidate principles that we will explore as necessary conditions for consciousness.
Our goal with isolating and studying these principles is not merely to strengthen our understanding of consciousness as a physical process, but also to identify simple traits of conscious matter that can help us tackle other open problems in physics. 
For example, the only property of consciousness that Hugh Everett needed to assume for his work on quantum measurement was that of the information principle: by applying the Schr\"odinger equation to systems that could record and store information, he inferred that they would perceive subjective randomness in accordance with the Born rule.
In this spirit, we might hope that adding further simple requirements such as in the integration principle, the independence principle and the dynamics principle might suffice to solve currently open problems related to observation. 
The last principle listed in Table~\ref{PrincipleTable}, the utility principle, is of a different character than the others: we consider it not as a necessary condition for consciousness, but as a potential unifying evolutionary explanation of the others.

In this paper, we will pay particular attention to what I will refer to as the {\it  quantum factorization problem}:
why do conscious observers like us perceive the particular Hilbert space factorization corresponding to classical space (rather than Fourier space, say), and more generally, why do we perceive the world around us as a dynamic hierarchy of objects that are strongly integrated and relatively independent?
This fundamental problem has received almost no attention in the literature \cite{Schwindt12}.
We will see that this problem is very closely related to the one Tononi confronted for the brain, merely on a larger scale. 
Solving it would also help solve the {\it ``physics-from-scratch'' problem} \cite{toe2}: 
If the Hamiltonian $\H$ and the total density matrix $\rho$ fully specify our physical world, how do we extract 3D space and the rest of our semiclassical world from nothing more than two Hermitian matrices, which come without any {\it a priori} physical interpretation or additional structure such as a physical space, quantum observables, quantum field definitions, an ``outside'' system, \etc? 
Can some of this information be extracted even from $\H$ alone, which is fully specified by nothing more than its eigenvalue spectrum? 
We will see that a generic Hamiltonian cannot be decomposed using tensor products, which would correspond to a decomposition of the cosmos into non-interacting parts  --- instead, there is an optimal factorization of our universe into integrated and relatively independent parts. 
Based on Tononi's work, we might expect that this factorization, or some generalization thereof, is what conscious observers perceive, because an integrated and relatively autonomous information complex is fundamentally what a conscious observer is! 


\bigskip
The rest of this paper is organized as follows. In \Sec{IntegrationSec}, we explore the integration principle by quantifying integrated information in physical systems, finding encouraging results for classical systems and interesting challenges introduced by quantum mechanics.
In \Sec{IndependenceSec}, we explore the independence principle, finding that at least one additional principle is required to account for the observed factorization of our physical world into an object hierarchy in three-dimensional space.
In \Sec{DynamicsSec}, we explore the dynamics principle and other possibilities for reconciling quantum-mechanical theory with our observation of  a semiclassical world. 
We discuss our conclusions in \Sec{ConclusionsSec}, including applications of the utility principle, and cover various mathematical details in the three appendices.
Throughout the paper, we mainly consider finite-dimensional Hilbert spaces that can be viewed as collections of qubits; as explained in \App{EmergenceAppendix}, this appears to cover standard quantum field theory with its infinite-dimensional Hilbert space as well.

\section{Integration}
\label{IntegrationSec}

\subsection{Our physical world as an object hierarchy}
\label{ObjectHierarchySec}

\begin{figure*}[pbt]
\centerline{\includegraphics[width=160mm]{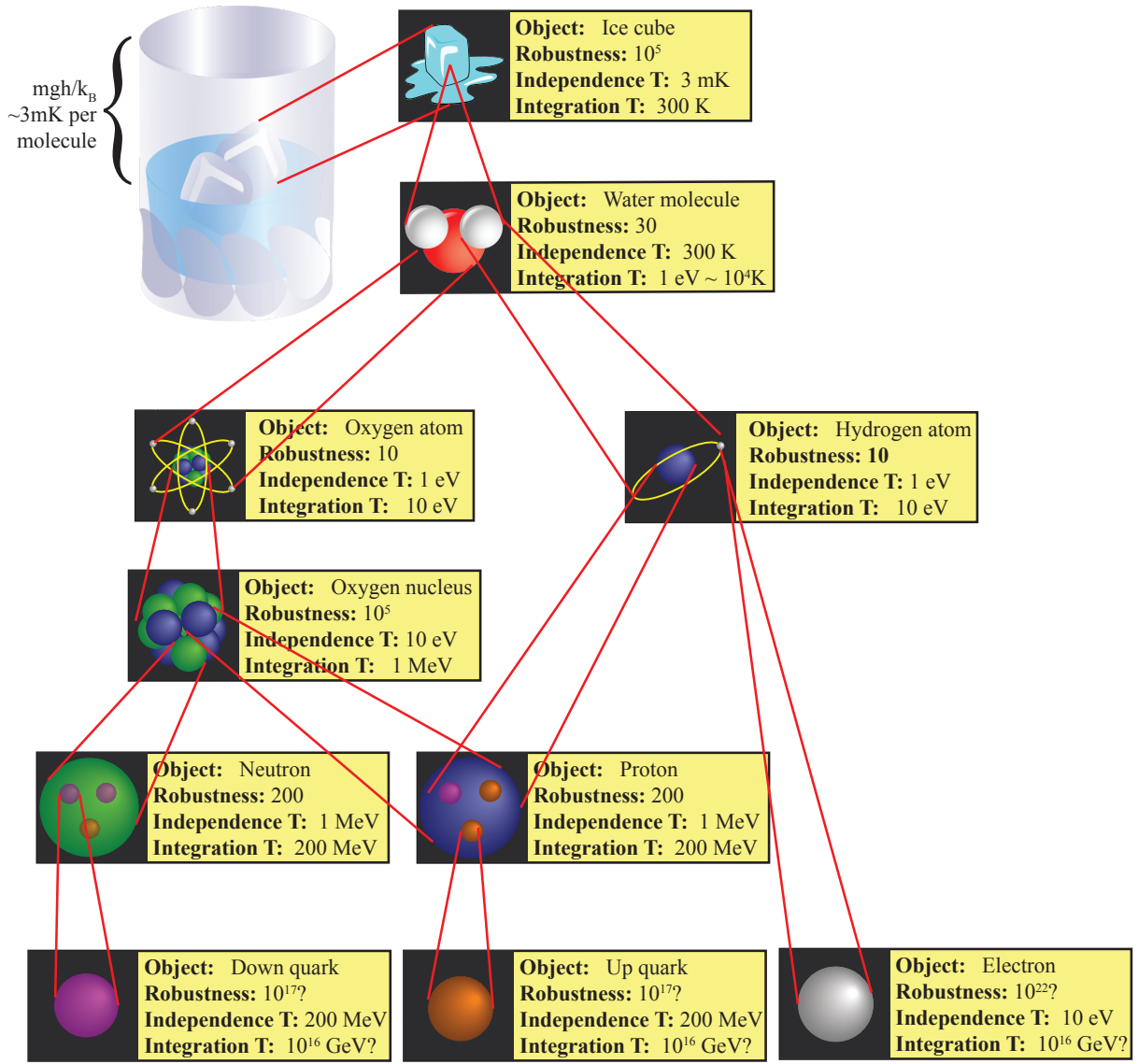}}
\caption{
We perceive the external world as a hierarchy of objects,  whose parts are more strongly connected to one another than to the outside.
The robustness of an object is defined as the ratio of the integration temperature (the energy per part needed to separate them) to the independence temperature (the energy per part needed to separate the parent object in the hierarchy).
}
\label{HierarchyFig}
\end{figure*}


The problem of identifying consciousness in an arbitrary collection of moving particles is similar to the simpler problem of identifying objects there.
One of the most striking features of our physical world is that we perceive it as an object hierarchy, as illustrated in \fig{HierarchyFig}.
If you are enjoying a cold drink, you perceive ice cubes in your glass as separate objects because they are both fairly integrated and fairly independent, \eg, 
their parts are more strongly connected to one another than to the outside. The same can be said about each of their constituents, ranging from water molecules all the way down to electrons and quarks. Zooming out, you similarly perceive the macroscopic world as a dynamic hierarchy of objects that are strongly integrated and relatively independent, all the way up to planets, solar systems and galaxies.
Let us quantify this by defining the {\it robustness} 
of an object as the ratio of the integration temperature (the energy per part needed to separate them) to the independence temperature (the energy per part needed to separate the parent object in the hierarchy). 
\Fig{HierarchyFig} illustrates that all of the ten types of objects shown have robustness of ten or more.
A highly robust object preserves its identity (its integration and independence) over a wide range of temperatures/energies/situations. The more robust an object is, the more useful it is for us humans to perceive it as an object and coin a name for it, as per the above-mentioned utility principle.

Returning to the {\it ``physics-from-scratch'' problem}, how can we identify this object hierarchy if all we have to start with are two Hermitian matrices, the density matrix $\rho$ encoding the state of our world and the Hamiltonian $\H$ determining its time-evolution? Imagine that we know only these mathematical objects $\rho$ and $\H$ and have no information whatsoever about how to interpret the various degrees of freedom or anything else about them.
A good beginning is to study integration. Consider, for example, $\rho$ and $\H$ for a single deuterium atom, whose Hamiltonian is 
(ignoring spin interactions for simplicity)
\beqa{deuteriumEq}
&&\H(\r_p,\p_p,\r_n,\p_n,\r_e,\p_e) =\\
&&= \H_1(\r_p,\p_p,\r_n,\p_n)+\H_2(\p_e)+\H_3(\r_p,\p_p,\r_n,\p_n,\r_e,\p_e),\nonumber
\eeqa
where $\r$ and $\p$ are position and momentum vectors, and the subscripts $p$, $n$ and $e$ refer to the 
proton, the neutron and  the electron.
On the second line, we have decomposed $\H$ into three terms: the internal energy of the proton-neutron nucleus, the internal (kinetic) energy of the electron, and the electromagnetic electron-nucleus interaction. 
This interaction is tiny, on average involving  much less energy than those within the nucleus:
\beq{DeuteriumRobustnessEq}
{\tr\H_3\rho\over\tr\H_1\rho}\sim 10^{-5},
\eeq
which we recognize as the inverse robustness for a typical nucleus in \fig{IntegrationFig}.
We can therefore fruitfully approximate the nucleus and the electron as separate objects that are almost independent, interacting only weakly with one another.
The key point here is that we could have performed this object-finding exercise of dividing the variables into two groups to find the greatest independence (analogous to what Tononi calls ``the cruelest cut'') based on the functional form of $\H$ alone, without even having heard of electrons or nuclei, thereby identifying their degrees of freedom through a purely mathematical exercise. 

\subsection{Integration and mutual information}
\label{IntegrationInformationSec}

If the interaction energy $\H_3$ were so small that we could neglect it altogether, then $\H$ would be decomposable into two parts 
$\H_1$ and $\H_2$, each one acting on only one of the two sub-systems (in our case the nucleus and the electron).
This means that any thermal state would be {\it factorizable}:
\beq{ThermalSeparabilityEq}
\rho\propto e^{-\H/kT} = e^{-\H_1/kT}e^{-\H_2/kT}=\rho_1\rho_2,
\eeq
so the total state $\rho$ can be factored into a product of the subsystem states $\rho_1$ and $\rho_2$.
In this case, the mutual information 
\beq{IdefEq}
I\equiv S(\rho_1)+S(\rho_2)-S(\rho)
\eeq
vanishes, where
\beq{IdefEq}
S(\rho)\equiv -\tr\rho\log_2\rho 
\eeq
is the von Neumann entropy (in bits) --- which is simply the Shannon entropy of eigenvalues of $\rho$.
Even for non-thermal states, the time-evolution operator $\U$ becomes separable:
\beq{EvolSeparabilityEq}
\U\equiv e^{i\H t/\hbar} = e^{i\H_1 t/\hbar}e^{i\H_2 t/\hbar}=\U_1\U_2,
\eeq
which (as we will discuss in detail in \Sec{IndependenceSec}) implies that the mutual information stays constant over time and no information is ever exchanged between the objects.
In summary, if a Hamiltonian can be decomposed without an interaction term (with $\H_3=0$), then it describes two perfectly independent systems.\footnote{Note that in this paper, we are generally considering $\H$ and $\rho$ for the entire cosmos, so that there is no ``outside'' containing observers
{\etc} 
If $\H_3=0$, entanglement between the two systems thus cannot have any observable effects. 
This is in stark contrast to most textbook quantum mechanics considerations, where one studies a small subsystem of the world.
}

Let us now consider the opposite case, when a system cannot be decomposed into independent parts.
Let us define the {\it integrated information} $\Phi$ as the mutual information $I$ for the ``cruelest cut'' (the cut minimizing $I$) in some class of cuts that subdivide the system into two (we will discuss many different classes of cuts below).
Although our $\Phi$-definition is slightly different from Tononi's \cite{TononiManifesto}\footnote{Tononi's definition of $\Phi$ \cite{TononiManifesto} applies only for classical systems, whereas we wish to study the quantum case as well. Our $\Phi$ is measured in bits and can grow with system size like an extrinsic variable, whereas his is an intrinsic variable akin representing a sort of average integration per bit.}, 
it is similar in spirit, and we are reusing his $\Phi$-symbol for its elegant symbolism (unifying the shapes of $I$ for information and $O$ for integration). 

\subsection{Maximizing integration}

\begin{figure*}[pbt]
\centerline{\includegraphics[width=180mm]{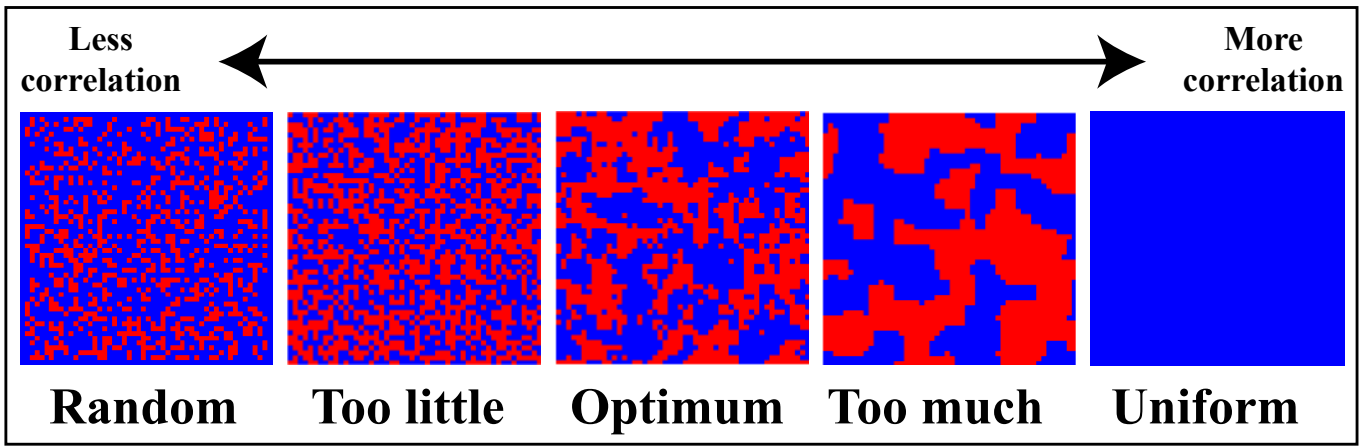}}
\caption{The panels show simulations of the 2D Ising model on a $50\times 50$ lattice, with the temperature 
progressively decreasing from left to right. The integrated information $\Phi$ drops to zero bits at $T\to\infty$ (leftmost panel) and to one bit as $T\to 0$ (rightmost panel), taking a maximum at an intermediate temperature near the phase transition temperature.}
\label{IsingFig}
\end{figure*}

We just saw that if two systems are dynamically independent ($\H_3=0$), then $\Phi=0$ at all time 
both for thermal states and for states that were independent ($\Phi=0$) at some point in time.
Let us now consider the opposite extreme. How large can the integrated information $\Phi$ get?
A as warmup example, let us consider the familiar 2D Ising model in \fig{IsingFig} where $n=2500$ magnetic dipoles (or spins) that can point up or down are placed on a square lattice, and $\H$ is such that they prefer aligning with their nearest neighbors.
When $T\to\infty$, $\rho\propto e^{-\H/kT}\to\I$, so all $2^n$ states are equally likely, all $n$ bits are statistically independent, and $\Phi=0$. When $T\to 0$, all states freeze out except the two degenerate ground states (all spin up or all spin down), so all spins are perfectly correlated and $\Phi=1$ bit.
For intermediate temperatures, long-range correlations are seen to exist such that typical states have contiguous spin-up or spin-down patches. On average, we get about one bit of mutual information for each such patch crossing our cut (since a spin on one side ``knows'' about at a spin on the other side), so for bipartitions that cut the system into two equally large halves, the mutual information will be proportional to the length of the cutting curve. The ``cruelest cut'' is therefore a vertical or horizontal straight line of length $n^{1/2}$, giving 
$\Phi\sim n^{1/2}$ at the temperature where typical patches are only a few pixels wide.
We would similarly get a maximum integration $\Phi\sim n^{1/3}$ for a 3D Ising system and $\Phi\sim 1$ bit for a 1D Ising system. 

Since it is the spatial correlations that provide the integration, it is interesting to speculate about whether the conscious subsystem of our brain is a system near its critical temperature, close to a phase transition. 
Indeed, Damasio has argued that to be in homeostasis, a number of physical parameters of our brain need to be kept within a narrow range of values \cite{Damasio10} --- this is precisely what is required of any condensed matter system to be near-critical, exhibiting correlations that are long-range (providing integration) but not so strong that the whole system becomes correlated like in the right panel or in a brain experiencing an epileptic seizure.

\subsection{Integration, coding theory and error correction}
\label{CodeSubsec}

 \begin{figure}[phbt]
\centerline{\includegraphics[width=88mm]{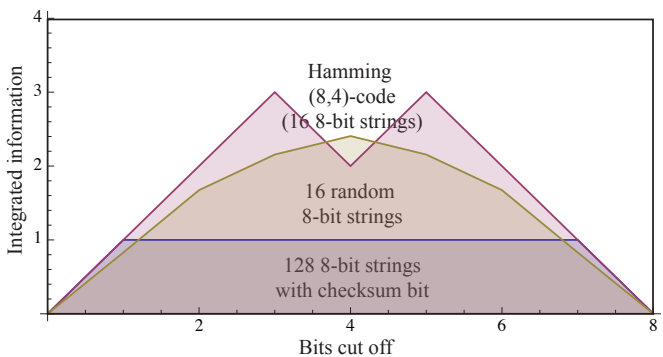}}
\caption{
For various 8-bit systems, the integrated information is plotted as a function of the number of bits cut off into a sub-system with the ``cruelest cut''.
The Hamming (8,4)-code is seen to give classically optimal integration except for a bipartition into $4+4$ bits:
an arbitrary subset containing no more than three bits is completely determined by the remaining bits. The code consisting of the half of all 8-bit strings whose bit sum is even 
(\protect\ie, each of the 128 7-bit strings followed by a parity checksum bit) has Hamming distance $d=2$ and gives $\Phi=1$ however many bits are cut off. 
A random set of 16 8-bit strings is seen to outperform the Hamming (8,4)-code for 4+4-bipartitions, but not when fewer bits are cut off.
\label{IntegrationFig}
}
\end{figure}

Even when we tuned the temperature to the most favorable value in our 2D Ising model example, the integrated information never exceeded $\Phi\sim n^{1/2}$ bits, which is merely a fraction $n^{-1/2}$ of the $n$ bits of information that $n$ spins can potentially store.
So can we do better? Fortunately, a closely related question has been carefully studied in the branch of mathematics known as coding theory, with the aim of optimizing error correcting codes.
Consider, for example, the following set of $m=16$ bit strings, each written as a column vector of length $n=8$:
$$
M=\left(\begin{tabular}{cccccccccccccccc}
0&0&0&0&1&1&1&1&0&0&0&0&1&1&1&1\\
0&0&0&0&0&0&0&0&1&1&1&1&1&1&1&1\\
0&1&1&0&1&0&0&1&0&1&1&0&1&0&0&1\\
0&1&0&1&0&1&0&1&1&0&1&0&1&0&1&0\\
0&1&0&1&1&0&1&0&0&1&0&1&1&0&1&0\\
0&0&1&1&0&0&1&1&1&1&0&0&1&1&0&0\\
0&0&1&1&1&1&0&0&0&0&1&1&1&1&0&0\\
0&1&1&0&0&1&1&0&1&0&0&1&1&0&0&1\\
\end{tabular}
\right)
$$
This is known as the Hamming(8,4)-code, and has Hamming distance $d=4$, which means that at least 4 bit flips are required to change one string into another \cite{Hamming50}.
It is easy to see that for a code with Hamming distance $d$, any $(d-1)$ bits can always be reconstructed from the others: You can always reconstruct $b$ bits as long as erasing them does not make two bit strings identical, which would cause ambiguity about which the correct bit string is. This implies that reconstruction works when the Hamming distance $d>b$.

To translate such codes of $m$ bit strings of length $n$ into physical systems, we simply created a state space with $n$ bits
(interpretable as $n$ spins or other two-state systems) and construct a Hamiltonian which has an $m$-fold degenerate ground state, with
one minimum corresponding to each of the $m$ bit strings in the code. In the low-temperature limit, all bit strings will receive the same probability weight $1/m$, giving an entropy $S=\log_2 m$.
The corresponding integrated information $\Phi$ of the ground state is plotted in \fig{IntegrationFig}
for a few examples,  as a function of cut size $k$ (the number of bits assigned to the first subsystem).
To calculate $\Phi$ for a cut size $k$ in practice, we simply minimize the mutual information $I$ over all $\left({n\atop k}\right)$ ways of partitioning the $n$ bits into $k$ and $(n-k)$ bits.

We see that, as advertised, the Hamming(8,4)-code gives gives $\Phi=3$ when 3 bits are cut off. However, it gives only $\Phi=2$ for bipartitions; the $\Phi$-value for bipartitions is not simply related to the Hamming distance, and is not a quantity that most popular bit string codes are optimized for. Indeed, \fig{IntegrationFig} shows that for bipartitions, it underperforms a code consisting of 16 random unique bit strings of the same length.
A rich and diverse set of codes have been published in the literature, and the state-of-the-art in terms of maximal Hamming distance for a given $n$ is continually updated \cite{HammingDistanceSite}.
Although codes with arbitrarily large Hamming distance $d$ exist, there is (just as for our Hamming(8,4)-example above) no guarantee that $\Phi$ will be as large as $d-1$ when the smaller of the two subsystems contains more than $d$ bits. Moreover, although Reed-Solomon codes are sometimes billed as classically optimal erasure codes (maximizing $d$ for a given $n$), their fundamental units are generally not bits but groups of bits (generally numbers modulo some prime number), and the optimality is violated if we make cuts that do not respect the boundaries of these bit groups. 

 \begin{figure}[pbt]
\centerline{\includegraphics[width=88mm]{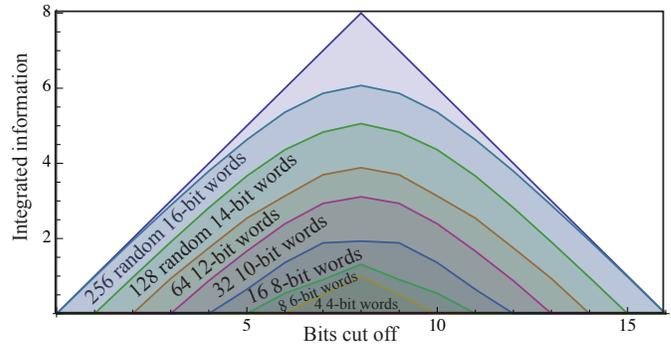}}
\caption{
Same as for previous figure, but for random codes with progressively longer bit strings, consisting of a random subset containing $\sqrt{2^n}$ of the $2^n$ possible bit strings. For better legibility, the vertical axis has been re-centered for the shorter codes.
\label{IntegrationFig2}
 }
\end{figure}

Although further research on codes maximizing $\Phi$ would be of interest, it is worth noting that simple random codes appear to give $\Phi$-values within a couple of bits of the theoretical maximum in the limit of large $n$, as illustrated in \fig{IntegrationFig2}.
When cutting off $k$ out of $n$ bits, the mutual information in classical physics clearly cannot exceed the number of bits in either subsystem, \ie, $k$ and $n-k$, so the $\Phi$-curve for a code must lie within the shaded triangle in the figure. (The quantum-mechanical case is more complicated, and we well see in the next section that it in a sense integrates both better and worse.)
The codes for which the integrated information is plotted simply consist of a random subset 
containing $2^{n/2}$ of the $2^n$ possible bit strings, so roughly speaking, half the bits encode fresh information and the other half provide the redundancy giving near-perfect integration.

\begin{figure}[pbt]
\centerline{\includegraphics[width=88mm]{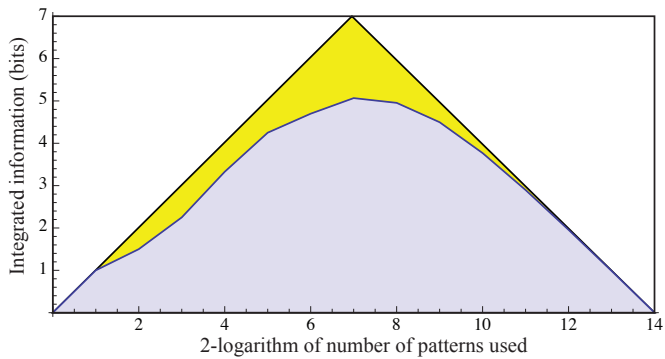}}
\caption{
The integrated information is shown for random codes using progressively larger random subsets of the $2^{14}$ possible strings of 14 bits. 
The optimal choice is seen to be using about $2^7$ bit strings, \protect\ie, using about
half the bits to encode information and the other half to integrate it.
\label{IntegrationOptimizationFig}
}
\end{figure}

Just as we saw for the Ising model example, these random codes show a tradeoff between entropy and redundancy, as illustrated in \fig{IntegrationOptimizationFig}. 
When there are $n$ bits, how many of the $2^n$ possible bit strings should we use to maximize the integrated information $\Phi$? 
If we use $m$ of them, we clearly have $\Phi\le \log_2 m$, since in classical physics, $\Phi$ cannot exceed the entropy if the system (the mutual information is $I=S_1+S_2-S$, where 
$S_1\le S$ and $S_2\le S$ so  $I\le S$). Using very few bit strings is therefore a bad idea. 
On the other hand, if we use all $2^n$ of them, we lose all redundancy, the bits become independent, and $\Phi=0$, so being greedy and using too many bit strings in an attempt to store more information is also a bad idea. 
\Fig{IntegrationOptimizationFig} shows that the optimal tradeoff is to use $\sqrt{2^n}$ of the codewords, \ie, to use half the bits to encode information and the other half to integrate it.
Taken together, the last two figures therefore suggest that $n$ physical bits can be used to provide about $n/2$ bits of integrated information in the large-n limit.

\begin{figure}[pbt]
\centerline{\includegraphics[width=88mm]{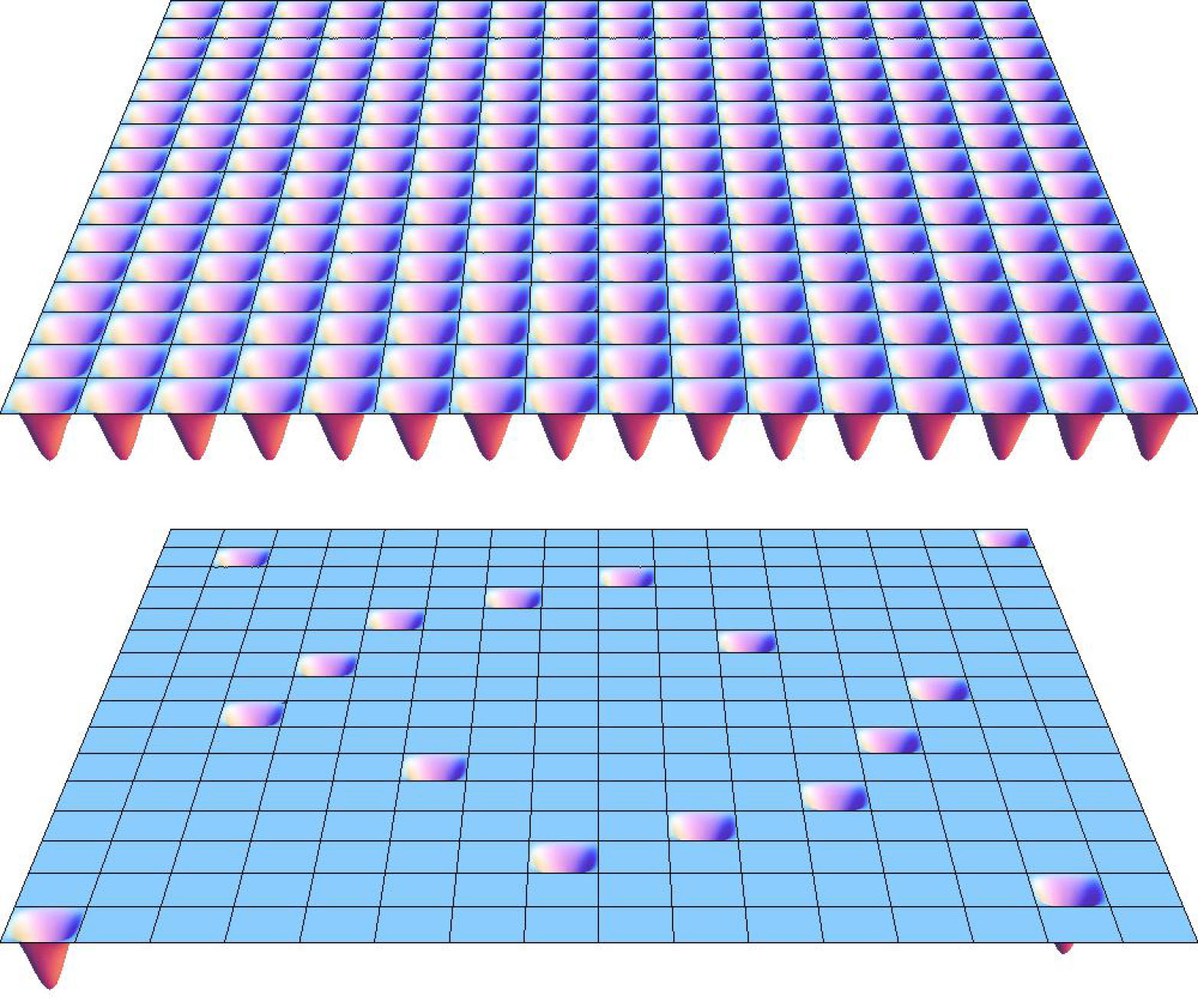}}
\caption{
A particle in the egg-crate potential energy landscape (top panel) stably encodes 8 bits of information that are completely independent of one another and therefore not integrated. In contrast, a particle in a Hamming(8,4) potential (bottom panel)  encodes only 4 bits of information, but with excellent integration. Qualitatively, a hard drive is more like the top panel, while a neural network is more like the bottom panel.
}
\label{CratesFig}
\end{figure}

\subsection{Integration in physical systems}

Let us explore the consequences of these results for physical systems described by a Hamiltonian $\H$ and a state $\rho$. 
As emphasized by Hopfield \cite{Hopfield82}, any physical system with multiple attractors can be viewed as an information storage device, since its state permanently encodes information about which attractor it belongs to.
\Fig{CratesFig} shows two examples of $\H$ interpretable as potential energy functions for a a single particle in two dimensions. 
They can both be used as information storage devices, by placing the particle  in a potential well and keeping the system cool enough that the particle stays in the same well indefinitely.
The egg crate potential $V(x,y)=\sin^2(\pi x) \sin^2(\pi y)$  (top) has 256 minima and hence a ground state entropy (information storage capacity) $S=8$ bits, whereas the lower potential has only 16 minima and  $S=4$ bits. 

The basins of attraction in the top panel are seen to be the squares shown in the bottom panel.
If we write the $x-$ and $y-$ coordinates as binary numbers with $b$ bits each, then the first 4 bits of $x$ and $y$ encode which square $(x,y)$ is in.
The information in the remaining bits encodes the location within this square; these bits are not useful for information storage because they can vary over time, as the particle oscillates around a minimum. If the system is actively cooled, these oscillations are gradually damped out and the particle settles toward the attractor solution at the minimum, at the center of its basin. 
This example illustrates that cooling is a physical example of error correction:
if thermal noise adds small perturbations to the particle position, altering the least significant bits, then cooling will remove these perturbations and push the particle back towards the minimum it came from. As long as cooling keeps the perturbations small enough that the particle never rolls out of its basin of attraction, all the 8 bits of information encoding its basin number are perfectly preserved. 
Instead of interpreting our $n=8$ data bits as positions in two dimensions, we can interpret them as positions in $n$ dimensions, where each possible state corresponds to a corner of the $n$-dimensional hypercube. This captures the essence of many computer memory devices, where each bit is stored in a system with two degenerate minima; the least significant and redundant bits that can be error-corrected via cooling now get equally distributed among all the dimensions. 

How integrated is the information $S$? For the top panel of \fig{CratesFig}, not at all: $\H$ can be factored as a tensor product of $8$ two-state systems, so
$\Phi=0$, just as for typical computer memory. In other words, if the particle is in a particular egg crate basin, knowing any one of the bits specifying the basin position tells us nothing about the other bits. 
The potential in the lower panel, on the other hand, gives good integration. This potential retains only 16 of the 256 minima, corresponding to the 16 bit strings of the Hamming(8,4)-code, which as we saw gives $\Phi=3$ for any 3 bits cut off and $\Phi=2$ bits for symmetric bipartitions. Since the Hamming distance $d=4$ for this code, at least 4 bits must be flipped to reach another minimum, which among other things implies that no two basins can share a row or column.

\subsection{The pros and cons of integration}

Natural selection suggests that self-reproducing information-processing systems will evolve integration if it is useful to them, regardless of whether they are conscious or not.
Error correction can obviously be useful, both to correct errors caused by thermal noise and to provide redundancy that improves robustness toward failure of individual physical components such as neurons. Indeed, such utility explains the preponderance of error correction built into human-developed devices, from RAID-storage to bar codes to forward error correction in telecommunications.
If Tononi is correct and consciousness requires integration, then this raises an interesting possibility: our human consciousness may have evolved as an accidental by-product of error correction. There is also empirical evidence that integration is useful for problem-solving: artificial life simulations of vehicles that have to traverse mazes and whose brains evolve by natural selection show that the more adapted they are to their environment, the higher the integrated information of the main complex in their brain \cite{Joshi13}.

However, integration comes at a cost, and as we will now see, near maximal integration appears to be prohibitively expensive.
Let us distinguish between the maximum amount of information that can be stored in a {\it state} defined by $\rho$ and the maximum amount of information that can be stored in a physical {\it system} defined by $\H$.
The former is simply $S(\rho)$ for the perfectly mixed ($\T=\infty$) state, \ie, $\log_2$ of the number of possible states (the number of bits characterizing the system).
The latter can be much larger, corresponding to $\log_2$ of the number of Hamiltonians that you could distinguish between given your  time and energy available for experimentation. Let us consider potential energy functions whose $k$ different minima can be encoded as bit strings (as in \fig{CratesFig}), and let us limit our experimentation to finding all the minima.
Then $\H$ encodes not a single string of $n$ bits, but a subset consisting of $k$ out of all $2^n$ such strings, one for each minimum.
There are $\left({2^n\atop k}\right)$ such subsets, so the information contained in $\H$ is 

\beqa{SHeq}
S(\H)&=&\log_2 \left({2^n\atop k}\right)=\log_2{2^n!\over k!(2^n-k)!}\approx\nonumber\\
&\approx&\log_2{(2^n)^k\over k^k}=k(n-\log_2 k)
\eeqa
for $k\ll 2^n$, where we used Stirling's approximation $k!\approx k^k$.
So crudely speaking, $\H$ encodes not $n$ bits but $kn$ bits.
For the near-maximal integration given by the random codes from the previous section, we had 
$k=2^{n/2}$, which gives $S(\H)\sim  2^{n/2}\>{n\over 2} $ bits.
For example, if the $n\sim 10^{11}$ neurons in your brain were maximally integrated in this way, 
then your neural network would require a dizzying
$10^{10000000000}$ bits to describe, vastly more information than can be encoded by all the $10^{89}$ particles in our universe combined.

The neuronal mechanisms of human memory are still unclear despite intensive experimental and theoretical explorations \cite{Barak13},
but there is significant evidence that the brain uses attractor dynamics in its integration and memory functions, where  discrete attractors may be used to represent discrete items \cite{Yoon13}.
The classic implementation of such dynamics as a simple symmetric and asynchronous Hopfield neural network \cite{Hopfield82} can be conveniently interpreted in terms of potential energy functions: 
the equations of the continuous Hopfield network are identical to a set of mean-field equations that minimize a potential energy function, so 
this network always converges to a basin of attraction \cite{McKayBook}.
Such a Hopfield network gives a dramatically lower information content $S(\H)$ of only about 0.25 bits per synapse\cite{McKayBook}, and we have only about $10^{14}$ synapses, suggesting that our brains can store only on the order of a few Terabytes of information.

The {\it integrated} information of a Hopfield network is even lower.
For a Hopfield network of $n$ neurons with Hebbian learning, the total number of attractors is bounded by $0.14n$ \cite{McKayBook}, so the maximum information capacity is merely $S\approx \log_2 0.14 n\approx \log_2 n\approx 37$ bits for $n=10^{11}$ neurons.
Even in the most favorable case where these bits are maximally integrated, our 
$10^{11}$ neurons thus provide a measly $\Phi\approx 37$ bits of integrated information, as opposed to about $\Phi\approx 5\times 10^{10}$ bits for a random coding. 

\subsection{The integration paradox}
\label{IntegrationParadoxSec}

This leaves us with an {\it integration paradox}: why does the information content of our conscious experience appear to be vastly larger than 37 bits?
If Tononi's information and integration principles from \Sec{IntroSec} are correct, the integration paradox 
forces us\footnote{Can we sidestep the integration paradox by simply dismissing the idea that integration is necessary? 
Although it remains controversial whether integrated information is a {\it sufficient} condition for consciousness as asserted by IIT, it appears rather obvious that it is
a {\it necessary} condition if the conscious experience is unified: if there were no integration,  the conscious mind would consist of two separate parts 
that were independent of one another and hence unaware of each other.
} to draw at least one of the following three conclusions:
\begin{enumerate}
\item Our brains use some more clever scheme for encoding our conscious bits of information, which allows dramatically larger $\Phi$ than Hebbian Hopfield networks.
\item  These conscious bits are much fewer than we might naively have thought from introspection, implying that we are only able to pay attention to a very modest amount of information at any instant.
\item To be relevant for consciousness, the definition of integrated information that we have used must be modified or supplemented by at least one additional principle. 
\end{enumerate}
We will see that the quantum results in the next section bolster the case for conclusion 3.
Interestingly, there is also support for conclusion 2 in the large psychophysical literature on the
illusion of the perceptual richness of the world. For example, there is evidence suggesting that of the roughly $10^7$ bits of information that enter our brain each second from our sensory organs, we can only be aware of  a tiny fraction, with estimates ranging from 10 to 50 bits \cite{Kupfmuller62,NorretrandersBook}.

The fundamental reason why a Hopfield network is specified by much less information than a near-maximally integrated network is that it involves only pairwise couplings between neurons, thus requiring only $\sim n^2$ coupling parameters to be specified --- as opposed to 
$2^n$ parameters giving the energy for each of the $2^n$ possible states.
It is striking how $\H$ is similarly simple for the standard model of particle physics, with the energy involving only sums of pairwise interactions between particles supplemented with occasional 3-way and 4-way couplings. $\H$ for the brain and $\H$ for fundamental physics thus both appear to belong to an extremely simple sub-class of all Hamiltonians, that require an unusually small amount of information to describe.
Just as a system implementing near-maximal integration via random coding is too complicated to fit inside the brain, it is also too complicated to work in fundamental physics: 
Since the information storage capacity $S$ of a physical system is approximately bounded by its number of particles \cite{Lloyd99} or by its area in Planck units by the Holographic principle \cite{tHooft93}, it cannot be integrated by physical dynamics that itself requires storage of the exponentially larger information quantity $S(H)\sim 2^{S/2}\>{S\over 2}$ unless the Standard Model Hamiltonian is replaced by something dramatically more complicated.
 
An interesting theoretical direction for further research (pursuing resolution 1 to the integration paradox) is therefore to investigate what maximum amount of integrated information $\Phi$ can be feasibly stored in a physical system using codes that are algorithmic (such as RS-codes) rather than random.
An interesting experimental direction would be to search for concrete implementations of error-correction algorithms in the brain.

 
In summary, we have explored the integration principle by quantifying integrated information in physical systems.
We have found that although excellent integration is possible in principle, it is more difficult in practice. 
In theory, random codes provide nearly maximal integration, with about half of all $n$ bits coding for data and the other half providing 
$\Psi\approx n$ bits of integration), but in practice, the dynamics required for implementing them is too complex for our brain or our universe.
Most of our exploration has focused on classical physics, where cuts into subsystems have corresponded to partitions of classical bits. 
As we will see in the next section, finding systems encoding large amounts of integrated information is even more challenging when we turn to the quantum-mechanical case.

\section{Independence}
\label{IndependenceSec}


\subsection{Classical versus quantum independence}

 \begin{figure}[pbt]
\centerline{\includegraphics[width=88mm]{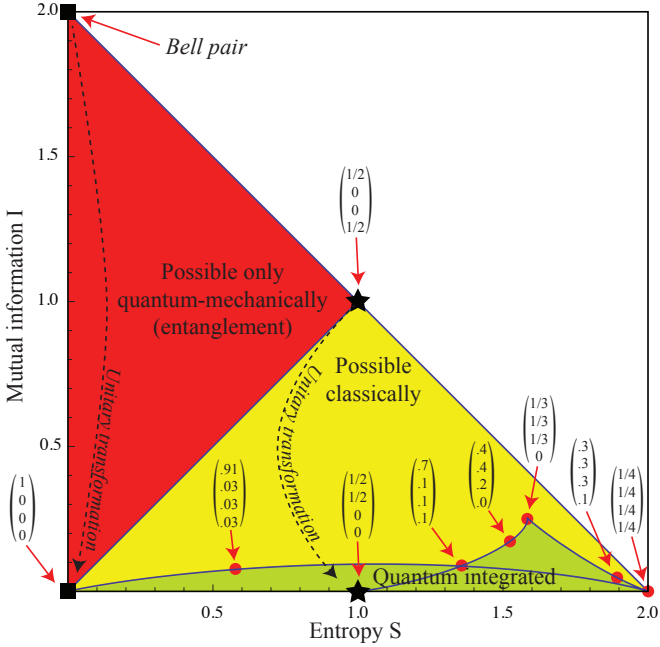}}
\caption{Mutual information versus entropy for various 2-bit systems.
The different dots, squares and stars correspond to different states, which in the classical cases are defined by the probabilities for the four basis states 00, 01 10 and 11.
Classical states can lie only in the pyramid below the upper black star with 
$(S,I)= (1,1)$, whereas entanglement allows quantum states to extend all the way up to 
the upper black square at $(0,2)$. However, the integrated information $\Phi$ for a quantum state cannot lie above the shaded green/grey region, into which any other quantum state can be brought by a unitary transformation. Along the upper boundary of this region, either three of the four probabilities are equal, or to two of them are equal while one vanishes.  
}
\label{QuantumIntegrationFig}
\end{figure}

How cruel is what Tononi calls ``the cruelest cut'', dividing a system into two parts that are maximally independent?
The situation is quite different in classical physics and quantum physics, as \fig{QuantumIntegrationFig} illustrates for a simple 2-bit system. In classical physics, the state is specified by a $2\times 2$ matrix giving the probabilities for the four states 00, 01, 10 and 11, which define an entropy $S$ and mutual information $I$. Since there is only one possible cut, the integrated information $\Phi=I$.
The point defined by the pair $(S,\Phi)$ can lie anywhere in the ``pyramid'' in the figure, who's top at $(S,\Phi)=(1,1)$ (black star) gives maximum
integration, and corresponds to perfect correlation between the two bits: 50\% probability for 00 and 11. Perfect anti-correlation gives the same point.
The other two vertices of the classically allowed region are seen to be $(S,\Phi)=(0,0)$ (100\% probability for a single outcome) and $(S,\Phi)=(2,0)$ (equal probability for all four outcomes). 

In quantum mechanics, where the 2-qubit state is defined by a $4\times 4$ density matrix, the available area in the
$(S,I)$-plane doubles to include the entire shaded triangle, with the classically unattainable region opened up because of entanglement. The extreme case is a Bell pair state such as 
\beq{BellEq}
\psiket={1\over\sqrt{2}}\left(\up\up+\down\down\right),
\eeq
which gives $(S,I)=(0,2)$. However, whereas there was only one possible cut for 2 classical bits, there are now infinitely many possible cuts because in quantum mechanics, all Hilbert space bases are equally valid, and we can choose to perform the factorization in any of them. Since $\Phi$ is defined as $I$ after the cruelest cut, it is the $I$-value minimized over all possible factorizations. For simplicity, we use the notation where $\tensormult$ denotes factorization in the coordinate basis, so the integrated information is 
\beq{QuantumPhiDefEq}
\Phi=\min_\U \> I(\U\rho\U^\dagger), 
\eeq
\ie, the mutual information minimized over all possible unitary transformations $\U$.
Since the Bell pair of \eq{BellEq} is a pure state $\rho=\psiket\psibra$, we can unitarily transform it into a basis where the first basis vector is $\psiket$, making it factorizable:
\beq{BellFactorizationEq}
\U\left(
\begin{tabular}{cccc}
${1\over 2}$		&$0$	&$0$	&${1\over 2}$\\
$0$		&$0$	&$0$	&$0$\\
$0$		&$0$	&$0$	&$0$\\
${1\over 2}$		&$0$	&$0$	&${1\over 2}$
\end{tabular}
\right)\U^\dagger
=\left(
\begin{tabular}{cccc}
$1$		&$0$	&$0$	&$0$\\
$0$		&$0$	&$0$	&$0$\\
$0$		&$0$	&$0$	&$0$\\
$0$		&$0$	&$0$	&$0$
\end{tabular}
\right)
=\left(
\begin{tabular}{cccc}
$1$		&$0$\\
$0$		&$0$
\end{tabular}
\right)
\tensormult
\left(
\begin{tabular}{cccc}
$1$		&$0$\\
$0$		&$0$
\end{tabular}
\right).
\eeq
This means that $\Phi=0$, so in quantum mechanics, the cruelest cut can be very cruel indeed:
the most entangled states possible in quantum mechanics have no integrated information at all! 

The same cruel fate awaits the most integrated 2-bit state from classical physics:
the perfectly correlated mixed state
$\rho ={1\over 2}\up\upbra + {1\over 2}\down\downbra$. 
It gave $\Phi=1$ bit classically above (upper black star in the figure), but a unitary transformation permuting its diagonal elements makes it factorable:
\beq{MixtureFactorizationEq}
\U\left(
\begin{tabular}{cccc}
${1\over 2}$		&$0$	&$0$	&$0$\\
$0$		&$0$	&$0$	&$0$\\
$0$		&$0$	&$0$	&$0$\\
$0$		&$0$	&$0$	&${1\over 2}$
\end{tabular}
\right)\U^\dagger
=\left(
\begin{tabular}{cccc}
${1\over 2}$		&$0$	&$0$	&$0$\\
$0$		&${1\over 2}$	&$0$	&$0$\\
$0$		&$0$	&$0$	&$0$\\
$0$		&$0$	&$0$	&$0$
\end{tabular}
\right)
=\left(
\begin{tabular}{cccc}
$1$		&$0$\\
$0$		&$0$
\end{tabular}
\right)
\tensormult
\left(
\begin{tabular}{cccc}
${1\over 2}$		&$0$\\
$0$		&${1\over 2}$
\end{tabular}
\right),
\eeq
so $\Phi=0$ quantum-mechanically (lower black star in the figure).

\subsection{Canonical transformations, independence and relativity}

The fundamental reason that these states are more separable quantum-mechanically is clearly that more cuts are available, making the cruelest one crueler. Interestingly, the same thing can happen also in classical physics. Consider, for example, our example of the deuterium atom from \eq{deuteriumEq}. 
When we restricted our cuts to simply separating different degrees of freedom, we found  
that the group $(\r_p,\p_p,\r_n,\p_n)$ was quite (but not completely) independent of the group  
$(\r_e,\p_e)$, and that there was no cut splitting things into perfectly independent pieces.
In other words, the nucleus was fairly independent of the electron, but none of the three particles was completely independent of the other two.
However, if we allow our degrees of freedom to be transformed before the cut, then things can be split into two perfectly independent parts! 
The classical equivalent of a unitary transformation is of course a canonical transformation (one that preserves phase-space volume). If we perform the canonical transformation where the new coordinates 
are the center-of-mass position $\r_M$ and the relative displacements $\r_p'\equiv\r_p-\r_M$ and 
$\r_e'\equiv\r_e-\r_M$, and correspondingly define $\p_M$ as the total momentum of the whole system, \etc,
 then we find that $(\r_M,\p_M)$ is completely independent of the rest. 
In other words, the average motion of the entire deuterium atom is completely decoupled from the internal motions around its center-of-mass. 

Interestingly,  this well-known possibility of decomposing any isolated system into average and relative motions 
(the {\it ``average-relative decomposition"}, for short) is equivalent to relativity theory in the following sense.
The core of relativity theory is that all laws of physics (including the speed of light) are the same in all inertial frames. 
This implies the average-relative decomposition, since the laws of physics governing the relative motions of the system are the same in all inertial frames and hence independent of the (uniform) center-of-mass motion.
Conversely, we can view relativity as a special  case of the average-relative decomposition. 
If two systems are completely independent, then they can gain no knowledge of each other, so a conscious observer in one will be unaware of the other. The average-relative decomposition therefore implies that an observer in an isolated system has no way of knowing whether she is at rest or in uniform motion, because these are simply two different allowed states for the center-of-mass subsystem, which is completely independent from (and hence inaccessible to) the internal-motions subsystem of which her consciousness is a part.
 


\subsection{How integrated can quantum states be?}

We saw in \fig{QuantumIntegrationFig} that some seemingly integrated states, such as a Bell pair or a pair of classically perfectly correlated bits, are in fact not integrated at all.
But the figure also shows that some states are truly integrated even quantum-mechanically, with $I>0$ even for the cruelest cut. How integrated can a quantum state be? 
The following theorem, proved by Jevtic,  Jennings \& Rudolph \cite{Jevtic11}, enables the answer to be straightforwardly 
calculated\footnote{The converse of the $\rho$DC is straightforward to prove: if $\Phi=0$ (which is equivalent to the state being factorizable; $\rho=\rho_1\tensormult\rho_2$), then it is factorizable also in its eigenbasis where both $\rho_1$ and $\rho_2$ are diagonal.}:
\smallskip
\centerline{\framebox{\parbox{7cm}{
{\bf $\rho$-Diagonality Theorem ($\rho$DC):}\\ {\it The mutual information always takes its minimum in a basis where $\rho$ is diagonal}
}}}
\smallskip


The first step in computing the integrated information $\Phi(\rho)$ is thus to diagonalize the $n\times n$ density matrix $\rho$. If all $n$ eigenvalues are different, then there are $n!$ possible ways of doing this, corresponding to the $n!$ ways of permuting the eigenvalues, so the $\rho$DC simplifies the continuous minimization problem of \eq{QuantumPhiDefEq} to a discrete minimization problem over these $n!$ permutations.
Suppose that $n=l\times m$, and that we wish to factor the $n$-dimensional Hilbert space into factor spaces of dimensionality $l$ and $m$, so that $\Phi=0$. 
It is easy to see that this is possible if the $n$ eigenvalues of $\rho$ can be arranged into an $l\times m$ matrix that is multiplicatively {\it separable} (rank 1), \ie, the product of a column vector and a row vector. Extracting the eigenvalues for our example from \eq{MixtureFactorizationEq} where $l=m=2$ and $n=4$, we see that 
$$
\left(
\begin{tabular}{cc}
${1\over 2}$		&${1\over 2}$\\
$0$		&$0$
\end{tabular}
\right)\hbox{ is separable, but}
\quad
\left(
\begin{tabular}{cc}
${1\over 2}$		&$0$\\
$0$		&${1\over 2}$
\end{tabular}
\right)\hbox{ is not},
$$
and that the only difference is that the order of the four numbers has been permuted.
More generally, we see that to find the ``cruelest cut'' that defines the integrated information $\Phi$, we want to find the permutation that makes the matrix of eigenvalues as separable as possible.
It is easy to see that when seeking the permutation giving maximum separability, we 
can without loss of generality place the largest eigenvalue first (in the upper left corner) and the smallest one last (in the lower right corner). If there are only 4 eigenvalues (as in the above example), the ordering of the remaining two has no effect on $I$. 

\subsection{The quantum integration paradox}

We now have the tools in hand to answer the key question from the last section: which state $\rho$ maximizes the integrated information $\Phi$?
Numerical search suggests that the most integrated state is a rescaled {\it projection matrix}
satisfying $\rho^2\propto\rho$. This means that some number $k$ of the $n$ eigenvalues equal $1/k$ and the remaining ones vanish.\footnote{A heuristic way of understanding why having many equal eigenvalues is advantageous is that it helps eliminate the effect of the eigenvalue permutations that we are minimizing over. If the optimal state has two distinct eigenvalues, then if swapping them changes $I$, it must by definition increase $I$ by some finite amount. This suggests that we can increase the integration $\Phi$ by bringing the eigenvalues infinitesimally closer or further apart, and repeating this procedure lets us further increase $\Phi$ until all eigenvalues are either zero or equal to the same positive constant.}
For the $n=4$ example from \fig{QuantumIntegrationFig}, $k=3$ is seen to give the best integration, with eigenvalues (probabilities)  $1/3$, $1/3$, $1/3$ and $0$, giving $\Phi=\log(27/16)/\log(8)\approx 0.2516$.

For classical physics, we saw that the maximal attainable $\Phi$ grows roughly linearly with $n$. Quantum-mechanically, however, it {\it decreases} 
as $n$ increases!\footnote{One finds that $\Phi$ is maximized when the $k$ identical nonzero eigenvalues are arranged in a Young Tableau, which corresponds to a partition of $k$ as a sum of positive integers $k_1+k_2+...$, giving $\Phi=S(\p)+S(\p^*)-\log_2 k$, where the probability vectors $\p$ and $\p^*$ are defined by $p_i=k_i/k$ and $p^*_i=k_i^*/k$. Here $k_i^*$ denotes the conjugate partition. For example, if we cut an even number of qubits into two parts with $n/2$ qubits each, then $n=2, 4, 6, ..., 20$ gives $\Phi\approx 0.252, 0.171, 0.128, 0.085, 0.085, 0.073, 0.056, 0.056, 0.051$ and $0.042$ bits, respectively. 
}

In summary, no matter how large a quantum system we create, its state can never contain more than about a quarter of a bit of integrated information! This exacerbates the integration paradox from \Sec{IntegrationParadoxSec}, eliminating both of the first two resolutions: 
you are clearly aware of more than $0.25$ bits of information right now, and this quarter-bit maximum applies not merely to states of Hopfield networks, but to {\it any} quantum states of {\it any system}.
Let us therefore begin exploring the third resolution: that our definition of integrated information must be modified or supplemented by at least one additional principle. 

\subsection{How integrated is the Hamiltonian?}

An obvious way to begin this exploration is to consider the state $\rho$ not merely at a single fixed time $t$, but  as a function of time. 
After all, it is widely assumed that consciousness is related to information {\it processing}, not mere information {\it storage}. Indeed, Tononi's original $\Phi$-definition \cite{TononiManifesto} (which applies to classical neural networks rather than general quantum systems) involves time, depending on the extent to which current events affect future ones.

Because the time-evolution of the state $\rho$ is determined by the Hamiltonian $\H$ via the 
Schr\"odinger equation
\beq{SchrodingerEq}
\dot\rho=i[\H,\rho],
\eeq
whose solution is
\beq{UnitaryEvolEq}
\rho(t) = e^{i\H t}\rho e^{-i\H t},
\eeq
we need to investigate the extent to which the cruelest cut can decompose not merely $\rho$ but the pair $(\rho,H)$ into independent parts. 
(Here and throughout, we often use units where $\hbar=1$ for simplicity.)

\subsection{Evolution with separable Hamiltonian}

As we saw above, the key question for $\rho$ is whether it it is {\it factorizable} (expressible as {\it product}  $\rho=\rho_1\tensormult\rho_2$ of matrices acting on the two subsystems), whereas the key question for $\H$ is whether it is what we will call {\it additively separable}, being a {\it sum} of matrices acting on the two subsystems, \ie, expressible in the form 
\beq{HseparabilityDefEq}
\H=\H_1\tensormult\I + \I\tensormult\H_2
\eeq
for some matrices $\H_1$ and $\H_2$.
For brevity, we will often write simply {\it separable} instead of additively separable. 
As mentioned in \Sec{IntegrationInformationSec}, a separable Hamiltonian $\H$ implies that both the thermal state $\rho\propto e^{-\H/kT}$ and the time-evolution operator $\U\equiv e^{i\H t/\hbar}$ are factorizable.
An important property of density matrices which was pointed out already by von Neumann when he invented them 
\cite{vonNeumann32}
is that if $\H$ is separable, 
then
\beq{SeparablerhodotEq}
\dot\rho_1=i[\H_1,\rho_1],
\eeq
\ie, the time-evolution of the state of the first subsystem,
$\rho_1\equiv\tr_2\rho$, 
is independent of the other subsystem and of any entanglement with it that may exist.
This is easy to prove:
Using the identities\eqn{identity10} and\eqn{identity12} shows that
\beqa{rho1dotProofEq1}
\trace_2[\H_1\tensormult\I,\rho]&=&\trace_2\{(\H_1\tensormult\I)\rho\} - \trace_2\{\rho(\H_1\tensormult\I)\}\nonumber\\
&=&\H_1\rho_1-\rho_1\H_2=[\H_1,\rho_1].
\eeqa
Using the identity\eqn{identity8} shows that
\beq{rho1dotProofEq2}
\trace_2[\I\tensormult\H_2,\rho]=0.
\eeq
Summing equations\eqn{rho1dotProofEq1} and\eqn{rho1dotProofEq2} completes the proof. 

\subsection{The cruelest cut as the maximization of separability}
\label{PhysicsFromSeparabilitySec}

Since a general Hamiltonian $\H$ cannot be written in the separable form of \eq{HseparabilityDefEq}, it will 
also include a third term $\H_3$ that is non-separable.
The independence principle from \Sec{IntroSec} therefore suggests an interesting mathematical approach to the physics-from-scratch problem of analyzing the total Hamiltonian $\H$ for our physical world:
\begin{enumerate}
\item Find the Hilbert space factorization giving the ``cruelest cut'', decomposing $\H$ into parts with the smallest interaction Hamiltonian $\H_3$ possible.
\item Keep repeating this subdivision procedure for each part until only relatively integrated parts remain that cannot be further decomposed with a small interaction Hamiltonian.
\end{enumerate}
The hope would be that applying this procedure to the Hamiltonian of our standard model would reproduce the full observed object hierarchy from \fig{HierarchyFig}, with the factorization corresponding to the objects, and the various non-separable terms $\H_3$ describing the 
interactions between these objects. 
Any decomposition with $\H_3=0$ would correspond to two parallel universes unable to communicate with one another. 

We will now formulate this as a rigorous mathematics problem, solve it, and derive the observational consequences. We will find that this 
approach fails catastrophically when confronted with observation, giving interesting hints regarding further physical principles needed for understanding why we perceive our world as an object hierarchy. 

\subsection{The Hilbert-Schmidt vector space}
\label{HilbertSchmidtSec}

To enable a rigorous formulation of our problem, let us first briefly review the Hilbert-Schmidt vector space, a convenient inner-product space
where the vectors are not wave functions $\psiket$ but matrices such as $\H$ and $\rho$.
For  any two matrices $\A$ and $\B$, the Hilbert-Schmidt inner product is defined by 
\beq{HilbertSchmidtEq}
(\A,\B)\equiv\tr\A^\dagger\B.
\eeq
For example, the trace operator can be written as an inner product with the identity matrix:
\beq{HilbertSchmidtTraceEq}
\tr\A = (\I,\A).
\eeq
This inner product defines the Hilbert-Schmidt norm (also known as the Frobenius norm)
\beq{HilbertSchmidtNormEq}
||\A||\equiv (\A,\A)^{1\over 2} = (\tr\A^\dagger\A)^{1\over 2} = \left(\sum_{ij} |A_{ij}|^2\right)^{1\over 2}.
\eeq
If $\A$ is Hermitian ($\A^\dagger=\A$), then $||\A||^2$ is simply the sum of the squares of its eigenvalues.

Real symmetric and antisymmetric matrices form orthogonal subspaces under the Hilbert-Schmidt inner product, since 
$(\S,\A )=0$
for any symmetric matrix $\S$ (satisfying $\S^t=\S)$ and any antisymmetric matrix $\A$  (satisfying $\A^t=-\A$).
Because a Hermitian matrix (satisfying $\H^\dagger=\H$) can be written in terms of real symmetric and antisymmetric matrices as $\H=\S+i\A$, we have 
$$(\H_1,\H_2)=(\S_1,\S_2)+(\A_1,\A_2),$$
which means that the inner product of two Hermitian matrices is purely real.

\subsection{Separating $\H$ with orthogonal projectors}
\label{ProjectorSec}

By viewing $\H$ as a vector in the Hilbert-Schmidt vector space, we can rigorously define an decomposition of it into orthogonal components, two of which are the separable terms from \eq{HseparabilityDefEq}.
Given a factorization of the Hilbert space where the matrix $\H$ operates, we define four linear superoperators\footnote{Operators on the Hilbert-Schmidt space are usually called {\it superoperators} in the literature, to avoid confusions with operators on the underlying Hilbert space, which are mere vectors in the Hilbert-Schmidt space.}
 $\PP_i$ as follows:  
\beqa{ProjectorEq}
\PP_0 \H &\equiv&{1\over n}(\tr \H)\>\I\\
\PP_1 \H &\equiv&\left({1\over n_2}\trace_2 \H\right)\tensormult\I_2 -  \PP_0\H\\
\PP_2 \H &\equiv&\I_1\tensormult\left({1\over n_1}\trace_1 \H\right) -\PP_0\H\\
\PP_3 \H &\equiv&(\I-\PP_1-\PP_2-\PP_3) \H\label{Projector3Eq}
\eeqa
It is straightforward to show that these four linear operators $\PP_i$ form a complete set of orthogonal projectors, \ie, that
\beqa{OrthogonalityEq}
\sum_{i=0}^3\PP_i &=& \I,\\
\PP_i \PP_j&=&\PP_i \delta_{ij},\\
(\PP_i \H,\PP_j \H)&=& ||\PP_i \H||^2 \delta_{ij}.
\eeqa
This means that any Hermitian matrix $\H$ can be decomposed as a sum of  four orthogonal components $\H_i\equiv\PP_i\H$, so that its squared Hilbert-Schmidt norm can be decomposed as a sum of contributions from the four components:
\beqa{HnormDecompEq1}
\H&=&\H_0 + \H_1 + \H_2 + \H_3,\label{DecompDefEq}\\ 
\H_i&\equiv&\PP_i\H,\\
(\H_i,\H_j)&=& ||\H_i||^2 \delta_{ij},\\
||\H||^2&=&||\H_0||^2  \ns +\ns  ||\H_1||^2 \ns +\ns ||\H_2||^2 \ns+\ns ||\H_3||^2.\label{NormDecompEq}
\eeqa
We see that $\H_0\propto\I$ picks out the trace of $\H$, whereas the other three matrices are trace-free.
This trace term is of course physically uninteresting, since it can be eliminated by simply adding an unobservable constant zero-point energy to the Hamiltonian. 
$\H_1$ and $\H_2$ corresponds to the two separable terms in \eq{HseparabilityDefEq} (without the trace term, which could have been arbitrarily assigned to either), and $\H_3$ corresponds to the non-separable residual.
A Hermitian matrix $\H$ is therefore separable if and only if $\PP_3\H=0$.
Just as it is customary to write the norm or a vector $\r$ by $r\equiv|\r|$ (without boldface), we will denote 
the Hilbert-Schmidt norm of a matrix $\H$ by $H\equiv ||\H||$. For example, with this notation we can rewrite \eq{NormDecompEq} as simply $H^2=H_0^2+H_1^2+H_2^2+H_3^2$.

Geometrically, we can think of $n\times n$ Hermitian matrices $\H$ as points in the $N$-dimensional vector space $R^N$, where $N=n\times n$ (Hermiteal matrices have $n$ real numbers on the diagonal and $n(n-1)/2$ complex numbers off the diagonal, constituting a total of 
$n+2\times n(n-1)/2=n^2$ real parameters).
Diagonal matrices form a hyperplane of dimension $n$ in this space. 
The projection operators $\PP_0$, $\PP_1$, $\PP_2$ and $\PP_3$ project onto hyperplanes of dimension
$1$, $(n-1)$, $(n-1)$ and $(n-1)^2$, respectively, so separable matrices form a hyperplane in this space of dimension $2n-1$.
For example, a general $4\times 4$ Hermitian matrix can be parametrized by 10 numbers (4 real for the diagonal part and 6 complex for the off-diagonal part), and its decomposition from \eq{DecompDefEq} can be written as follows:
\beqa{HdecompExampleEq1}
\H&=&\left(
\begin{tabular}{cccc}
$t\ms+\ms a\ms+\ms b\ms+\ms v$	&$d\ms+\ms w$	&$c\ms+\ms x$		&$y$\\
$d^*\ms+\ms w^*$	&$t\ms+\ms a\ms-\ms b\ms-\ms v$	&$z$		&$c\ms-\ms x$\\
$c^*\ms+\ms x^*$		&$z^*$		&$t\ms-\ms a\ms+\ms b\ms-\ms v$	&$d\ms-\ms w$\\
$y^*$		&$c^*\ms-\ms x^*$	&$d^*\ms-\ms w^*$	&$t\ms-\ms a\ms-\ms b\ms+\ms v$\\
\end{tabular}
\right)=\nonumber\\
&=&
\left(\ms
\begin{tabular}{cccc}
$t$&$0$&$0$&$0$\\
$0$&$t$&$0$&$0$\\
$0$&$0$&$t$&$0$\\
$0$&$0$&$0$&$t$
\end{tabular}
\ms\right)
\ns+\ns
\left(\ms
\begin{tabular}{cccc}
$a$&$0$&$c$&$0$\\
$0$&$a$&$0$&$c$\\
$c^*$&$0$&$-a$&$0$\\
$0$&$c^*$&$0$&$-a$
\end{tabular}
\ms\right)
\ns+\ns
\left(\ms
\begin{tabular}{cccc}
$b$&$d$&$0$&$0$\\
$d^*$&$-b$&$0$&$0$\\
$0$&$0$&$b$&$d$\\
$0$&$0$&$d^*$&$-b$
\end{tabular}
\ms\right)
\ns+\ns
\nonumber\\
&+&
\left(
\begin{tabular}{cccc}
$v$		&$w$	&$x$		&$y$\\
$w^*$	&$-v$	&$z$		&$-x$\\
$x^*$	&$z^*$	&$-v$	&$-w$\\
$y^*$	&$-x^*$	&$-w^*$	&$v$\\
\end{tabular}
\right)
\eeqa
We see that $t$ contributes to the trace (and $\H_0$) while the other three components $\H_i$ are traceless. We also see that $\tr_1\H_2=\tr_2\H_1 = 0$, and that both partial traces vanish for $\H_3$.

\subsection{Maximizing separability}

We now have all the tools we need to rigorously maximize separability and test the physics-from-scratch approach described in \Sec{PhysicsFromSeparabilitySec}.
Given a Hamiltonian $\H$, we simply wish to minimize the norm of its non-separable component $\H_3$ over all possible Hilbert space factorizations, \ie, over all possible unitary transformations.
In other words, we wish to compute
\beq{EnergyIntegrationEq}
\Eint\equiv \min_U ||\PP_3\H||,
\eeq
where we have defined the {\it integration energy} $\Eint$ by analogy with the 
integrated information $\Phi$.
If $\Eint=0$, then there is a basis where our system separates into two parallel universes, otherwise 
$\Eint$ quantifies the coupling between the two parts of the system under the cruelest cut.

\begin{figure}[phbt]
\centerline{\includegraphics[width=88mm]{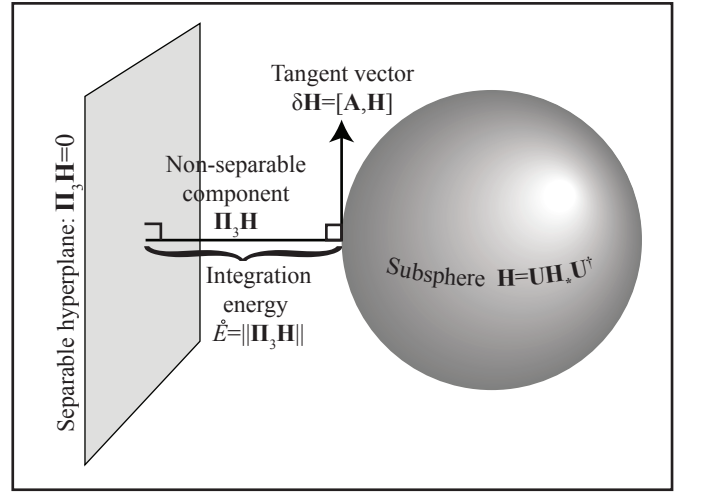}}
\caption{Geometrically, we can view the integration energy as the shortest distance (in Hilbert-Schmidt norm) between the hyperplane of separable Hamiltonians and a subsphere of Hamiltonians that can be unitarily transformed into one another. The most separable Hamiltonian $\H$ on the subsphere is such that its non-separable component $\PP_3$ is orthogonal to all subsphere tangent vectors $[\A,\H]$ generated by anti-Hermitian matrices $\A$.
}
\label{SubsphereFig}
\end{figure}

The Hilbert-Schmidt space allows us to interpret the minimization problem of \eq{EnergyIntegrationEq} geometrically, as illustrated in \fig{SubsphereFig}.
Let $\H^*$ denote the Hamiltonian in some given basis, and 
consider its orbit $\H=\U\H\U^\dagger$ under all unitary transformations $\U$.
This is a curved hypersurface whose dimensionality is generically $n(n-1)$, \ie, $n$ lower than that of the full space of Hermitian matrices,  since unitary transformation leave all $n$ eigenvalues invariant.\footnote{$n\times n$-dimensional Unitary matrices $\U$ are known to form an $n\times n$-dimensional manifold:
they can always be written as $\U=e^{i\H}$ for some Hermitian matrix $\H$, so they are parametrized by the same number of real parameters ($n\times n$) as Hermitian matrices.}
We will refer to this curved hypersurface as a {\it subsphere}, because it is a subset of the full $n^2$-dimensional sphere:  the radius $H$ (the Hilbert-Schmidt norm $||\H||$) is invariant under unitary transformations, but the subsphere may have a more complicated topology than a hypersphere; for example, the 3-sphere is known to topologically be the double cover of SO(3), the matrix group of $3\times 3$ orthonormal transformations.

We are interested in finding the most separable point $\H$ on this subsphere, \ie, the point on the subsphere that is closest to the $(2n-1)$-dimensional separable hyperplane. 
In our notation, this means that we want to find the point $\H$ on the subsphere that minimizes $||\PP_3\H||$, 
the Hilbert-Schmidt norm of the non-separable component.
If we perform infinitesimal displacements along the subsphere, $||\PP_3\H||$ thus remains constant to first order (the gradient vanishes at the minimum), 
so all tangent vectors of the subsphere are
orthogonal to $\PP_3\H$, the vector from the separable hyperplane to the subsphere.

Unitary transformations are generated by anti-Hermitian matrices, so the most general tangent vector 
$\delta \H$ is of the form 
\beq{TangentVectorEq}
\delta\H=[\A,\H]\equiv \A\H-\H\A
\eeq
for some anti-Hermitian  $n\times n$ matrix $\A$ (any matrix satisfying $\A^\dagger=-\A$).
We thus obtain the following simple condition for maximal separability:
\beq{OptimalityConditionEq}
(\PP_3\H, [\A,\H]) = 0
\eeq
for any anti-Hermitian matrix $\A$.
Because the most general anti-Hermitian matrix can be written 
as $\A=i\B$ for a Hermitian matrix $\B$,
\eq{OptimalityConditionEq} is equivalent to the condition 
$(\PP_3\H, [\B,\H]) = 0$ for all Hermitian matrices $\B$.
Since there are $n^2$ anti-Hermitian matrices, \eq{OptimalityConditionEq} is a system of $n^2$ coupled quadratic equations that the components of $\H$ must obey.


\subsection{The Hamiltonian diagonality theorem}

Analogously to the above-mentioned $\rho$-diagonality theorem, we will now prove that maximal separability is attained in the eigenbasis.

\smallskip
\centerline{\framebox{\parbox{7cm}{
{\bf $\H$-Diagonality Theorem (HDT):}\\ {\it The Hamiltonian is always maximally separable (minimizing $||\H_3||$) in the energy eigenbasis where it is diagonal.}
}}}
\smallskip

As a preliminary, let us first prove the following:
\smallskip

{\bf Lemma 1:} {\it
For any Hermitian positive semidefinite matrix $\H$, there is a diagonal matrix $\H_*$ 
giving the same subsystem eigenvalue spectra,
$\lambdavec(\PP_1\H_*)=\lambdavec(\PP_1\H)$, 
$\lambdavec(\PP_2\H_*)=\lambdavec(\PP_2\H)$, and 
whose eigenvalue spectrum is majorized by that of $\H$, \ie, 
$\lambdavec(\H)\succ\lambdavec(\H_*)$.
}

{\bf Proof:} Define the matrix $\H'\equiv\U\H\U^\dagger$, where $\U\equiv\U_1\tensormult\U_2$, and $\U_1$ and $\U_2$ are unitary matrices diagonalizing the partial trace matrices 
$\tr_2\H$ and $\tr_1\H$, respectively. This implies that $\tr_1\H'$ and $\tr_2\H'$ are diagonal, and 
$\lambdavec(\H')=\lambdavec(\H)$.
Now define the matrix $\H_*$ to be $\H'$ with all off-diagonal elements set to zero.
Then $\tr_1\H_*=\tr_1\H'$ and $\tr_2\H_*=\tr_2\H'$, so 
$\lambdavec(\PP_1\H_*)= \lambdavec(\PP_1\H)$ and $\lambdavec(\PP_2\H_*)= \lambdavec(\PP_2\H)$.
Moreover, since the eigenvalues of any Hermitian positive semidefinite matrix majorize its diagonal elements \cite{MarshallBook}, $\lambdavec(\H_*)\prec \lambdavec(\H')= \lambdavec(\H)$, which completes the proof.

\smallskip

{\bf Lemma 2:} 
{\it The set $S(\H)$ of all diagonal matrices whose diagonal elements are majorized by the vector $\lambdavec(\H)$ is a convex subset of the subsphere, with boundary points on the surface of the subsphere that are diagonal matrices with all permutations of $\lambdavec(\H)$.}

{\bf Proof:}
Any matrix $\H_*\in S(\H)$ must lie either on the subsphere surface or in its interior, because of the well-known result that for any two positive semidefinite Hermitian matrices of equal trace, 
the majorization condition
$\lambdavec(\H_*)\prec\lambdavec(\H)$ 
is equivalent to the former lying in the convex hull of the unitary orbit of the latter \cite{Bravyi03}:
$\H_*=\sum_i p_i\U_i\H\U_i^\dagger$, $p_i\ge 0$, $\sum_i p_i=1$, $\U_i\U_i^\dagger=\I$.
$S(\H)$ contains the above-mentioned boundary points, because they can be written as 
$\U\H\U^\dagger$ for all unitary matrices $\U$ that diagonalize $\H$, and for a diagonal 
matrix, the corresponding $\H_*$ is simply the matrix itself.
The set $S(\H)$ is convex, because the convexity condition that $p\lambdavec_1+(1-p)\lambdavec_2\succ\lambdavec$
if $\lambdavec_1\succ\lambdavec$, $\lambdavec_2\succ\lambdavec$, $0\le p\le 1$
follows straight from the definition of $\succ$.

\smallskip

{\bf Lemma 3:} {\it
The function $f(\H)\equiv ||\PP_1\H||^2 + ||\PP_2\H||^2$ is convex, \ie, satisfies 
$f(p_a\H_a+p_b\H_b)\le p_a f(\H_a)+p_b f(\H_b)$ for any constants satisfying 
$p_a\ge 0$, $p_b\ge 0$, $p_a+p_b=1$.
}

{\bf Proof:}
If we arrange the elements of $\H$ into a vector $\h$ and denote the action of the superoperators 
$\PP_i$ on $\h$ by matrices $\P_i$, then
$f(\H)=|\P_1\h|^2+|\P_2\h|^2 = \h^\dagger(\P_1^\dagger\P_1+\P_2^\dagger\P_2)\h$. Since the matrix in parenthesis is symmetric and positive semidefinite, the function $f$ is a positive semidefinite quadratic form and hence convex.

\bigskip

We are now ready to prove the $\H$-diagonality theorem.
This is equivalent to proving that $f(\H)$ takes its maximum value on the subsphere in 
\fig{SubsphereFig} for a diagonal $\H$:
since both $||\H||$ and $||\H_0||$ are unitarily invariant, 
minimizing $||\H_3||^2=||\H||^2-||\H_0||^2-f(\H)$ is equivalent to maximizing $f(\H)$.

Let $O(\H)$ denote the subphere, \ie, the unitary orbit of $\H$.
By Lemma 1, for every $\H\in O(H)$, there is an $\H_*\in S(\H)$ such that 
$f(\H)=f(\H_*)$.
If $f$ takes its maximum over $S(\H)$ at a point $\H_*$ which also belongs to $O(\H)$, then
this is therefore also the maximum of $f$ over $O(\H)$.
Since the function $f$ is convex (by Lemma 3) and the set $S(\H)$ is convex (by Lemma 2), $f$ cannot have any local maxima within the set and must take its maximum value at at least one point on the boundary of the set.
As per Lemma 2, these boundary points are diagonal matrices with all permutations of the eigenvalues of $\H$, 
so they also belong to $O(\H)$ and therefore constitute maxima of $f$ over the subsphere.
In other words, the Hamiltonian is always maximally separable in its energy eigenbasis, q.e.d. 

This result holds also for Hamiltonians with negative eigenvalues, since we can  make all eigenvalues positive by adding an $\H_0$-component without altering the optimization problem. 
In addition to the diagonal optimum, there will generally be other bases with identical values of $||\H_3||$, corresponding to separable unitary transformations of the diagonal optimum.

We have thus proved that separability is always maximized in the energy eigenbasis, 
where the $n\times n$ matrix $\H$ is diagonal and the projection operators $\PP_i$ defined by equations\eqn{ProjectorEq}-(\ref{Projector3Eq})
greatly simplify.
If we arrange the $n=lm$ diagonal elements of $\H$ into an $l\times m$ matrix $H$, then the action of the linear operators $\PP_i$ is given by simple matrix operations:
\beqa{DiagProjectorEq1}
H_0&\equiv Q_l H Q_m,\\
H_1&\equiv P_l H Q_m,\\
H_2&\equiv Q_l H P_m,\\
H_3&\equiv P_l H P_m,
\eeqa
where
\beqa{DiagProjectorEq2}
P_k&\equiv& I-Q_k,\\
(Q_k)_{ij}&\equiv& {1\over k}
\eeqa
are $k\times k$ projection matrices satisfying 
$P_k^2= P_k$, $Q_k^2=Q_k$, $P_k Q_k=Q_k P_k=0$, $P_k+Q_k=I$.
(To avoid confusion, we are using boldface for $n\times n$ matrices and plain font for smaller matrices involving only the eigenvalues.)
For the $n=2\times 2$ example of  \eq{HdecompExampleEq1}, we have 
\beq{DiagProjectorExampleEq1}
\bgroup
\def\arraystretch{1.3}
P_2=\left(
\begin{tabular}{cccc}
${1\over 2}$		&${1\over 2}$\\
${1\over 2}$		&${1\over 2}$
\end{tabular}
\right),
\quad
Q_2={1\over 2}\left(
\begin{tabular}{cccc}
${1\over 2}$		&$-{1\over 2}$\\
$-{1\over 2}$		&${1\over 2}$
\end{tabular}
\right),
\egroup
\eeq
and a general diagonal $\H$ is decomposed into four terms $H=H_0+H_1+H_2+H_3$ as follows:
\beq{DiagProjectorExampleEq2}
H=
\left(
\begin{tabular}{cccc}
$t$		&$t$\\
$t$		&$t$
\end{tabular}
\right)
+
\left(
\begin{tabular}{cccc}
$a$		&$a$\\
$-a$		&$-a$
\end{tabular}
\right)
+
\left(
\begin{tabular}{cccc}
$b$		&$-b$\\
$b$		&$-b$
\end{tabular}
\right)
+
\left(
\begin{tabular}{cccc}
$v$		&$-v$\\
$-v$		&$v$
\end{tabular}
\right).
\eeq
As expected, only the last matrix is non-separable, and  the row/column sums vanish for the two previous matrices, corresponding to vanishing partial traces.

Note that we are here choosing the $n$ basis states of the full Hilbert space to be products of basis states from the two factor spaces.
This is without loss of generality, since any other basis states can be transformed into such product states by a unitary transformation.

Finally, note that the theorem above applies only to exact finite-dimensional Hamiltonians, not to approximate discretizations of infinite-dimensional ones such as are frequently employed in physics.  If $n$ is not factorizable, the $\H$-factorization problem can be rigorously mapped onto a physically indistinguishable one with a slightly larger factorizable $n$ by setting the corresponding new rows and columns of the density matrix $\rho$ equal to zero, so that the new degrees of freedom are all frozen out --- we will discuss this idea in more detail in in \Sec{BumpEqSec}.

\subsection{Ultimate independence and the Quantum Zeno paradox}
\label{ZenoSec}

\begin{figure}[phbt]
\centerline{\includegraphics[width=88mm]{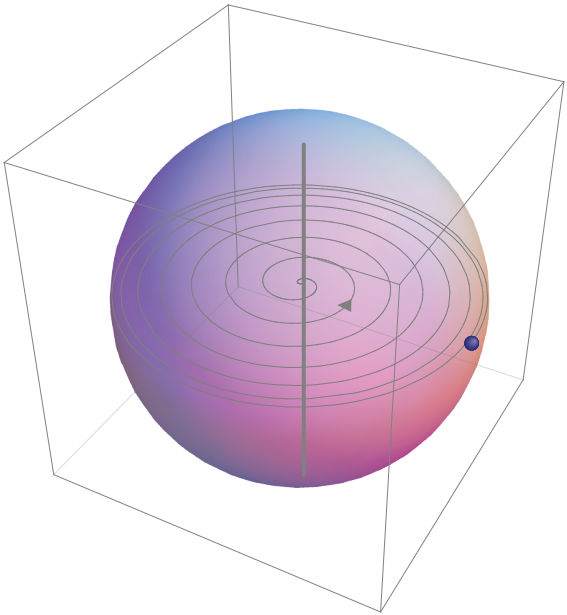}}
\caption{If the Hamiltonian of a system commutes with the interaction Hamiltonian ($[\H_1,\H_3]=0$), then decoherence drives the system toward a time-independent state $\rho$ where nothing ever changes.
The figure illustrates this for the Bloch Sphere of a single qubit starting in a pure state and ending up in a fully mixed state $\rho=\I/2$. More general initial states end up somewhere along the $z$-axis. Here $\H_1\propto\sigma_z$, generating a simple precession around the $z$-axis.
}
\label{ZenoFig}
\end{figure}

In \Sec{PhysicsFromSeparabilitySec}, we began exploring the idea that if we divide the world into maximally independent parts (with minimal interaction Hamiltonians), then the observed object hierarchy from \fig{HierarchyFig} would emerge.
The HDT tells us that this decomposition (factorization) into maximally independent parts can be performed in the energy eigenbasis of the total Hamiltonian. 
This means that all subsystem Hamiltonians and all interaction Hamiltonians commute with one another, corresponding to an essentially classical world where none of the quantum effects associated with non-commutativity manifest themselves! 
In contrast, many systems that we customarily refer to as objects in our classical world do {\it not} commute with their interaction Hamiltonians: for example, the Hamiltonian governing the dynamics of a baseball
involves its momentum, which does not commute with the position-dependent potential energy due to external forces.

As emphasized by Zurek \cite{Zurek01},
states commuting with the interaction Hamiltonian form a ``pointer basis'' of classically observable states, playing
an important role in understanding the emergence of a classical world.
The fact that the independence principle automatically leads to commutativity with interaction Hamiltonians might therefore be taken as an encouraging indication that we are on the right track. However, whereas the pointer states in Zurek's examples evolve over time due to the system's own Hamiltonian $\H_1$, those in our independence-maximizing decomposition do not, because they commute also with $\H_1$.
Indeed, the situation is even worse, as illustrated in \fig{ZenoFig}: 
any time-dependent system will evolve into a time-independent one, as environment-induced decoherence \cite{Zeh70,JZ85,ZurekHabibPaz93,ZehBook,Zurek09,SchlosshauerBook}
drives it towards an eigenstate of the interaction Hamiltonian, \ie, an energy eigenstate.\footnote{For a system with a finite environment, the entropy will eventually decrease again, causing the resumption of time-dependence, but this Poincar\'e recurrence time grows exponentially with environment size and is normally large enough that decoherence can be approximated as permanent.}

The famous Quantum Zeno effect, whereby a system can cease to evolve in the limit where it is 
arbitrarily strongly coupled to its environment \cite{Sudarshan77}, thus has a stronger and more pernicious cousin, which we will term the {\it Quantum Zeno Paradox} or the {\it Independence Paradox}.

\smallskip
\centerline{\framebox{\parbox{7cm}{
{\bf Quantum Zeno Paradox:}\\ {\it If we decompose our universe into maximally independent objects, then all change grinds to a halt.}
}}}
\smallskip

In summary, we have tried to understand the emergence of our observed semiclassical world, with its hierarchy of moving objects, by decomposing the world into maximally independent parts, but our attempts have failed dismally, producing merely a timeless world reminiscent of heat death.
In \Sec{IntegrationParadoxSec}, we saw that using the integration principle alone led to a similarly embarrassing failure, with no more than a quarter of a bit of integrated information possible. 
At least one more principle is therefore needed.


\section{Dynamics and autonomy}
\label{DynamicsSec}

Let us now explore the implications of the {\it dynamics principle} from \Tab{PrincipleTable}, according to which a conscious system has the capacity to not only {\it store} information, but also to {\it process} it. 
As we  just saw above, there is an interesting tension between this principle and the independence principle, whose Quantum Zeno Paradox gives the exact opposite: no dynamics and no information processing at all.

We will term the synthesis of these two competing principles the {\it autonomy principle}:
a conscious system has substantial dynamics {\it and} independence.
When exploring autonomous systems below, we can no longer study the state $\rho$ and the Hamiltonian $\H$ separately, since their interplay is crucial. Indeed, we well see that there are interesting classes of states $\rho$ that provide substantial dynamics and near-perfect independence even when the interaction Hamiltonian $\H_3$ is {\it not} small.
In other words, for certain preferred classes of states, the independence principle no longer pushes us to simply minimize $H_3$ and face the Quantum Zeno Paradox. 

\subsection{Probability velocity and energy coherence}

To obtain a quantitative measure of dynamics, let us first define 
the {\it probability velocity} $\vv\equiv\dot\p$, where the probability vector $\p$ is given by $p_i\equiv\rho_{ii}$.
In other words, 
\beq{vEq}
v_k=\dot\rho_{kk}=i[\H,\rho]_{kk}.
\eeq
Since $\vv$ is basis-dependent, we are interested in finding the basis where 
\beq{v2Eq}
v^2\equiv\sum_k v_k^2 = \sum_k(\dot\rho_{kk})^2
\eeq
is maximized, \ie, the basis where the sums of squares of the diagonal elements of $\dot\rho$ is maximal.
It is easy to see that this basis is the eigenbasis of $\dot\rho$:
\beqa{ProofEq}
v^2&=&
 \sum_k(\dot\rho_{kk})^2=\sum_{jk} (\dot\rho_{jk})^2 -\sum_{j\ne k} (\dot\rho_{jk})^2\nonumber\\
&=&  ||\dot\rho||^2 -  \sum_{j\ne k} (\dot\rho_{jk})^2
\eeqa
is clearly maximized in the eigenbasis where all off-diagonal elements in the last term vanish, since
the Hilbert-Schmidt norm $||\dot\rho||$ is the same in every basis; 
$||\dot\rho||^2 = \tr\dot\rho^2$, which is simply the sum of the squares of the eigenvalues of $\dot\rho$.

Let us define the {\it energy coherence} 
\beqa{EnergyCoherenceEq}
\delta H&\equiv&{1\over\sqrt{2}}||\dot\rho||
=
{1\over\sqrt{2}}||i[\H,\rho]|| =\sqrt{-\tr\{[\H,\rho]^2\}\over 2}\nonumber\\
&=& \sqrt{\tr[\H^2\rho^2-\H\rho\H\rho]}.
\eeqa
For a pure state $\rho=\ket{\psi}\bra{\psi}$, this definition implies that 
$\delta H\equiv\Delta H$, 
where $\Delta H$ is the {\it energy uncertainty}
\beq{EnergyUncertaintyEq}
\Delta H = \left[\bra{\psi}\H^2\ket{\psi}- \bra{\psi}\H\ket{\psi}^2\right]^{1/2},
\eeq
so we can think of $\delta H$ as the coherent part of the energy uncertainty, 
\ie, as the part that is due to quantum rather than classical uncertainty.

Since $||\dot\rho||=||[\H,\rho]||=\sqrt{2}\delta H$, we see that 
the maximum possible probability velocity $v$ is simply 
\beq{vmaxEq}
\vmax=\sqrt{2}\>\delta H,
\eeq
so we can equivalently use either of $v$ or $\delta H$ as convenient measures of quantum 
dynamics.\footnote{The {\it fidelity} between the state $\psi(t)$ and the initial state $\psi_0$ is defined as
\beq{FidelityEq}
F(t)\equiv \langle\psi_0\ket{\psi(t)},
\eeq
and it is easy to show that $\dot F(0)=0$ and $\ddot F(0)=-(\Delta H)^2$, so 
the energy uncertainty is a good measure of dynamics in that it also determines the fidelity evolution to lowest order, for pure states. For a detailed review of related measures of dynamics/information processing capacity, see \cite{Lloyd99}.
}
Whimsically speaking, the dynamics principle thus implies that energy eigenstates are as unconscious as things come, and that if you know your own energy exactly, you're dead.

\begin{figure*}[pht]
\centerline{\includegraphics[width=180mm]{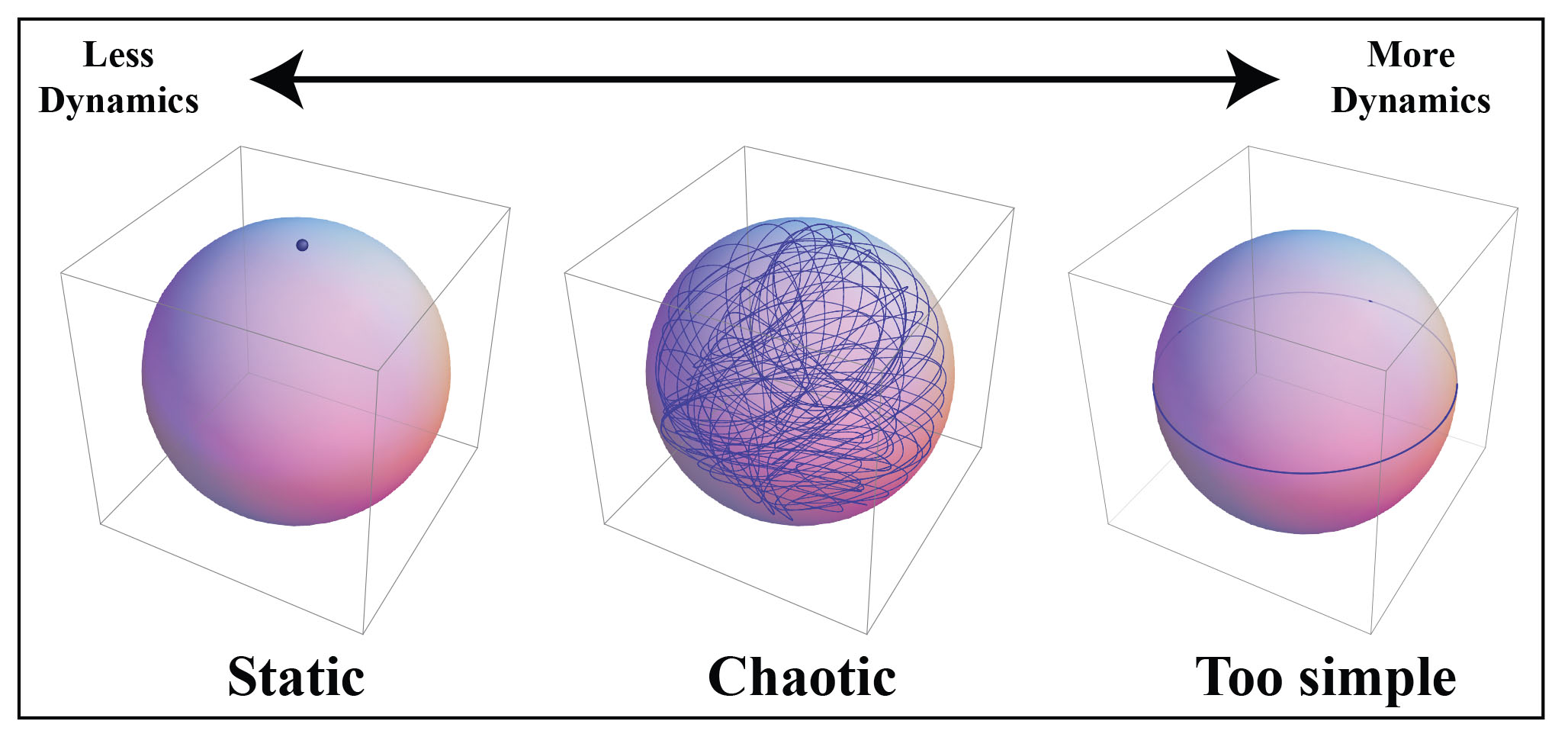}}
\caption{Time-evolution of Bloch vector $\tr\sigmab\dot\rho_1$ for a single qubit subsystem.
We saw how minimizing $H_3$ leads to a static state with no dynamics, such as the left example.
Maximizing $\delta H$, on the other hand, produces extremely simple dynamics such as the right example.
Reducing $\delta H$ by a modest factor of order unity can allow complex and chaotic dynamics (center); shown here is a 2-qubit system where the second qubit is traced out.
}
\label{DynamicsFig}
\end{figure*}

Although it is not obvious from their definitions, these quantities $\vmax$ and $\delta H$ are independent of time (even though $\rho$ generally evolves).
This is easily seen in the energy eigenbasis, where
\beq{EnergyEigenbasisEq} 
-i\dot\rho_{mn} = [\H,\rho]_{mn}=\rho_{mn}(E_m-E_n),
\eeq
where the energies $E_n$ are the eigenvalues of $\H$.
In this basis, $\rho(t)=e^{i\H t}\rho(0) e^{-i\H t}$ simplifies to
\beq{EnergyEigenbasisEvolEq}
\rho(t)_{mn}=\rho(0)_{mn}e^{i(E_m-E_n)t},
\eeq
This means that in the energy eigenbasis, the probabilities $p_n\equiv\rho_{nn}$
are invariant over time. 
These quantities constitute the energy spectral density for the state:
\beq{SpectralDensityEq}
p_n=\langle E_n|\rho|E_n\rangle.
\eeq
In the energy eigenbasis, \eq{EnergyUncertaintyEq} reduces to 
\beq{DHinvarianceEq}
\delta H^2=\Delta H^2=\sum_n p_n E_n^2 - \left(\sum_n p_n E_n\right)^2,
\eeq
which is time-invariant because the spectral density $p_n$  is. For general states, 
\eq{EnergyCoherenceEq} simplifies to 
\beq{dHinvarianceEq}
\delta H^2=\sum_{mn}|\rho_{mn}|^2 E_n (E_n - E_m).
\eeq
This is time-independent because \eq{EnergyEigenbasisEvolEq} shows that $\rho_{mn}$ changes merely by a phase factor, leaving $|\rho_{mn}|$ invariant.
In other words, when a quantum state evolves unitarily in the Hilbert-Schmidt vector space, 
both the position vector $\rho$ and the velocity vector $\dot\rho$ retain their lengths: 
both $||\rho||$ and $||\dot\rho||$ remain invariant over time.

\subsection{Dynamics versus complexity}

Our results above show that if all we are interested in is maximizing the maximal probability velocity $\vmax$, then we should find the two most widely separated eigenvalues of $\H$, $\Emin$ and $\Emax$, and choose a pure state that involves a coherent superposition of the two:
\beq{DynMaxEq}
\psiket=c_1\ket{\Emin}+c_2\ket{\Emax},
\eeq
where $|c_1|=|c_2|=1/\sqrt{2}$.
This gives $\delta H=(\Emax-\Emin)/2$, the largest possible value, but produces an extremely simple and boring solution $\rho(t)$. Since the spectral density $p_n=0$ except for these two energies, the dynamics is effectively that of a 2-state system (a single qubit) no matter how large the dimensionality of $\H$ is, corresponding to a simple periodic solution with frequency $\omega=\Emax-\Emin$ (a circular trajectory in the Bloch sphere as in the right panel of \fig{DynamicsFig}). This violates the dynamics principle as defined in \Tab{PrincipleTable}, since no substantial information processing capacity exists: the system is simply performing the trivial computation that flips a single bit repeatedly. 

To perform interesting computations, the system clearly needs to exploit a significant part of its energy spectrum. As can be seen from \eq{EnergyEigenbasisEvolEq}, 
if the eigenvalue differences are irrational multiples of one another, then the time evolution will never repeat, and 
$\rho$ will eventually evolve through all parts of Hilbert space allowed by the invariants 
$|\langle E_m|\rho|E_n\rangle|$.
The reduction of $\delta H$ required to transition from simple periodic motion to such complex aperiodic motion is quite modest. For example, if the eigenvalues are roughly equispaced, then changing the spectral density $p_n$ 
from having all weight at the two endpoints to having approximately equal weight for all eigenvalues will only reduce the energy coherence $\delta H$ by about a factor $\sqrt{3}$, since
the standard deviation of a uniform distribution is $\sqrt{3}$ times smaller than its half-width.

\subsection{Highly autonomous systems: sliding along the diagonal}

\begin{figure}[pbt]
\centerline{\includegraphics[width=84mm]{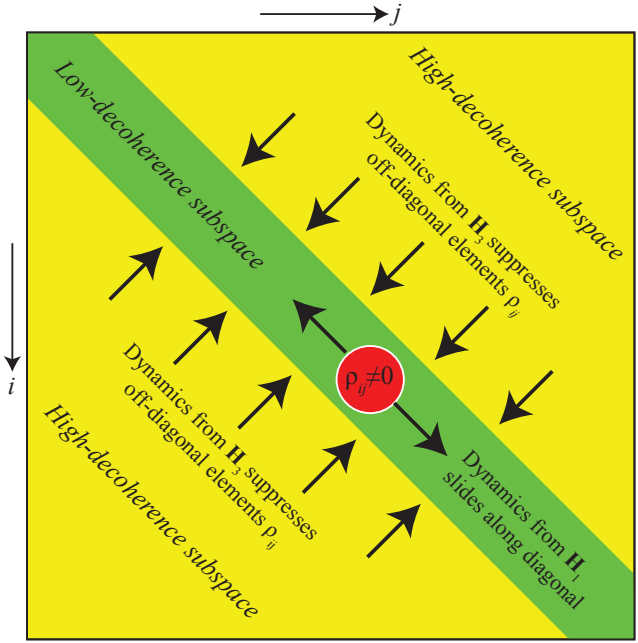}}
\caption{Schematic representation of the time-evolution of the density matrix $\rho_{ij}$ for a highly autonomous subsystem. $\rho_{ij}\approx 0$ except for a single region around the diagonal (red/grey dot), and this region slides along the diagonal under the influence of the subsystem Hamiltonian $\H_1$.
Any $\rho_{ij}$-elements far from the diagonal rapidly approach zero because of environment-decoherence caused by the interaction Hamiltonian $\H_3$.
}
\label{SlidingFig}
\end{figure}

What combinations of $\H$, $\rho$ and factorization produce highly autonomous systems? 
A broad and interesting class corresponds to macroscopic objects around us that move classically 
to an excellent approximation.

The states that are most robust toward environment-induced decoherence are those that approximately commute with the interaction Hamiltonian \cite{ZurekHabibPaz93}.
As a simple but important example, let us consider an interaction Hamiltonian 
of the factorizable form 
\beq{ProductInteractionEq}
\H_3 = \A\tensormult\B,
\eeq
and work in a system basis where the interaction term $\A$ is diagonal.
If $\rho_1$ is approximately diagonal in this basis, then
$\H_3$ has little effect on the dynamics, which becomes dominated by the internal subsystem Hamiltonian 
$\H_1$.
The Quantum Zeno Paradox we encountered in \Sec{ZenoSec} involved a situation where $\H_1$ was also diagonal in this same basis, so that we ended up with no dynamics. 
As we will illustrate with examples below, classically moving objects in a sense constitute the opposite limit:
the commutator $\dot\rho_1 = i[\H_1,\rho_1]$ is essentially as large as possible instead of as small as possible,
continually evading decoherence by concentrating $\rho$ around a single point that continually slides along 
the diagonal, as illustrated in \fig{SlidingFig}.
Decohererence rapidly suppresses off-diagonal elements far from this diagonal, but leaves the diagonal elements completely unaffected, so there exists a low-decoherence band around the diagonal.  
Suppose, for instance, that our subsystem is the center-of-mass position $x$ of a macroscopic object experiencing a position-dependent potential $V(x)$ caused by coupling to the environment, so that \fig{SlidingFig} represents the density matrix $\rho_1(x,x')$ in the position basis. If the potential $V(x) $ has a flat ($V'=0$) bottom of width $L$, then 
$\rho_1(x,x')$ will be completely unaffected by decoherence for the band $|x'-x|<L$.
For a generic smooth potential $V$, the decoherence suppression of off-diagonal elements grows only quadratically with the distance $|x'-x|$ from the diagonal \cite{JZ85,brain}, again making decoherence much slower than the internal dynamics in a narrow diagonal band.

As a specific example of this highly autonomous type, let us consider a subsystem with a uniformly spaced energy spectrum. 
Specifically, consider an $n$-dimensional Hilbert space and a Hamiltonian with spectrum
\beq{SHOspectrumEq}
E_k=\left[k-{n-1\over 2}\right]\hbar\omega=k\hbar\omega+E_0,
\eeq
$k=0,1,...,n-1$. We will often set $\hbar\omega=1$ for simplicity.
For example, $n=2$ gives the spectrum $\{-{1\over 2},{1\over 2}\}$ like the Pauli matrices divided by two,
$n=5$ gives $\{-2,-1,0,1,2\}$ and $n\to\infty$ gives the simple Harmonic oscillator (since the zero-point energy is physically irrelevant, we have chosen it so that $\tr\H=\sum E_k=0$, whereas the customary choice for the harmonic oscillator is such that the ground state energy is $E_0=\hbar\omega/2$).

If we want to, we can define the familiar position and momentum operators $x$ and $p$, and interpret this system as a Harmonic oscillator. However, the probability velocity $v$ is not maximized in either the position or the momentum basis, except twice per oscillation --- when the oscillator has only kinetic energy, $v$ is maximized in the $x$-basis, and when it has only potential energy, $v$ is maximized in the $p$-basis, and when it has only potential energy. If we consider the Wigner function $W(x,p)$, which simply rotates uniformly with frequency $\omega$, it becomes clear that the observable which is always changing with the maximal probability velocity is instead the {\it phase}, the Fourier-dual of the energy. 
Let us therefore define the phase operator 
\beq{PhaseOperatorEq}
\Phimatrix\equiv\F\H\F^\dagger,
\eeq
where $\F$ is the unitary Fourier matrix.

\begin{figure*}[tpb]
\centerline{\includegraphics[width=180mm]{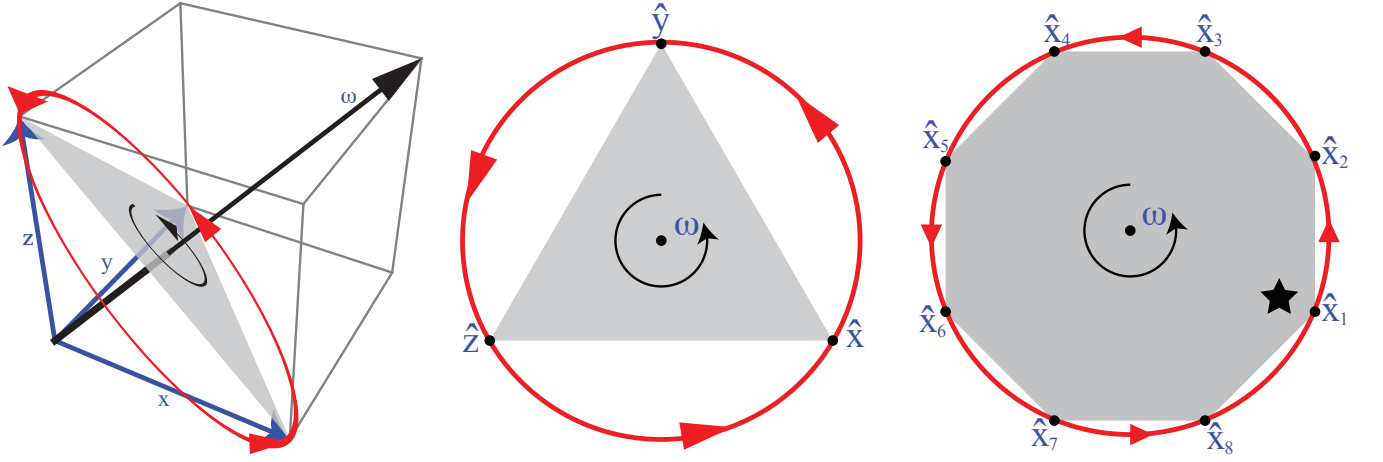}}
\vskip-10mm
\caption{For a system with an equispaced energy spectrum (such as a truncated harmonic oscillator or a massless particle in a discrete 1-dimensional periodic space), the time-evolution has a simple geometric interpretation in the space spanned by the eigenvectors ${\hat x_k}$ of the phase operator $\F\H\F$, the Fourier dual of the Hamiltonian, corresponding to unitarily rotating the entire space with frequency $\omega$, where $\hbar\omega$ is the energy level spacing.
After a time $2\pi/n\omega$, each basis vector has been rotated into the subsequent one, as schematically illustrated above. (The orbit in Hilbert space is only planar for $n\le 3$, so the figure should not be taken too literally.)
The black star denotes the $\alpha=1$ apodized state described in the text, which is more robust toward
decoherence.
}
\label{RotationFig}
\end{figure*}

Please remember that none of the systems $\H$ that we consider have any {\it a priori} physical interpretation; rather, the ultimate goal of the physics-from-scratch program is to derive any interpretation from the mathematics alone. Generally, any thus emergent interpretation of a subsystem will depend on its interactions with other systems. Since we have not yet introduced any interactions for our subsystem, we are free to interpret it in whichever way is convenient. In this spirit, an equivalent and sometimes more convenient way to interpret our Hamiltonian from \eq{SHOspectrumEq} is as a massless one-dimensional scalar particle, for which the momentum equals the energy,  so the momentum operator is $\p=\H$.
If we interpret the particle as existing in a discrete space with $n$ points and a toroidal topology (which we can think of as $n$ equispaced points on a ring), then the position operator is related to the momentum operator by a discrete Fourier transform:
\beq{JuggernautxEq}
\x=\F\p\F^\dagger,\quad F_{jk}\equiv {1\over \sqrt{N}}e^{i{jk\over 2\pi n}}.
\eeq
Comparing equations\eqn{PhaseOperatorEq} and \eqn{JuggernautxEq}, we see that $\x=\Phimatrix$.
Since $\F$ is unitary, the operators $\H$, $\p$, $\x$  and $\Phimatrix$ all have the same spectrum: the evenly spaced grid of \eq{SHOspectrumEq}.

As illustrated in \fig{RotationFig}, the time-evolution generated by $\H$ has a simple geometric interpretation in the space spanned by the position eigenstates $\ket{x_k}$, $k=1,...n$: the space is unitarily rotating with frequency $\omega$, so after a time $t=2\pi/n\omega$, a state $\ket{\psi(0)}=\ket{x_k}$ has been rotated such that it equals the next eigenvector: $\ket{\psi(t)}=\ket{x_{k+1}}$, where the addition is modulo $n$.
This means that the system has period $T\equiv2\pi/\omega$, and that 
$\psiket$ rotates through each of the $n$ basis vectors during each period.  

\begin{figure}[pbt]
\centerline{\includegraphics[width=84mm]{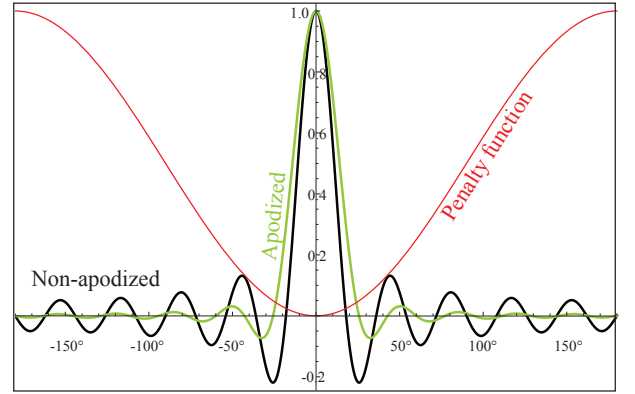}}
\caption{The wiggliest (heavy black) curve shows the inner product of a position eigenstate with what it evolves into a time $t=\phi/\omega$ later due to our $n=20$-dimensional Hamiltonian with energy spacings $\hbar\omega$. When optimizing to minimize the square of this curve using the $1-\cos\phi$ penalty function shown, corresponding to apodization in the Fourier domain, we instead obtain the green/light grey curve, resulting in much less decoherence.}
\label{WavepacketFig}
\end{figure}

Let us now quantify the autonomy of this system, starting with the dynamics. 
Since a position eigenstate is a Dirac delta function in position space, it is a plane wave in momentum space --- and in energy space, since $\H=\p$. This means that the spectral density is $p_n=1/n$ for a position eigenstate.
Substituting \eq{SHOspectrumEq} into \eq{DHinvarianceEq} gives 
an energy coherence 
\beq{JuggernautEnergyCoherenceEq}
\delta H=\hbar\omega\sqrt{n^2 - 1\over 12}.
\eeq
For comparison, 
\beq{JuggernautHnormEq}
||\H||=\left(\sum_{k=0}^{n-1} E_k^2\right)^{1/2} =\hbar\omega\sqrt{n(n^2-1)\over 12}=\sqrt{n}\>\delta H.
\eeq
Let us now turn to quantifying independence and decoherence.
The inner product between the unit vector $\ket{\psi(0)}$
and the vector $\ket{\psi(t)}\equiv e^{i\H t}\ket{\psi(0)}$ into which it evolves after a time $t$ is 
\beqa{rotationDotEq}
f_n(\phi)&\equiv&\langle\psi|e^{i\H{\phi\over\omega}} |\psi\rangle={1\over n}\sum_{k=0}^{n-1} e^{iE_k\phi}=
e^{-i{n-1\over 2}\phi}\sum_{k=0}^{n-1} e^{ik\phi}\nonumber\\
&=&{1\over n}e^{-i{n-1\over 2}\phi}{1-e^{in\phi}\over 1-e^{i\phi}}={\sin n\phi\over n\sin\phi}
,
\eeqa
where $\phi\equiv\omega t$.
This inner product $f_n$ is plotted in \fig{WavepacketFig}, and is seen to be a sharply peaked even function satisfying $f_n(0)=1$, $f_n(2\pi k/n)=0$ for $k=1,...,n-1$ and exhibiting one small oscillation between each of these zeros.
The angle $\theta\equiv\cos^{-1} f_n(\phi)$ between an initial vector $\phi$ and its time evolution thus grows
rapidly from $0^\circ$ to $90^\circ$, then oscillates close to $90^\circ$ until returning to $0^\circ$ after a full period $T$.
An initial state $\ket{\psi(0)}=\ket{x_k}$ therefore evolves as 
$$\psi_j(t)=f_n(\omega t - 2\pi [j-k]/n)$$
in the position basis, \ie, a wavefunction $\psi_j$ sharply peaked for $j\sim k+n\omega t/2\pi$ (mod $n$).
Since the density matrix evolves as $\rho_{ij}(t)=\psi_i(t)\psi_j(t)^*$, 
it will therefore be small except for $i\sim j\sim k+n\omega t/2\pi$ (mod $n$),
corresponding to the round dot on the diagonal in \fig{SlidingFig}.
In particular, the decoherence-sensitive elements $\rho_{jk}$ will be small far from the diagonal, corresponding to the small values that $f_n$ takes far from zero. How small will the decoherence be? Let us now develop the tools needed to quantify this.

\subsection{The exponential growth of autonomy with system size}

Let us return to the most general Hamiltonian $\H$ and study how an initially separable state 
$\rho=\rho_1\tensormult\rho_2$ evolves over time.
Using the orthogonal projectors of \Sec{ProjectorSec}, we can decompose $\H$ as
\beq{HseparabilityDefEq2}
\H=\H_1\tensormult\I + \I\tensormult\H_2 + \H_3,
\eeq
where $\tr_1\H_3=\tr_2\H_3=0$.
By substituting \eq{HseparabilityDefEq2} into  the evolution equation 
$\dot\rho_1=\tr_2\dot\rho=i\tr_2[\H,\rho]$ and using 
various partial trace identities from \Sec{IdentitySec} to simplify the resulting three terms, we obtain
\beq{rhodotEq}
\dot\rho_1 = i\,\trace_2[\H,\rho_1\tensormult\rho_2] = i\,[\H_1+\H_*,\rho_1],
\eeq
where what we will term the {effective interaction Hamiltonian}
\beq{HstarDefEq}
\H_*\equiv \trace_2\{(\I\tensormult\rho_2)\H_3\}
\eeq
can be interpreted as an average of the interaction Hamiltonian $\H_3$, weighted by the environment state $\rho_2$.
A similar effective Hamiltonian is studied in \cite{Omnes01,Gemmer01,Durt04}.
\Eq{rhodotEq} implies that the evolution of $\rho_1$ remains unitary to first order in time, the only effect of the interaction $\H_3$ being to replace $\H_1$ from \eq{SeparablerhodotEq} by 
an effective Hamiltonian $\H_1+\H_*$. 

The second time derivative is given by $\ddot\rho_1=\tr_2\dot\rho=-\tr_2[\H,[\H,\rho]]$, and by analogously 
substituting \eq{HseparabilityDefEq2}  and using partial trace identities from \Sec{IdentitySec} to simplify the resulting nine terms, we obtain 
\beqa{rhodotdoteq}
-\ddot\rho_1&=&\,\tr_2[\H,[\H,\rho_1\tensormult\rho_2]] =\nonumber\\
&=&[\H_1,[\H_1,\rho_1]]-i\,[\K,\rho_1]+\nonumber\\
&+&[\H_1,[\H_*,\rho_1]]+[\H_*,[\H_1,\rho_1]]+\nonumber\\
&+&\tr_2[\H_3,[\H_3,\rho_1\tensormult\rho_2]],
\eeqa
where we have defined the Hermitian matrix 
\beq{KdefEq}
\K\equiv i\,\tr_2\{(\I\tensormult[\H_2,\rho_2])\H_3\}.
\eeq

To qualify independence and autonomy, we are interested in the extent to which $\H_3$ causes entanglement and makes the time-evolution of $\rho_1$ non-unitary. 
When thinking of $\rho$ as a vector in the Hilbert-Schmidt vector space that we reviewed in \Sec{HilbertSchmidtSec}, unitary evolution preserves its length $||\rho||$. To provide geometric intuition for this, let us define dot and cross product notation analogous to vector calculus.
First note that 
\beq{dotcrossEq}
(\A^\dagger,[\A,\B])=\tr\A\A\B-\tr\A\B\A=0,
\eeq
since a trace of a product is invariant under cyclic permutations of the factors.
This shows that a commutator $[\A,\B]$ is orthogonal to both $\A^\dagger$ and $\B^\dagger$ under the Hilbert-Schmidt inner product, and a Hermitian matrix $\H$ is orthogonal to its commutator with any matrix.

This means that it we restrict ourselves to the Hilbert-Schmidt vector space of {\it Hermitian} matrices, 
we obtain an interesting generalization of the standard dot and cross products for 3D vectors.
Defining
\beqa{dotcrossDefEq}
\A\cdot\B&\equiv&(\A,\B),\\\
\A\times\B&\equiv&i[\A,\B],
\eeqa
we see that these operations satisfy all the same properties as their familiar 3D analogs:
the scalar (dot) product is symmetric ($\B\cdot\A=\tr\B^\dagger\A=\tr\A\B^\dagger=\A\cdot\B)$,
while the vector (cross) product is antisymmetric ($\A\times\B=\B\times\A$), 
orthogonal to both factors ($[\A\times\B]\cdot\A=[\A\times\B]\cdot\B=0$), 
and produces a result of the same type as the two factors (a Hermitian matrix).

In this notation, the products of an arbitrary Hermitian matrix $\A$ with the identity matrix $\I$ are
\beqa{Ieqs}
\I\cdot\A&=&\tr\A,\\
\I\times\A&=&0,
\eeqa
and the Schr\"odinger equation $\dot\rho=i[\H,\rho]$ becomes simply
\beq{SchroedingerEq}
\dot\rho = \H\times\rho.
\eeq
Just as in the 3D vector analogy, we can think of this as generating rotation of the vector $\rho$ that preserves its length:
\beq{rhonormdotEq}
{d\over dt}||\rho||^2 = {d\over dt}\rho\cdot\rho=2\dot\rho\cdot\rho=2(\H\times\rho)\cdot\rho=0.
\eeq

A simple and popular way of quantifying whether evolution is non-unitary is to compute the linear entropy
\beq{LinearEntropyEq}
\Slin\equiv 1-\tr\rho^2=1-||\rho||^2,
\eeq
and repeatedly differentiating \eq{LinearEntropyEq} tells us that 
\beqa{SlinDerivEq}
{\dot S}^{\rm lin}&=&-2\rho\cdot\dot\rho,\\
{\ddot S}^{\rm lin}&=&-2(||\dot\rho||^2+\rho\cdot\ddot\rho),\label{SlinDerivEq2}\\
{\dddot S}^{\rm lin}&=&-6\dot\rho\cdot\ddot\rho  - 2\rho\cdot\dddot\rho.\label{SlinDerivEq3}
\eeqa
Substituting equations\eqn{rhodotEq} and\eqn{rhodotdoteq} into equations\eqn{SlinDerivEq}
and\eqn{SlinDerivEq2} for $\rho_1$, we find that almost all terms cancel, leaving us with the simple result
\beqa{SlindotEq}
{\dot S}^{\rm lin}_1&=&0,\\
{\ddot S}^{\rm lin}_1&=&2\,\tr\{\rho_1\trace_2[\H_3,[\H_3,\rho]]\}-2||[\H_*,\rho_1]||^2.\label{SlinddotEq}
\eeqa
This means that, to second order in time, the entropy production is completely independent of $\H_1$ and $\H_2$, 
depending only on quadratic combinations of $\H_3$, weighted by quadratic
combinations of $\rho$. 
%
We find analogous results for the Shannon entropy $S$: 
If the density matrix is initially separable, then $\dot S_1=0$ and $\ddot S_1$
depends not on the full Hamiltonian $\H$, but only on its non-separable component $\H_3$, quadratically.

We now have the tools we need to compute the autonomy of our ``diagonal-sliding'' system from the previous subsection.
As a simple example, let us take $\H_1$ to be our Hamiltonian from \eq{SHOspectrumEq} with its equispaced energy spectrum, 
with $n=2^b$, so that we can view the Hilbert space as that of  $b$ coupled qubits.
\Eq{JuggernautEnergyCoherenceEq} then gives an energy coherence 
\beq{JuggernautEnergyCoherenceEq2}
\delta H\approx{\hbar\omega\over\sqrt{12}}\> 2^b,
\eeq
so the probability velocity grows exponentially with the system size $b$.

We augment this Hilbert space with one additional ``environment'' qubit that begins in the state $\up$, with internal dynamics given by $\H_2=\hbar\omega_2\sigma_x$,
and couple it to our subsystem with an interaction
\beq{JuggernautH3eq}
\H_3=V(\x)\tensormult\sigma_x
\eeq
for some potential $V$; $x$ is the position operator from \eq{JuggernautxEq}.
As a first example, we use the sinusoidal potential $V(\x) = \sin(2\pi\x/n)$, start the first subsystem in the position eigenstate 
$\ket{x_1}$ and compute the linear entropy $\Slin_1(t)$ numerically.

\begin{figure}[phbt]
\centerline{\includegraphics[width=88mm]{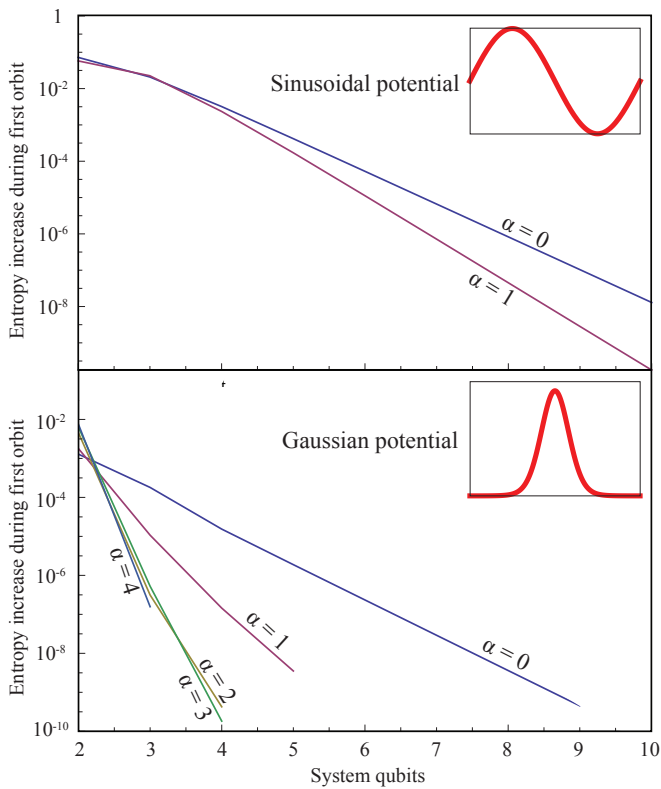}}
\caption{The linear entropy increase during the first orbit, 
${\ddot S}_1^{\rm lin}(2\pi/\omega)$, is plotted for as a function of the subsystem size (number of qubits $b$). The interaction potential $V(x)$ is sinusoidal (top) and Gaussian (bottom), and the different apodization schemes used to select the initial state are labeled by their corresponding $\alpha$-value, where $\alpha=0$ corresponds to no apodization (the initial state being a position eigenstate). Some lines have been terminated in the bottom panel due to insufficient numerical precision.
\label{ApodizationFig}
}
\end{figure}

As expected  from our qualitative arguments of the previous section, $\Slin_1(t)$ grows only very slowly, and we find that it can be accurately approximated by its Taylor expansion around $t=0$ for many orbital periods $T\equiv 2\pi/\omega$:
$\Slin_1(t)\approx\ddot{S}^{\rm lin}_1(0)\, t^2/2$, where $\ddot{S}^{\rm lin}_1(0)$ is given by \eq{SlinddotEq}.
\Fig{ApodizationFig} shows the linear entropy after one orbit, $\Slin_1(T)$, as a function of the number of qubits $b$ in our subsystem
(top curve in top panel).
Whereas \eq{JuggernautH3eq} showed that the dynamics {\it increases} exponentially with system size (as $2^b$), the figure shows that $\Slin_1(T)$
{\it decreases} exponentially with system size, asymptotically falling as $2^{-4b}$ as $b\to\infty$.

Let us define the 
{\it dynamical timescale} $\tdyn$
and the
{\it independence timescale} $\tind$
as
\beqa{TimescaleDefEq}
\tdyn&=&{\hbar\over \delta H},\\
\tind&=&[\ddot{S}^{\rm lin}_1(0)]^{-1/2}.
\eeqa
Loosely speaking,  we can think of $\tdyn$ as  the time our system requires to perform an elementary information processing operation such as a bit flip \cite{Lloyd99},  and 
$\tind$ as the time it takes for the linear entropy to change by of order unity, \ie, for significant 
information exchange with the environment to occur.
If we define the autonomy $A$ as the ratio
\beq{AutonomyDefEq}
A\equiv{\tind\over\tdyn},
\eeq
the autonomy of our subsystem thus grows exponentially with system size, asymptotically increasing as
$A\propto 2^{2b}/2^{-b}=2^{3b}$ as $b\to\infty$.

As illustrated by \fig{SlidingFig}, we expect this exponential scaling to be quite generic, independent of interaction details: the origin of the 
exponential is simply that the size of the round dot in the figure is of order $2^b$ times smaller than the size of the square representing the full density matrix.
The independence timescale $\tind$ is exponentially large because the dot, with its non-negligible elements $\rho_{ij}$, is exponentially close to the diagonal.
The dynamics timescale $\tdyn$ is exponentially small because it is roughly the time it takes the dot to traverse its own diameter as it moves around at some $b$-independent speed in the figure.

This exponential increase of autonomy with system size makes it very easy to have highly autonomous systems even 
if the magnitude $H_3$ of the interaction Hamiltonian is quite large. Although the environment continually ``measures'' the position of the subsystem
through the strong coupling $\H_3$, this measurement does not decohere the subsystem because it is (to an exponentially good approximation)
a non-demolition measurement, with the subsystem effectively in a position eigenstate. This phenomenon is intimately linked to the quantum Darwinism paradigm developed by Zurek and collaborators \cite{Zurek09}, where the environment mediates the emergence of a classical world by acting as a witness, storing large numbers of redundant copies of information about the system state in the basis that it measures.
We thus see that systems that have high autonomy via the ``diagonal-sliding'' mechanism are precisely objects that dominate quantum Darwinism's  
``survival of the fittest'' by proliferating imprints of their states in the environment.


\subsection{Boosting autonomy with optimized wave packets}

In our worked example above, we started our subsystem in a position eigenstate $\ket{x_1}$, which 
cyclically evolved though all other position eigenstates. The slight decoherence that did occur thus originated during the times when the state was between eigenstates, in a coherent superpositions of multiple eigenstates quantified by the most wiggly curve in \fig{WavepacketFig}.
Not surprisingly, these wiggles (and hence the decoherence) can be reduced by a better choice of initial state
$\psiket=\sum_k\psi_k\ket{x_k}=\sum_k\hat\psi_k\ket{E_k}$ for our subsystem,
where $\psi_k$ and $\hat{\psi}_k$ are the  wavefunction amplitudes in the position and energy bases, respectively. 
\Eq{rotationDotEq} then gets generalized to
\beq{rotationDotEq2}
g_n(\phi)\equiv\langle x_1|e^{i\H{\phi\over\omega}} |\psi\rangle
=
e^{-i{n-1\over 2}\phi}\sum_{k=0}^{n-1}\hat{\psi}_k e^{ik\phi}.
\eeq
Let us choose the initial state $\psiket$ that minimizes the quantity
\beq{PenaltyEq}
\int_{-\pi}^\pi |g_n(\theta)|^2 w(\theta) d\theta
\eeq
for some penalty function $w(\theta)$ that punishes states giving  large unwanted $|g(\theta)|$ far from $\theta=0$. 
This gives a simple quadratic minimization problem for the vector of coefficients $\hat{\psi}_k$, whose solution turns out to be the last (with smallest eigenvalue) eigenvector of the Toeplitz matrix whose first row is the Fourier series of $w(\theta)$.
A convenient choice  of penalty function $1-\cos\phi$ (see \fig{WavepacketFig}), which respects the periodicity of the problem and grows quadratically around its $\phi=0$ minimum.
In the $n\to\infty$ limit, the Toeplitz eigenvalue problem simplifies to Laplace's equation with a 
$\hat{\psi}(\phi)=\cos{\phi\over 2}$ winning eigenvector, giving
\beq{alpha1Eq}
\psi_k\equiv\int_{-\pi}^\pi \cos(k\phi) \hat{\phi}(\phi) d\phi={\cos(\pi k)\over 1 - 4 k^2}.
\eeq
The corresponding curve $g_n(\phi)$ is plotted is \fig{WavepacketFig}, and is seen to have significantly smaller wiggles away from the origin at the cost of a very slight widening of the central peak.
\Fig{ApodizationFig} (top panel, lower curve) shows that this choice significantly reduces decoherence. 

What we have effectively done is employ the
standard signal processing technique known as apodization.
Aside from the irrelevant phase factor, \eq{rotationDotEq2} is simply the Fourier transform
of $\hat{\psi}$, which can be made narrower by making  $\hat{\psi}$ smoothly approach zero at the two endpoints. In the $n\to\infty$ limit, our original choice corresponded to $\hat{\psi}=1$ for 
$-\pi\le\phi\le\pi$, which is discontinuous, whereas our replacement function $\hat{\psi}=\cos{\phi\over 2}$
vanishes at the endpoints and is continuous. This reduces the wiggling because 
Riemann-Lebesgue's lemma implies that the Fourier transform of a function whose first $d$ derivatives are continuous falls off faster than $k^{-d}$.
By instead using $\hat{\psi}^{(\alpha)}(\phi)=(\cos{\phi\over 2})^\alpha$ for some integer $\alpha\ge 0$,
we get $\alpha$ continuous derivatives, so the larger we choose $\alpha$, the smaller the decoherence-inducing wiggles, at the cost of widening the central peak.
The first five cases give
\beqa{ApodizationEq}
\psi^{(0)}_k&=&\delta_{0k},\\
\psi^{(1)}_k&=&{\cos(\pi k)\over 1 - 4 k^2},\\
\psi^{(2)}_k&=&\delta_{0k}+{1\over 2}\delta_{1,|k|},\\
\psi^{(3)}_k&=&{\cos(\pi k)\over (1 - 4 k^2) (1 - {4\over 9} k^2)},\\
\psi^{(4)}_k&=&\delta_{0k}+{2\over 3}\delta_{1,|k|}+{1\over 6}\delta_{2,|k|},
\eeqa
and it is easy to show that the $\alpha\to\infty$ limit corresponds to a Gaussian shape.

Which apodization is best? This depends on the interaction $\H_3$. 
For our sinusoidal interaction potential (\Fig{ApodizationFig}, top),
the best results are for $\alpha=1$, when the penalty function has a quadratic minimum.
When switching to the roughly Gaussian interaction potential 
$V(\x)\propto e^4\cos(2\pi\x/n)$ (\Fig{ApodizationFig}, bottom),
the results are instead seen to keep improving as we increase $\alpha$, producing dramatically less decoherence than for the sinusoidal potential, and suggesting that the optical choice is the $\alpha\to\infty$ state: a Gaussian wave packet.
Gaussian wave packets have long garnered interest as models of approximately classical states.
They correspond to generalized coherent states, which have shown to be maximally robust toward decoherence in important situations involving harmonic oscillator interactions \cite{ZHP93}. They have also been shown to emerge dynamically in harmonic oscillator environments, from the accumulation of many independent interactions, in much the same way as the central limit theorem gives a Gaussian probability distribution to sums of many independent contributions \cite{gaussians}.
Our results suggest that Gaussian wave packets may also emerge as the most robust states towards decoherence from short-range interactions with exponential fall-off.

\subsection{Optimizing autonomy when we can choose the state: factorizable effective theories}
\label{BumpEqSec}

Above we explored specific examples of highly autonomous systems, motivated by approximately classical systems that we find around us in nature. 
We found that there are combinations of $\rho$, $\H$ and Hilbert space factorization that provide excellent autonomy even when the interaction $H_3$ is not small.
We will now see that, more generally, given {\it any} $\H$ and factorization, there are states $\rho$ that give perfect factorization and infinite autonomy.
The basic idea is that for states such that some of the spectral density invariants $p_k$ vanish, it makes no difference if we replace the corresponding unused eigenvalues of $\H$ by others to make the Hamiltonian separable.

Consider a subspace of the full Hilbert space defined by a projection operator $\PP$. 
A projection operator satisfies $\PP^2=\PP=\PP^\dagger$, so its eigenvalues are all zero or one, and the latter correspond to our subspace of interest.
Let us define the symbol $\bumpeq$ to denote that operator equality holds in this subspace.
For example, 
\beq{BumpeqDefEq1}
\A-\B\bumpeq 0
\eeq
means that 
\beq{BumpeqDefEq2}
\PP(\A-\B)\PP=0.
\eeq
Below will often chose the subspace to correspond to low-energy states,
so the wave symbol in $\bumpeq$ is intended to remind us that equality holds in the long wavelength limit.

We saw that the energy spectral density $p_n$ of \eq{SpectralDensityEq}
remains invariant under unitary time evolution, so any energy levels for which $p_n=0$ will never have any  physical effect, and the corresponding dimensions of the Hilbert space can simply be ignored as ``frozen out''. This remains true even considering observation-related state projection as described in the next subsection.
Let us therefore define 
\beq{PPdefEq}
\PP=\sum_k\theta(p_n) |E_n\rangle\langle E_n|,
\eeq
where $\theta$ is the Heaviside step function ($\theta(x)=1$ if $x>0$, vanishing otherwise)  
\ie, summing only over those energy eigenstates for which the probability $p_n$ is non-zero.
Defining new operators in our subspace by
\beqa{SubspaceOperatorDefEq} 
 \rho'&\equiv&\PP\rho\PP,\\
\H'&\equiv&\PP\H\PP,\\
\eeqa
\eq{PPdefEq} implies that
\beqa{rhoprimeeq}
\rho'&=&\sum_{mn}\theta(p_m) \theta(p_n) |E_m\rangle\langle E_m|\rho |E_n\rangle\langle E_n|\nonumber\\
        &=&\sum_{mn} |E_m\rangle\langle E_m|\rho |E_n\rangle\langle E_n|=\rho,
\eeqa
Here the second equal sign follows from the fact that 
$|\langle E_m|\rho |E_n\rangle|^2\le \langle E_m|\rho|E_m\rangle \langle E_n|\rho|E_n\rangle$\footnote{This last inequality follows because $\rho$ is Hermitian and positive semidefinite, so the determinant must be non-negative for the $2\times 2$ matrix $\langle E_i|\rho |E_j\rangle$ where $i$ and $j$ each take the two values $k$ and $l$.
},
so that the left hand side must vanish whenever either $p_m$ or $p_n$ vanishes --- the Heaviside step functions
therefore have no effect in \eq{rhoprimeeq} and can be dropped. 

Although $\H'\neq\H$, we do have $\H'\bumpeq\H$, and this means that the time-evolution of $\rho$ can be correctly computed using $\H'$ in place of the full Hamiltonian $\H$:
$$\rho(t) = \PP\rho(t)\PP=\PP\e^{i\H t}\PP\rho(0)\PP e^{-i\H t}\PP = e^{i\H't}\rho(0) e^{-i\H't}.$$
The frozen-out part of the Hilbert space is therefore completely unobservable, and 
we can act as though the subspace is the only Hilbert space that exists, and as if $\H'$ is the true Hamiltonian.
By working only with $\rho'$ and $\H'$ restricted to the subspace, we have also simplified things by reducing the dimensionality of these matrices.


Sometimes, $\H'$ can possess more symmetry than $\H$. 
Sometimes, $\H'$ can be separable even if $\H$ is not:
\beq{EffectiveSeparabilityEq}
\H\bumpeq\H' =\H_1\tensormult\I+\I\tensormult\H_2\eeq
To create such a situation for an arbitrary $n\times n$ Hamiltonian, where $n=n_1 n_2$, simply pick a state $\rho$ such that 
the spectral densities $p_k$ vanish for all except $n_1+n_2-1$ energy eigenvectors. This means that in the energy eigenbasis, with the eigenvectors sorted to place these $n_1+n_2-1$ special ones first, $\rho$ is a block-diagonal matrix vanishing outside of the upper left 
$(n_1+n_2-1)\times(n_1+n_2-1)$ block.
\Eq{EnergyEigenbasisEvolEq} shows that $\rho(t)$ will retain this block form for all time, and that changing the energy eigenvalues
$E_k$ with $k>n_1+n_2-1$ leaves the time-evolution of $\rho$ unaffected.
We can therefore choose these eigenvalues so that $\H$ becomes separable.
For example, for the case where the Hilbert space dimensionality $n=9$, suppose that $p_k$ vanishes for all energies except $E_0$, $E_1$, $E_2$, $E_3$, $E_4$, and 
adjust the irrelevant zero-point energy so that $E_0=0$.
Then define $\H'$ whose 9 eigenvalues are 
\beq{EffectiveSeparabilityEq}
\left(
\begin{tabular}{c@{\hskip 1cm}c@{\hskip 1cm}c}
$0$		&$E_1$	&$E_2$\\
$E_3$	&$E_1+E_3$	&$E_2+E_3$\\
$E_4$	&$E_1+E_4$	&$E_2+E_4$
\end{tabular}
\right).
\eeq
Note that $\H'\bumpeq\H$, and that although $\H$ is generically not separable, 
$\H'$ is separable, with subsystem Hamiltonians $\H'_1=\diag\{0, E_1,E_2\}$ 
and $\H'_2=\diag\{0, E_3, E_4\}$.  
Subsystems 1 and 2 will therefore evolve as a parallel universes governed by $\H'_1$
and $\H'_1$, respectively.


\subsection{Minimizing quantum randomness}
\label{TraceYourselfSec}

When we attempted to maximize the independence for a subsystem above, we implicitly wanted to maximize the ability to predict the subsystems future state from its present state.
The source of unpredictability that we considered was influence from outside the subsystem, from the environment, which caused decoherence and increased subsystem entropy.

Since we are interested in modeling also conscious systems, there is a second independent source of unpredictability that we need to consider, which can occur even if there is no interaction with the environment: ``quantum randomness''.
If the system begins in a single conscious state and unitarily evolves into a superposition of subjectively distinguishable conscious states, then the observer in the initial state has no way of uniquely predicting her future perceptions.

A comprehensive framework for treating such situations is given in \cite{secondlaw}, and in the interest of brevity, we will not review it here, merely use the results. 
To be able to state them as succinctly as possible, let us first introduce notation for a projection process ``$\pr$'' that is in a sense dual to partial-tracing.

For a Hilbert space that is factored into two parts, we define the following notation.
We indicate the tensor product structure by splitting a single index $\alpha$ into an
index pair  $ii'$. For example, if the Hilbert space is the tensor product of an $m$-dimensional and an $n$-dimensional space, then $\alpha = n(i-1)+i'$, $i=1,...,m$, $i'=1,...,n$, $\alpha=1,...,mn$,
and if $\A=\B\tensormult\C$, then
\beq{TensorProductDefEq}
\A_{\alpha\beta}=\A_{ii'jj'}=\B_{ij}\C_{i'j'}.
\eeq
We define $\star$ as the operation exchanging subsystems 1 and 2:
\beq{StarDefEq}
(\A^\star)_{ii'jj'}=\A_{i'ij'j}
\eeq
We define $\pr_k\A$ as the $k^{th}$ diagonal block of $\A$:
$$(\proj_k\A)_{ij} = \A_{kikj}$$
For example, $\pr_1\A$ is the $m\times m$ upper left  corner of $\A$.

As before $\tr_i\A$, denotes the partial trace over the $i^{th}$ subsystem:
\beqa{PartialTraceDefEq}
(\trace_1\A)_{ij}&=&\sum_k\A_{kikj}\\
(\trace_2\A)_{ij}&=&\sum_k\A_{ikjk}
\eeqa
The following identities are straightforward to verify:
\beqa{StarIdentities}
\trace_1\A^\star&=&\trace_2\A\\
\trace_2\A^\star&=&\trace_1\A\\
\trace_1\A&=&\sum_k\proj_k\A\label{ProjectionIdentity3}\\
\trace_2\A&=&\sum_k\proj_k\A^\star\\
\tr\proj_k\A&=&(\trace_2\A)_{kk}\\
\tr\proj_k\A^\star&=&(\trace_1\A)_{kk}
\eeqa


Let us adopt the framework of \cite{secondlaw} and decompose the full Hilbert space into three parts corresponding to the subject (the conscious degrees of freedom of the observer), the object (the external degrees of freedom that the observer is interested in making predictions about)
and the environment (all remaining degrees of freedom).
 
If the subject knows the object-environment density matrix to be $\rho$, 
it obtains its density matrix for the object by tracing out the environment:
$$\rho_o=\trace_e\rho.$$
If the subject-object density matrix is $\rho$, then the subject may be in a superposition of having many different perceptions $|s_k\rangle$. Take the $|s_k\rangle$ to form a basis of the subject Hilbert space.
The probability that the subject finds itself in the state $|s_k\rangle$ is
\beq{BayespEq}
p_k=(\trace_2\rho)_{kk},
\eeq
and for a subject finding itself in this state $|s_k\rangle$, the object density matrix is
\beq{BayesrhoEq}
\rho_o^{(k)}={\proj_k\rho\over p_k}.
\eeq
If $\rho$ refers to a future subject-object state, and the subject wishes to predict its future knowledge of the object, it takes the weighted average of these density matrices, obtaining
$$\rho_o = \sum_k p_k \rho_o^{(k)} = \sum_k\proj_k\rho=\trace_s\rho,$$
\ie, it traces out itself! (We used the identity\eqn{ProjectionIdentity3} in the last step.)
Note that this simple result is independent of whatever basis is used for the object-space, so all issues related to how various states are perceived become irrelevant.

As proven in \cite{tripartite}, any unitary transformation of a separable $\rho$ will increase the entropy of $\tr_1\rho$. This means that the subject's future knowledge of $\rho_o$ is {\it more} uncertain than its present knowledge thereof. However, as proven in \cite{secondlaw}, the future subject's knowledge of $\rho_o$ will on average be {\it less} uncertain than it presently is, at least if the time-evolution is restricted to be of the measurement type.

The result $\rho_o=\trace_1\rho$ also holds if you measure the object and then forget what the outcome was. In this case, you are simply playing the role of an environment, resulting in the exact same partial-trace equation.

In summary,  for a conscious system to be able to predict the future state of what it cares about ($\rho_o$) as well as possible, we must minimize uncertainty introduced both by the interactions with the environment (fluctuation, dissipation and decoherence) and by measurement (``quantum randomness''). 
The future evolution can be better predicted for certain object states than for others, because they are more stable against both of the above-mentioned sources of unpredictability.
The utility principle from \Tab{PrincipleTable} suggests that it is precisely these most stable and predictable states that conscious observers will perceive. 
The successful ``predictability sieve'' idea of Zurek and collaborators \cite{Zurek05} involves precisely this idea when the source of unpredictability is environment-induced decoherence, so the utility principle lets us generalize this idea to include the second unpredictability source as well: to minimize apparent quantum randomness, we should pay attention to states whose dynamics lets them remain relatively diagonal in the eigenbasis of the subject-object interaction Hamiltonian, so that our future observations of the object are essentially quantum non-demolition measurements.

A classical computer is a flagship example of a such a maximally causal system, minimizing its uncertainty about its future.
By clever design, a small subset of the degrees of freedom in the computer, interpreted as bits, deterministically determine their future state with virtually no uncertainty.  For my laptop, each bit corresponds to the positions of certain electrons in its memory (determining whether a micro-capacitor is charged). An ideal computer with zero error rate thus has not only complex dynamics (which is Turing-complete modulo resource limitations), but also perfect autonomy, with its future state determined entirely by its own state, independently of the environment state. 
The Hilbert space factorization that groups the bits of this computer into a subsystem is therefore optimal, in the sense that any other factorization 
would reduce the autonomy. Moreover, this optimal solution to the quantum factorization problem is quite sharply defined:
considering infinitesimal unitary transformations away from this optimum, any transformation that begins rotating an environment bit into the system will cause a sharp reduction of the autonomy, because the decoherence rate for environment qubits 
(say a thermal collision frequency $\sim 10^{15}$ Hz) is orders of magnitude larger than the dynamics rate (say the clock frequency $\sim10^9$ Hz).
Note that $\H_3$ is far from zero in this example; the pointer basis corresponds to classical bit strings of which the 
environment performs frequent quantum non-demolition measurements.

This means that if artificial intelligence researchers one day succeed in making a classical computer conscious, and if we turn off any input devices though which our outside world can affect its
information processing, then  it will subjectively perceive itself as existing in a parallel universe completely disconnected from ours, even though we can probe its internal state from outside.
If a future quantum computer is conscious, then it will feel like in a parallel universe evolving under the Hamiltonian $\H_1(t)$ that we have designed for it --- until the readout stage, 
when we switch on an interaction $\H_3$.

\subsection{Optimizing autonomy when the state is given}
\label{SnipSec}

Let us now consider the case where both $\H$ and $\rho$ are treated as given, and we want to vary the Hilbert space factorization to attain maximal separability. $\H$ and $\rho$ together determine the full time-evolution $\rho(t)$ via the Schr\"odinger equation, so we seek the unitary transformation $\U$ that makes $\U\rho(t)\U^\dagger$ as factorizable as possible. For a pure initial state, exact factorability is equivalent to $\rho_1(t)$ being pure, with $||\rho_1||=1$ and
vanishing linear entropy $\Slin=1-||\rho_1(t)||^2$, so 
let us minimize the linear entropy averaged over a range of times.
As a concrete example, we minimize the function 
\beq{SnipEq}
f(\U)\equiv 
1-{1\over m}\sum_{i =1}^m||\trace_1\U\rho(t_i)\U^\dagger||^2,
\eeq
using 9 equispaced times $t_i$ ranging from $t=0$ and $t=1$, a random $4\times 4$ Hamiltonian $\H$, and a random pure state $\rho(0)$. 

 \begin{figure}[phbt]
\centerline{\includegraphics[width=88mm]{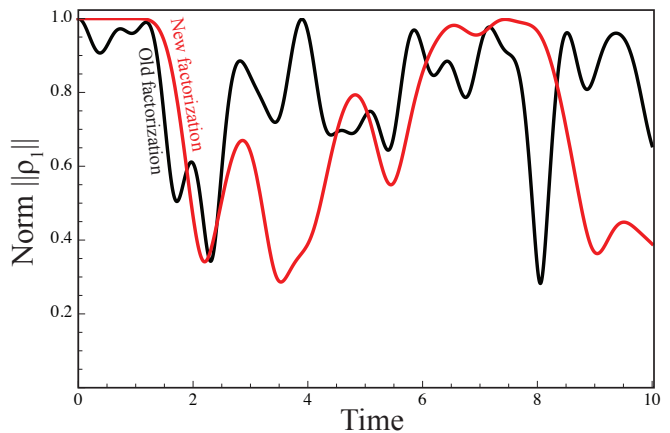}}
\caption{
The Hilbert-Schmidt norm $||\rho_1||$ is plotted for a random pure-state 2-qubit system when factorizing the Hilbert space in the original basis (black curve) and after a unitary transformation optimized to keep 
$\rho_1$ as pure as possible for $t\le 1$ (red/grey curve).
\label{SnipFig}
}
\end{figure}

The result of numerically solving this optimization problem is shown in \fig{SnipFig},
and we see that the new factorization keeps the norm $||\rho_1||$ visually indistinguishable from unity for the entire time period optimized for.
The optimization reduced the average Shannon entropy over this period from 
$S\approx 1.1$ bits to $S=0.0009$ bits.

The reason that the optimization is so successful is presumably that it by adjusting 
$N=n^2-n_1^2-n_2^2=16-4-4=8$ real parameters\footnote{There are $n^2$ parameters for $\U$, but transformations within each of the two subspaces have no effect, wasting $n_1^2$ and $n_2^2$ parameters.}  in $\U$, it is able to approximately zero out the first $N$ terms in the Taylor expansion of $\Slin(t)$, whose leading terms are given by equations\eqn{SlinDerivEq}-\eqn{SlinDerivEq3}.
A series of similar numerical experiments indicated that such excellent separability could generally be found as long as the number of time steps $t_i$ was somewhat smaller than the number of free parameters $N$ but not otherwise, suggesting that separability can be extended over long time periods for large $n$.
However, because we are studying only unitary evolution here, neglecting the important projection effect from the previous section, it is unclear how relevant these results are to our underlying goal. We have therefore not extended these numerical optimizations, which are quite time-consuming, to larger $n$.


\section{Conclusions}
\label{ConclusionsSec}

In this paper, we have explored two problems that are intimately related.
The first problem is that of understanding consciousness as a state of matter, ``perceptronium".
We have focused not on solving this problem, but rather on exploring the implications of this viewpoint.
Specifically, we have explored four basic principles that may distinguish conscious matter from other physical systems: the information, integration,  independence and dynamics  principles.

The second one is the {\it physics-from-scratch problem}: If the total Hamiltonian $\H$ and the total density matrix $\rho$ fully specify our physical world, how do we extract 3D space and the rest of our semiclassical world from nothing more than two Hermitian matrices? Can some of this information be extracted even from $\H$ alone, which is fully specified by nothing more than its eigenvalue spectrum? 
We have focused on a core part of this challenge which we have termed the {\it  quantum factorization problem}: why do conscious observers like us perceive the particular Hilbert space factorization corresponding to classical space (rather than Fourier space, say), and more generally, why do we perceive the world around us as a dynamic hierarchy of objects that are strongly integrated and relatively independent?

These two problems go hand in hand, because a generic Hamiltonian cannot be decomposed using tensor products, which would correspond to a decomposition of the cosmos into non-interacting parts, so there is some optimal factorization of our universe into integrated and relatively independent parts. 
Based on Tononi's work, we might expect that this factorization, or some generalization thereof, is what conscious observers perceive, because an integrated and relatively autonomous information complex is fundamentally what a conscious observer is.

\subsection{Summary of findings}

We first explored the integration principle, and found that classical physics allows information  to be essentially fully integrated using error-correcting codes, so that
any subset containing up to about half the bits can be reconstructed from the remaining bits.
Information stored in Hopfield neural networks is naturally error-corrected, but $10^{11}$ neurons support 
only about 37 bits of integrated information. This leaves us with an integration paradox: why does the information content of our conscious experience appear to be vastly larger than 37 bits?
We found that generalizing these results to quantum information exacerbated this integration paradox, allowing no more than about a quarter of a bit of integrated information --- and this result applied not only to Hopfield networks of a given size, but to the state of {\it any} quantum system of {\it any} size.
This strongly implies that the integration principle must be supplemented by at least one additional principle.

We next explored the independence principle and the extent to which a Hilbert space factorization can decompose the Hamiltonian $\H$  (as opposed to the state $\rho$) into independent parts. 
We quantified this using projection operators in the Hilbert-Schmidt vector space where $\H$ and $\rho$ are viewed as vectors rather than operators, and proved that the best decomposition can always be
found in the energy eigenbasis, where $\H$ is diagonal. 
This leads to a more pernicious variant of the Quantum Zeno Effect that we termed the 
Quantum Zeno Paradox:  if we decompose our universe into maximally independent objects, then all change grinds to a halt. Since conscious observers clearly do not perceive reality as being static and unchanging, 
the integration and independence principles must therefore be supplemented by at least one additional principle.
 
We then explored the dynamics principle, according to which a conscious system has the capacity to not only store information, but also to process it. We found the {\it energy coherence}
$\delta H\equiv\sqrt{2\>\tr\dot{\rho}^2}$
to be a convenient measure of dynamics: it can be proven to be time-independent, and it reduces to the energy uncertainty $\Delta H$ for the special case of pure states.
Maximizing dynamics alone gives boring periodic solutions unable to support complex information processing, but reducing $\delta H$ by merely a modest percentage enables chaotic and complex dynamics that explores the full dimensionality of the Hilbert space.
We found that high autonomy (a combination of dynamics and independence) can be attained 
even if the environment interaction is strong. One class of examples involves the environment effectively performing quantum-non-demolition measurements of the autonomous system, whose internal dynamics causes the non-negligible elements of the density matrix $\rho$ to ``slide along the diagonal'' in the measured basis, remaining in the low-decoherence subspace.
We studied such an example involving a truncated harmonic oscillator coupled to an external spin, and saw that it is easy to find classes of systems whose autonomy grows exponentially with the system size (measured in qubits). Generalized coherent states with Gaussian wavefunctions appeared particularly robust toward interactions with steep/short-range potentials. 
We found that any given $\H$ can also be perfectly decomposed given a suitably chosen $\rho$
that assigns zero amplitude to some energy eigenstates. 
When optimizing the Hilbert space factorization for $\H$ and $\rho$ jointly, it appears possible to make a subsystem history $\rho_1(t)$ close to separable for a long time. However, it is unclear how relevant this is, because the state projection caused by observation also alters $\rho_1$.

\subsection{How does a conscious entity perceive the world?}


What are we to make of these findings? 
We have not solved the quantum factorization problem, but our results have brought it into sharper focus, and highlighted both concrete open sub-problems and various hints and clues from observation about paths forward. Let us first discuss some open problems, then turn to the hints.

For the physics-from-scratch problem of deriving how we perceive our world from merely $\H$, $\rho$ and the Schr\"odinger equation, there are two possibilities: either the problem is well-posed or it is not. If not, this would be very interesting, implying that some sort of additional structure beyond $\rho$ and $\H$ is needed at the fundamental level ---  some additional mathematical structure encoding properties of space, for instance, which would be surprising given that this appears unnecessary in lattice Gauge theory (see \App{EmergenceAppendix}).
Since we have limited our treatment to unitary non-relativistic quantum mechanics, obvious candidates for missing structure relate to 
relativity and quantum gravity, where the Hamiltonian vanishes, and to mechanisms causing non-unitary wavefunction collapse.
Indeed, Penrose and others have speculated that gravity is crucial for a proper understanding of quantum mechanics even on small scales relevant to brains and laboratory experiments, and that it causes non-unitary wavefunction collapse \cite{PenroseBook}. 
Yet the Occam's razor approach is clearly the commonly held view that neither relativistic, gravitational nor non-unitary effects are central to understanding consciousness or how conscious observers perceive their immediate surroundings: 
astronauts appear to still perceive themselves in a semiclassical 3D space even when they are effectively in a zero-gravity environment, seemingly independently of relativistic effects, Planck-scale spacetime fluctuations, black hole evaporation, cosmic expansion of astronomically distant regions, {\etc}

If, on the other hand, the physics-from-scratch problem {\it is} well-posed, we face crucial unanswered questions related to Hilbert space factorization. Why do we perceive electromagnetic waves as transferring information between different regions of space, rather than as completely independent harmonic oscillators that each stay put in a fixed spatial location? 
These two viewpoints correspond to factoring the Hilbert space of the electromagnetic field in either real space or Fourier space, 
which are simply two unitarily equivalent Hilbert space bases.
Moreover, how can  we perceive a harmonic oscillator as an integrated system when its Hamiltonian can, as reviewed in \App{FactorizationAppendix}, be separated into completely independent qubits?
Why do we perceive a magnetic system described by the 3D Ising model as integrated, when it separates into completely independent qubits after a unitary transformation?\footnote{If we write the Ising Hamiltonian as a quadratic function of $\sigma_x$-operators, then it is also quadratic in the annihilation and creation operators and can therefore be diagonalized after a Jordan-Wigner transform
\cite{Nielsen05}. Note that such diagonalization is impossible for the Heisenberg ferromagnet, whose couplings are quadratic in all three Pauli matrices, because $\sigma_z^2$-terms are quartic in the annihilation and creation operators.}
In all three cases, the answer clearly lies not within the system itself  (in its internal dynamics $\H_1$), but in its interaction $\H_3$ with the rest of the world. But $\H_3$ involves the factorization problem all over again: whence this distinction between 
the system itself and the rest of the world, when there are countless other Hilbert space factorizations that mix the two?

\subsection{Open problems}

Based on our findings, three specific problems stand in the way of solving the quantum factorization problem and answering these questions, and we will now discuss each of them in turn.

\subsubsection{Factorization and the chicken-and-egg problem}

What should we determine first: the state or the factorization?
If we are given a Hilbert space factorization and an environment state, we can use the predictability sieve formalism \cite{Zurek05} to find the states of our subsystem that are most robust toward decoherence. In some simple cases, they are eigenstates of the effective interaction Hamiltonian $\H_*$  from \eq{HstarDefEq}.
However, to find the best factorization, we need information about the state.
A clock is a highly autonomous system if we factor the Hilbert space so that the first factor corresponds to the spatial volume containing the clock, but if the state were different such that the clock were somewhere else, we should factor out a different volume. 
Moreover, if the state has the clock in a superposition of two macroscopically different locations, then there is no single optimal factorization, but instead a separate one for each branch of the wavefunction. An observer looking at the clock would use the clock position seen to project onto the appropriate branch using \eq{BayesrhoEq}, so the solution to the quantum factorization problem that we should be looking for is not 
a single unique factorization of the Hilbert space. Rather, we need a criterion for identifying conscious observers, and then a prescription that determines which factorization each of them will perceive.

\subsubsection{Factorization and the integration paradox}

A second challenge that we have encountered is the extreme separability possible for both $\H$ and $\rho$.
In the introduction, we expressed hope that the apparent integration of minds and external objects might trace back to 
the fact that for generic $\rho$ and $\H$, there is no Hilbert space factorization that makes $\rho$ factorizable or $\H$ additively separable. Yet by generalizing Tononi's ideas to quantum systems, we found that what he terms the ``cruelest cut'' is very cruel indeed, able to reduce the mutual information in $\rho$ to no more than about $0.25$ bits, and typically able to make the interaction Hamiltonian $\H_3$ very small as well. We saw in \Sec{SnipSec} that even the combined effects $\rho$ and $\H$ can typically be made  close to separable, in the sense that there is a Hilbert space factorization where a subsystem history $\rho_1(t)$ is close to separable for a long time.
So why do we nonetheless perceive out universe as being relatively integrated, with abundant information available to us from near and far? Why do we not instead perceive our mind as essentially constituting its own parallel universe, solipsism-style, with merely exponentially small  interactions with the outside world?
We saw that the origin of this integration paradox is the vastness of the group of unitary transformations that we are minimizing over, whose number of parameters scales like $n^2=2^{2b}$ with the number of qubits $b$ and thus grows exponentially with system size (measured in either volume or number of particles).

\subsubsection{Factorization and the emergence of time}

A third challenge involves the emergence of time. Although this is a famously thorny problem in quantum gravity, our results show that it appears even in non-relativistic unitary quantum mechanics. 
It is intimately linked with our factorization problem, because we are optimizing over all unitary transformations $\U$, and 
time evolution is simply a one-dimensional subset of these transformations, given by $\U=e^{i\H t}$.
Should the optimal factorization be determined separately at each time, or only once and for all?
In the latter case, this would appear to select only one special time when our universe is optimally separable, seemingly contrary to our observations that the laws of physics are time-translation invariant. 
In the former case, the continuous change in factorization will simply undo time evolution \cite{Schwindt12}, making you feel that time stands still!
Observationally, it is obvious that the optimal factorization can change at least somewhat with time, since our 
designation of objects is temporary: the atoms of a highly autonomous wooden bowling ball rolling down a lane were once dispersed 
(as CO$_2$ and H$_2$O in the air, \etc) and will eventually disperse again. 
 
An obvious way out of this impasse is to bring consciousness back to center-stage as in \Sec{TraceYourselfSec} and \cite{brain,tripartite,secondlaw}.
Whenever a conscious observer interacts with her environment and gains new information, the state $\rho$ with which she describes her world gets updated according to \eq{BayesrhoEq}, the quantum-mechanical version of Bayes Theorem \cite{tripartite}. This change in her $\rho$ is non-unitary and therefore evades our timelessness argument above. Because she always perceives herself in a pure state, knowing the state of her mind, the joint state or her and the rest of the world is always separable. It therefore appears that if we can one day solve the quantum factorization problem, 
then we will find that the emergence of time is linked to the emergence of consciousness: the former cannot be fully understood without the latter.

\subsection{Observational hints and clues}

In summary, the quantum factorization problem is both very interesting and very hard.
However, as opposed to the hard problem of quantum gravity, say, where we have few if any observational clues to guide us, physics research has produced many valuable hints and clues relevant to the quantum factorization problem.
The factorization of the world that we perceive and the quantum states that we find objects in have turned out to be exceptionally unusual and special in various ways, and for each such way that we can identify, quantify and understand the underlying principle responsible for, we will make another important stride towards solving the factorization problem. Let us now discuss the hints that we have identified upon so far.

\subsubsection{The universality of the utility principle}

The principles that we listed in \Tab{PrincipleTable} were for conscious systems. If we shift attention to non-conscious objects, we find that although 
dynamics, independence and integration still apply in many if not most cases, the utility principle is the only one that universally applies to all of them.
For example, a rain drop lacks significant information storage capacity, a boulder lacks dynamics, a cogwheel can lack independence, 
and a sand pile lacks integration. This universality of the utility principle is hardly surprising, since utility is presumably the reason we evolved consciousness in the first place.  This suggests that we examine all other clues below through the lens of utility, to see whether the unusual circumstances in question can be explained via some implication of the utility principle. In other words, if we find that useful consciousness can only exist given certain strict requirements on the quantum factorization, then this could explain why we perceive a factorization satisfying these requirements. 

\subsubsection{$\rho$ is exceptional}

The observed state $\rho$ of our universe is exceptional in that it is extremely cold, with most of the Hilbert space frozen out  --- what principles might require this? Perhaps this is useful for consciousness by allowing relatively stable information storage and by allowing large autonomous systems 
thanks to the large available dynamic range in length scales (universe/brain/atom/Planck scale)? Us being far from thermal equilibrium with our 300K planet dumping heat from our 6000K sun into our 3K space is clearly conducive to dynamics and information processing.

\subsubsection{$\H$ is exceptional}

The Hamiltonian $\H$ of the standard model of particle physics is of the very special form 
\beq{StandardModelHeq}
\H=\int\H_\r(\r)d^3 r,
\eeq
which is seen to be almost additively separable in the spatial basis, and in no other basis.
Although \eq{StandardModelHeq} superficially looks completely separable just as $\H=\sum_i \H_i$, there is a coupling 
between infinitesimally close spatial points due to spatial derivatives in the kinetic terms.  
If we replace the integral by a sum in \eq{StandardModelHeq} by discretizing space as in lattice gauge theory, we need couplings only between nearest-neighbor points. 
This is a strong hint of the independence principle at work; all this near-independence gets ruined by a generic unitary transformation, making the factorization corresponding to our 3D physical space highly special; indeed, 3D space and the exact form of \eq{StandardModelHeq} could presumably be inferred from simply knowing the spectrum of $\H$.

$\H$ from \eq{StandardModelHeq} is also exceptional in that it contains mainly quadratic, cubic and quartic functions of the fermion and boson fields, which can in turn be expressed linearly or quadratically in terms of qubit raising and lowering operators (see \App{EmergenceAppendix}).
A generic unitary transformation would ruin this simplicity as well, introducing polynomials of enormous degree. 
What principle might be responsible for this?

$\H$ from \eq{StandardModelHeq} is also exceptional by exhibiting tremendous symmetry: the form of $\H_\r$ in invariant under both space and time translation, and indeed under the full Poincare group; using a factorization other than 3D space would ruin this symmetry.

\subsubsection{The ubiquity of autonomy}

When discussing the integration paradox above, we worried about factorizations splitting the world into nearly independent parts.
If there is a factorization with $\H_3=0$, then the two subsystems are independent for {\it any} state, for all time, and will act as two parallel universes. This means that if the only way to achieve high independence were to make $\H_3$ tiny, the integration paradox would indeed be highly problematic. 
However, we saw in \Sec{DynamicsSec} that this is not at all the case: it is quite easy to achieve high independence for {\it some states}, at least temporarily, even when $\H_3$ is large. The independence principle therefore does not push us inexorably towards perceiving a more disconnected world than the one we are familiar with. 
The ease of approximately factoring $\rho_1(t)$ during a significant time period as in \Sec{SnipSec} also appears unlikely to be a problem:
as mentioned, our calculation answered the wrong question by studying only unitary evolution, neglecting projection. The take-away hint is thus that observation needs to be taken into account to address this issue properly, just as we argued that it must be taken into account to understand the emergence of time.

\subsubsection{Decoherence as enemy}

%
%
%
%

Early work on decoherence \cite{Zeh70,JZ85} portrayed it mainly as an enemy,
rapidly killing off most quantum states, with only a tiny minority surviving long enough to be observable.
For example, a bowling ball gets struck by about $10^{25}$ air molecules each second, and a single strike suffices to ruin any macrosuperposition of the balls position extending further than about an
angstrom, the molecular De Broglie wavelength \cite{JZ85,collapse}.
The successful {\it predictability sieve} idea of Zurek and collaborators \cite{Zurek05} states that we will only perceive those quantum states that are most robust towards decoherence, which in the case of macroscopic objects such as bowling balls selects roughly classical states with fairly well-defined positions. 
In situations where the position basis emerges as special, this might thus trace back to the environmental interactions $\H_3$ (with air molecules \etc) probing the position, which might in turn traces back to the fact that $\H$ from \eq{StandardModelHeq} 
is roughly separable in the position basis. 
Note, however, that the general situation is more complicated, since the predictability sieve depends also on the state $\rho$, which might contain long-distance entanglement built up over time by the kinetic terms in \eq{StandardModelHeq}. Indeed, $\rho$ can describe a laboratory where a system is probed in a non-spatial basis, causing the predictablity sieve to favor, say, energy eigenstates. 

In terms of  \Tab{PrincipleTable}, we can view the predictability sieve as an application of the utility principle, since there is clearly no utility in trying to perceive something that will be irrelevant $10^{-25}$ seconds later.
In summary, the hint from this negative view of decoherence is that we should minimize it, either by factoring to minimize $\H_3$ itself or by using robust states on which $\H_3$ essentially performs quantum non-demolition measurements.

\subsubsection{Decoherence as friend}

Although quantum computer builders still view decoherence as their enemy, 
more recent work on decoherence has emphasized that it also has a positive side:
the Quantum Darwinism framework \cite{Zurek09} emphasizes the role of environment interactions $\H_3$ as a valuable communication channel, repeatedly copying information about the states of certain systems 
into the environment\footnote{Charles Bennett has suggested that Quantum Darwinism would be more aptly named ``Quantum Spam",
since the many redundant imprints of the system's state are normally not further reproduced.}, 
thereby helping explain the emergence of a consensus reality \cite{mubook}.
Quantum Darwinism can also be viewed as an application of the utility principle: it is only useful for us to try to
be aware of things that we can get information about, \ie, about states that have quantum-spammed the environment with redundant copies of themselves.
A hint from this positive view of environmental interactions is that we should {\it not} try to minimize $\H_3$ after all, but should instead reduce decoherence by the second mechanism: 
using states that are approximate eigenstates of the effective interaction $\H_*$ and therefore get abundantly copied into the environment.

Further work on Quantum Darwinism has revealed that such situations are quite exceptional, reaching the following conclusion \cite{Riedel12}:
{\it ``A state selected at random from the Hilbert space of a many-body 
  system is overwhelmingly likely to exhibit highly non-classical correlations. 
  For these typical states, half of the environment must be measured by an 
  observer to determine the state of a given subsystem. 
  The objectivity of classical reality --- the fact that multiple observers can 
  agree on the state of a subsystem after measuring just a small 
  fraction of its environment --- implies that the correlations found in nature between 
  macroscopic systems and their environments are very exceptional.''}
This gives a hint that the particular Hilbert space factorization we observe  
might be very special and unique, so that using the utility principle to insist on the existence of a consensus reality may have large constraining power among the factorizations --- perhaps even helping nail down the one we actually observe.

\subsection{Outlook}

In summary, the hypothesis that consciousness can be understood as a state of matter leads to fascinating interdisciplinary questions spanning the range from 
neuroscience to computer science, condensed matter physics and quantum mechanics.
Can we find concrete examples of error-correcting codes in the brain? 
Are there brain-sized non-Hopfield neural networks that support much more than 37 bits of integrated information?
Can a deeper understanding of consciousness breathe new life into the century-old quest to understand the emergence of a classical world from quantum mechanics, and can it even help explain how two Hermitian matrices $\H$ and $\rho$ lead to the subjective emergence of time?
The quests to better understand the internal reality of our mind and the external reality of our universe will hopefully assist one another. 

\bigskip

													
{\bf Acknowledgments:}
The author wishes to thank Christoph Koch, Meia Chita-Tegmark, Russell Hanson, Hrant Gharibyan, Seth Lloyd, Bill Poirier, Matthew Pusey,  Harold Shapiro and Marin Solja\v{c}i\'c and for helpful information and discussions, and Hrant Gharibyan for  mathematical insights regarding the $\rho$- and $\H$-diagonality theorems.
This work was supported by NSF AST-090884 \& AST-1105835.

\appendix

\section{Useful identities involving tensor products}
\label{IdentitySec}

Below is a list of useful identities involving tensor multiplication and partial tracing, many of which are used in the main part of the paper. Although they are all straightforward to prove by writing them out in the index notation of \eq{TensorProductDefEq}, I have been unable to find many of them in the literature. The tensor product $\tensormult$ is also known as the Kronecker product.
\beqa{PartialTraceIdentitiesEq}
(\A\tensormult\B)\tensormult\C&=&\A\tensormult(\B\tensormult\C)\\ 
\A\tensormult(\B+\C)&=&\A\tensormult\B+\A\tensormult\C\\ 
(\B+\C)\tensormult\A&=&\B\tensormult\A+\C\tensormult\A\\
%
(\A\tensormult\B)^\dagger&=&\A^\dagger\tensormult\B^\dagger\\ 
(\A\tensormult\B)^{-1}&=&\A^{-1}\tensormult\B^{-1}\\ 
\tr[\A\tensormult\B]&=&(\tr\A)(\tr\B)\\
\trace_1[\A\tensormult\B] &=&(\trace\A)\B\\
\trace_2[\A\tensormult\B] &=&(\trace\B)\A\\
\label{identity7}
\trace_1[\A(\B\tensormult\I)]&=&\trace_1[(\B\tensormult\I)\A]\\
\label{identity8}
\trace_2[\A(\I\tensormult\B)]&=&\trace_2[(\I\tensormult\B)\A]\\
\label{identity9}
\trace_1[(\I\tensormult\A)\B] &=& \A(\trace_1\B)\\
\label{identity10}
\trace_2[(\A\tensormult\I)\B] &=& \A(\trace_2\B)\\
\trace_1[\A(\I\tensormult\B)] &=& (\trace_1\A)\B\\
\label{identity12}
\trace_2[\A(\B\tensormult\I)] &=& (\trace_2\A)\B\\
\label{identity13}
\trace_1[\A(\B\tensormult\C)] &=& \trace_1[\A(\B\tensormult\I)]\C\\
\label{identity14}
\trace_2[\A(\B\tensormult\C)] &=& \trace_2[\A(\I\tensormult\C)]\B\\
\label{identity15}
\trace_1[(\B\tensormult\C)\A] &=& \C\>\trace_1[(\A\tensormult\I)\B]\\
\label{identity16}
\trace_2[(\B\tensormult\C)\A] &=& \B\>\trace_2[(\I\tensormult\C)\A]\\
\label{identity17}
\trace\left\{[(\trace_2\A)\tensormult\I]\B\right\}&=&\tr[(\trace_2\A)(\trace_2\B)]\\
\label{identity18}
\trace\left\{[\I\tensormult(\trace_1\A)]\B\right\}&=&\tr[(\trace_1\A)(\trace_1\B)]\\
(\A\tensormult\B,\C\tensormult\D)&=&(\A,\C)(\B,\D)\\
||\A\tensormult\B||&=&||\A||\>||\B||
\eeqa
Identities~\ref{identity9}-\ref{identity12} are seen to be special cases of 
identities~\ref{identity13}-\ref{identity16}.
If we define the superoperators $\T_1$ and $\T_2$ by
\beqa{SuperoperatorT1T2defEq}
\T_1\A &\equiv&{1\over n_1}\I\tensormult(\tr_1\A),\\
\T_2\A &\equiv&{1\over n_2}(\tr_2\A)\tensormult\I,
\eeqa
then identities~\ref{identity17}-\ref{identity18} imply that they are self-adjoint:
$$(\T_1\A,\B)=(\A,\T_1\B),\quad (\T_2\A,\B)=(\A,\T_2\B).$$
They are also projection operators, since they satisfy $\T_1^2=\T_1$ and $\T_2^2=\T_2$.

\section{Factorization of Harmonic oscillator into uncoupled qubits}
\label{FactorizationAppendix}

If  the Hilbert space dimensionality $n=2^b$ for some integer $b$, then the truncated harmonic oscillator Hamiltonian of \eq{SHOspectrumEq} can be decomposed into $b$ independent qubits:
in the energy eigenbasis,
\beq{SHOfactorizationEq1}
\H=\sum_{j=0}^{b-1} \H_j,\quad \H_j
=2^j
\left(\ms
\begin{tabular}{cc}
${1\over 2}$&$0$\\
$0$&$-{1\over 2}$
\end{tabular}
\ms\right)_j
=2^{j-1}\sigma^z_j,
\eeq
where the subscripts $j$ indicate that an operator acts only on the $j^{\rm th}$ qubit, leaving the others unaffected.
For example, for $b=3$ qubits,
\beqa{3qubitExampleEq}
\H&=&\left(\ms
\begin{tabular}{cc}
$2$&$0$\\
$0$&$-2$
\end{tabular}
\ms\right)
\tensormult\I
\tensormult\I
+
\I\tensormult
\left(\ms
\begin{tabular}{cc}
$1$&$0$\\
$0$&$-1$
\end{tabular}
\ms\right)
\tensormult\I
+
\I\tensormult\I\tensormult
\left(\ms
\begin{tabular}{cc}
${1\over 2}$&$0$\\
$0$&$-{1\over 2}$
\end{tabular}
\ms\right)
\nonumber
\\
&=&
\left(\ms
\begin{tabular}{rrrrrrrr}
$-{7\over 2}$&$0$&$0$&$0$&$0$&$0$&$0$&$0$\\
$0$&$-{5\over 2}$&$0$&$0$&$0$&$0$&$0$&$0$\\
$0$&$0$&$-{3\over 2}$&$0$&$0$&$0$&$0$&$0$\\
$0$&$0$&$0$&$-{1\over 2}$&$0$&$0$&$0$&$0$\\
$0$&$0$&$0$&$0$&${1\over 2}$&$0$&$0$&$0$\\
$0$&$0$&$0$&$0$&$0$&${3\over 2}$&$0$&$0$\\
$0$&$0$&$0$&$0$&$0$&$0$&${5\over 2}$&$0$\\
$0$&$0$&$0$&$0$&$0$&$0$&$0$&${7\over 2}$
\end{tabular}
\ms\right),
\eeqa
in agreement with \eq{SHOspectrumEq}.
This factorization corresponds to the standard binary representation of integers,
which is more clearly seen when adding back the trace $(n-1)/2=(2^b-1)/2$:
\beqa{3qubitExampleEq2}
H+{7\over 2}&=&\left(\ms
\begin{tabular}{cc}
$4$&$0$\\
$0$&$0$
\end{tabular}
\ms\right)
\tensormult\I
\tensormult\I
+
\I\tensormult
\left(\ms
\begin{tabular}{cc}
$2$&$0$\\
$0$&$0$
\end{tabular}
\ms\right)
\tensormult\I
+
\I\tensormult\I\tensormult
\left(\ms
\begin{tabular}{cc}
$1$&$0$\\
$0$&$0$
\end{tabular}
\ms\right)\nonumber\\
&=&
\left(\ms
\begin{tabular}{rrrrrrrr}
$0$&$0$&$0$&$0$&$0$&$0$&$0$&$0$\\
$0$&$1$&$0$&$0$&$0$&$0$&$0$&$0$\\
$0$&$0$&$2$&$0$&$0$&$0$&$0$&$0$\\
$0$&$0$&$0$&$3$&$0$&$0$&$0$&$0$\\
$0$&$0$&$0$&$0$&$4$&$0$&$0$&$0$\\
$0$&$0$&$0$&$0$&$0$&$5$&$0$&$0$\\
$0$&$0$&$0$&$0$&$0$&$0$&$6$&$0$\\
$0$&$0$&$0$&$0$&$0$&$0$&$0$&$7$
\end{tabular}
\ms\right).
\eeqa
Here we use the ordering convention that the most significant qubit goes to the left.
If we write $k$ as  
$$k = \sum_{j=0}^{b-1} k_j 2^j,$$
where $k_j$ are the binary digits of $k$ and take values 0 or 1, 
then the energy eigenstates can be written
\beq{MultiQubitEigenvectorEq}
|E_k\rangle=\tensormultiplication_{j=0}^{b-1} (\sigma^\dagger)^{k_j} |0\rangle,
\eeq
where $|0\rangle$ is the ground state (all $b$ qubits in the down state),
the creation operator 
\beq{CreationOperatorEq}
\sigma^\dagger=
\left(\ms
\begin{tabular}{cc}
$0$&$1$\\
$0$&$0$
\end{tabular}
\ms\right)
\eeq
raises a qubit from the down state to the up state, and
$(\sigma^\dagger)^0$ is meant to be interpreted as the identity matrix $\I$.
For example, since the binary representation of 6 is ``110", we have
\beq{E6eq}
|E_6\rangle = \sigma^\dagger\tensormult\sigma^\dagger\tensormult\I |0\rangle= |110\rangle,
\eeq
the state where the first two qubits are up and the last one is down.
Since $(\sigma^\dagger)^{k_j}\left({0\atop 1}\right)$ is an eigenvector of $\sigma^z$ with eigenvalue 
$(2k_j-1)$, \ie, $+1$ for spin up and $-1$ for spin down, 
equations\eqn{SHOfactorizationEq1} and\eqn{MultiQubitEigenvectorEq} give
$\H|E_k\rangle=E_k|E_k\rangle$, where
\beq{EkEq}
E_k=\sum_{j=0}^{b-1}2^{j-1}(2k_j-1)|E_k\rangle
=k-{2^b-1\over 2}
\eeq
in agreement with \eq{SHOspectrumEq}.

The standard textbook harmonic oscillator corresponds to the limit $b\to\infty$, which remains completely separable. In practice, a number of qubits $b=200$ is large enough to be experimentally indistinguishable from $b=\infty$ for describing any harmonic oscillator ever encountered in nature, 
since it corresponds to a dynamic range of $2^{200}\sim 10^{60}$,
the ratio between the largest and smallest potentially measurable energies (the Planck energy versus the energy of a photon with wavelength equal to the diameter of our observable universe).
So far, we have never measured any physical quantity to better than 17 significant digits, corresponding to 56 bits.


\section{Emergent space and particles from nothing but qubits}
\label{EmergenceAppendix}

Throughout the main body of our paper, we have limited our discussion to a Hilbert space of finite dimensionality $n$, often interpreting it as $b$ qubits with $n=2^b$.
On the other hand, textbook quantum mechanics usually sets $n=\infty$ and contains plenty of structure additional to merely $\H$ and $\rho$, such as a continuous space and
various fermion and boson fields. The purpose of this appendix is to briefly review how the latter picture might emerge from the former. An introduction to this ``it's all qubits'' approach by one of its pioneers, Seth Lloyd, is given in \cite{LloydBook}, and an up-to-date technical review can be found in 
\cite{Gu09}.

As motivation for this emergence approach, note that a large number of 
quasiparticles have been observed such as
phonons, holes, magnons, rotons, plasmons and polarons, 
which are known  not to be fundamental particles, but instead mere excitations in some underlying substrate.
This raises the question of whether our standard model particles may be quasiparticles as well.
It has been shown that this is indeed a possibility for photons, electrons and quarks
\cite{Wen03,Levin05,Levin06}, and perhaps even for gravitons \cite{Gu09}, with the substrate being nothing more than a set of qubits without any space or other additional structure.

In \App{FactorizationAppendix}, we saw how to build a harmonic oscillator out of infinitely many qubits, and that a truncated harmonic oscillator built from merely 200 qubits is experimentally indistinguishable from an infinite-dimensional one. We will casually refer to such a qubit collection 
describing a truncated harmonic oscillator as a ``qubyte'', even if the number of qubits it contains is not precisely 8. As long as our universe is cold enough that the very highest energy level is never excited, a qubyte will behave identically to a true harmonic oscillator, and can be used to define position and momentum operators obeying the usual canonical commutation relations.


To see how space can emerge from qubits alone, consider a large set of coupled truncated harmonic oscillators (qubytes),
whose position operators $q_\r$ and momentum operators $p_\r$  are labeled by an index $\r=(i,j,k)$ consisting of a triplet of integers --- $\r$ has no {\it a priori} meaning or interpretation whatsoever except as a record-keeping device used to specify the Hamiltonian.
Grouping these operators into vectors $\p$ and $\q$, we choose the Hamiltonian
\beq{qubyteHeq}
\H={1\over 2}|\p|^2 + {1\over 2}\q^t\A\q,
\eeq
where the coupling matrix $\A$ is translationally invariant, \ie, $\A_{\r\r'}=\a_{\r'-\r}$, depending only on the difference $\r'-\r$ between two index vectors. 
For simplicity, let us treat the lattice of index vectors $\r$ as infinite, so that $\A$ is diagonalized by a 3D Fourier transform. 
(Alternatively, we can take the lattice to be finite and the matrix $\A$ to be circulant, in which case $\A$ is again diagonalized by a Fourier transform; this will lead to the emergence of a toroidal space.)

Fourier transforming our  qubyte lattice preserves the canonical commutation relations and corresponds to a unitary transformation that  decomposes $\H$ into independent harmonic oscillators.
As in \cite{steady}, the frequency of the oscillator corresponding to wave vector $\kappab$ is 
\beq{qbyteomegaEq}
\omega(\k) ^2= \sum_\r \a_\r e^{-i\kappa\cdot\r}.
\eeq
For example, consider the simple case where each oscillator has a self-coupling $\mu$ and  is only coupled to its 6 nearest neighbors by a coupling $\gamma$:
$\a_{1,0,0}=\a_{-1,0,0}=\a_{0,1,0}=\a_{0,-1,0}=\a_{0,0,1}=\a_{0,0,-1}=-\gamma^2$, $\a_{0,0,0}=\mu^2+6\gamma^2$.
Then 
\beq{qbyteomegaEq2}
\omega(\kappab) ^2= \mu^2+4\gamma^2\left(\sin^2{\kappa_x\over 2}+\sin^2{\kappa_y\over 2}+\sin^2{\kappa_z\over 2}\right),
\eeq
where $\kappa_x$, $\kappa_y$ and $\kappa_z$ lie in the interval $[\pi,\pi]$.
If we were to interpret  the lattice points as existing in a three-dimensional space with separation $a$ between neighboring lattice points, then the physical wave vector $k$ would be given by
\beq{qbytekDefEq}
k={\kappa\over a}.
\eeq
Let us now consider a state $\rho$ where all modes except long-wavelength ones with 
$|\kappab|\ll 1$ are frozen out, in the spirit of our own relatively cold universe.
Using the $\bumpeq$ symbol from \Sec{BumpEqSec}, we then have $\H\bumpeq\H'$, where 
$\H'$ is a Hamiltonian with the 
 isotropic dispersion relation
\beq{qbyteomegaEq3}
\omega ^2= \mu^2+\gamma^2\left(\kappa_x^2+\kappa_y^2+\kappa_z^2\right)= \mu^2+\gamma^2 \kappa^2,
\eeq
\ie, where the discreteness effects are absent. 
Comparing this with the standard dispersion relation for a relativistic particle, 
\beq{qbyteomegaEq4}
\omega ^2= \mu^2+ (ck)^2,
\eeq
where $c$ is the speed of light, 
we see that the two agree if the lattice spacing is 
\beq{qbyteaEq}
a={c\over\gamma}.
\eeq
For example, if the lattice spacing is the Planck length, then the coupling strength $\gamma$ is the inverse Planck time.
In summary, this Hilbert space built out of qubytes, with no structure whatsoever except for the Hamiltonian $\H$, is physically indistinguishable from 
a system with quantum particles (scalar bosons of mass $\mu$) propagating in a continuous 3D space with the same translational and rotational symmetry that we normally associate with infinite Hilbert spaces, so not only did space emerge, but continuous symmetries not inherent in the original qubit Hamiltonian emerged as well. The 3D structure of space emerged from the pattern of couplings between the qubits: if they had been presented in a random order, the graph of which qubits were coupled could have
been analyzed to conclude that everything could be simplified into a 3D rectangular lattice with nearest-neighbor couplings. 

Adding polarization to build photons and other vector particles is straightforward.
Building  simple fermion fields using qubit lattices is analogous as well, except that a unitary Jordan-Wigner transform is required for converting the qubits to fermions. Details on how to build photons, electrons, quarks and perhaps even gravitons are given in 
\cite{Wen03,Levin05,Levin06,Gu09}.
Lattice gauge theory works similarly, except that here, the underlying finite-dimensional Hilbert space is viewed not as the actual truth but as a numerically tractable approximation to the presumed true infinite-dimensional Hilbert space of quantum field theory.




\begin{thebibliography}{99}

\bibitem{Almheiri12}
\rf\nn Almheiri A, \nn Marolf D, \nn Polchinski J\multiand\nn Sully J;2013;JHEP;2;62

\bibitem{Banks14}
\rfprep\nn Banks T, \nn Fischler W, \nn Kundu S\multiand\nnn Pedraza J F;2014;{arXiv:1401.3341}

\bibitem{SaundersBook}
\rfbook\nn Saunders S, \nn Barrett J, \nn Kent A\multiand\nn Wallace D;2010;Many Worlds? Everett, Quantum Theory, \& Reality;{Oxford Univ. Press};{Oxford}

\bibitem{brain}
\rf\nn Tegmark M;2000;PRE;61;4194

\bibitem{Chalmers95}
\rf\nnn Chalmers D J;1995;J.  Consc. Studies;2;200

\bibitem{mmm}
\rf\nn Hut P, \nn Alford M\multiand\nn Tegmark M;{2006, physics/0510188};Found.~Phys.;36;765	 
 
 \bibitem{toe2}
\rf\nn Tegmark M;2007;Found.Phys.;11/07;116

\bibitem{TononiManifesto}
\rf\nn Tononi G;2008;Biol. Bull.;215;{216, \url{http://www.biolbull.org/content/215/3/216.full}}

\bibitem{Dehaene11}
\rf\nn Dehaene S;2011;Neuron;70;200

\bibitem{TononiBook}
\rfbook\nn Tononi G;2012;Phi: A Voyage from the Brain to the Soul;Pantheon;{New York}

\bibitem{Casali13}
\rf\nn Casali A {\etal};2013;Sci. Transl. Med;198;1

\bibitem{IIT3}
\rfe\nn Oizumi M, \nn Albantakis L\multiand Tononi G;2014;PLoS comp. bio;e1003588

\bibitem{Dehaene14}
\rf\nn Dehaene S {\etal};2014;Current opinion in neurobiology;25;76

\bibitem{Wilson95}
\rfproc\nnn Wilson B A\dualand\nn Wearing D;1995;Broken memories: Case studies in memory impairment;\nn Campbell R\dualand\nnn Conway M A;Malden;Blackwell


\bibitem{computronium}
\rf\nn Amato I;1991;Science;253;856

\bibitem{Lloyd99}
\rf\nn Lloyd S;2000;Nature;406;1-47

\bibitem{tHooft93}
\rfprep\nn {t'Hooft} G;1993;arXiv:gr-qc/9310026

\bibitem{Schwindt12}
\rfprep\nn Schwindt J;2012;{arXiv:1210.8447 [quant-ph]}

\bibitem{Damasio10}
\rfbook\nn Damasio A;2010;Self Comes to Mind: Constructing the Conscious Brain;Vintage;{New York}

\bibitem{Hamming50}
\rf\nnn Hamming R W;1950;The Bell System Technical Journal;24;2

\bibitem{HammingDistanceSite}
\rn\nn Grassl M, \url{http://i20smtp.ira.uka.de/home/grassl/codetables/}


\bibitem{Hopfield82}
\rf\nnn Hopfield J J;1982;Proc. Natl. Acad. Sci.;79;2554

\bibitem{Joshi13}
\rf\nnn Joshi N J, \nn Tononi G\multiand\nn Koch C;2013;PLOS Comp. Bio.;9;e1003111

\bibitem{Barak13}
\rf\nn Barak O {\etal};2013;Progr. Neurobio.;103;214

\bibitem{Yoon13}
\rf Yoon K {\etal};2013;Nature Neuroscience;16;1077

\bibitem{McKayBook}
\rfbook\nnnn McKay D J C;2003;Information Theory, Inference, and Learning Algorithms;Cambridge University Press;Cambridge

\bibitem{Kupfmuller62}
\rn\nn K\"{u}pfm\"{u}ller K, {\it Nachrichtenverarbeitung im Menschen}, in {\it Taschenbuch der Nachrichtenverarbeitung},
\nn Steinbuch K, Ed., 1481-1502 (1962).

\bibitem{NorretrandersBook}
\rfbook\nn N{\o}rretranders T;1991;The User Illusion: Cutting Consciousness Down to Size;Viking;{New York}

\bibitem{Jevtic11}
\rf\nn Jevtic S, Jennings D\multiand\nn Rudolph T;2012;PRL;108;110403

\bibitem{vonNeumann32}
\rfbook\nn {von Neumann} J.;1932;Die mathematischen Grundlagen der Quantenmechanik;Springer;Berlin.

\bibitem{MarshallBook}
\rfbook\nnn Marshall A W, \nn Olkin I\multiand\nnn Arnold B;2011;Inequalities: Theory of Majorization and Its Applications, 2nd ed;Springer;{New York}

\bibitem{Bravyi03}
\rf\nn Bravyi S;2004;Quantum Inf. and Comp.;4;12
	
\bibitem{Zurek01}
\rfprep\nnn Zurek W H;2001;quant-ph/0111137



\bibitem{Zeh70}
\rf\nnn Zeh H D;1970;Found.Phys.;1;69

\bibitem{JZ85}
\rf\nn Joos E\dualand\nnn Zeh H D;1985;Z. Phys. B;59;223

\bibitem{ZurekHabibPaz93}
\rf\nnn Zurek W H, \nn Habib S\multiand\nnn Paz J P;1993;PRL;70;1187

\bibitem{ZehBook}
\rfbook\nn Giulini D, \nn Joos E, \nn Kiefer C, \nn Kupsch J,
\nnn Stamatescu I O\multiand\nnn Zeh H D;1996;Decoherence and the Appearance
of a Classical World in Quantum Theory;Berlin;Springer

\bibitem{Zurek09}
\rf\nnn Zurek W H;2009;Nature Physics;5;181

\bibitem{SchlosshauerBook}
\rfbook\nn Schlosshauer M;2007;Decoherence and the Quantum-To-Classical Transition;Springer;Berlin

\bibitem{Zurek09}
\rf\nnn Zurek W H;2009;Nature Physics;5;181

\bibitem{Sudarshan77}
\rf\nnnn Sudarshan E C G\dualand\nn Misra B;1977;J. Math. Phys.;18;756 

\bibitem{Omnes01}
\rfprep\nn Omn\`{e}s R;2001;quant-ph/0106006

\bibitem{Gemmer01}
\rf\nn Gemmer J\dualand\nn Mahler G;2001;Eur. Phys J. D;17;385

\bibitem{Durt04}
\rf\nn Durt T;2004;Z. Naturforsch.;59a;425

\bibitem{ZHP93}
\rf\nnn Zurek W H, \nn Habib S\multiand\nnn Paz J P;1993;PRL;70;1187

\bibitem{gaussians}
\rf\nn Tegmark M\dualand\nnn Shapiro H S;1994;Phys. Rev. E;50;2538

\bibitem{secondlaw}
\rfprep\nn Gharibyan H\dualand\nn Tegmark M;2013;{arXiv:1309.7349 [quant-ph]}

\bibitem{tripartite}
\rf\nn Tegmark M;2012;PRD;85;123517

\bibitem{Nielsen05}
\rn Nielsen 2005, \url{http://michaelnielsen.org/blog/archive/notes/fermions_and_jordan_wigner.pdf}

\bibitem{Zurek05}
\rf\nnnn Dalvit D A R, \nn Dziarmaga J\multiand\nnn Zurek W H;2005;PRA;72;062101

\bibitem{PenroseBook}
\rfbook\nn Penrose R;1989;The Emperor's New Mind;Oxford Univ. Press;Oxford

\bibitem{collapse}
\rf\nn Tegmark M;1993;Found. Phys. Lett.;6;571


\bibitem{mubook}
\rfbook\nn Tegmark M;2014;Our Mathematical Universe: My Quest for the Ultimate Nature of Reality;Knopf;{New York}

\bibitem{Riedel12}
\rf\nnn Riedel C J, \nnn Zurek W H\multiand\nn Zwolak M;2012;New J. Phys.;14;083010




 \bibitem{LloydBook}
 \rfbook\nn Lloyd S;2006;Programming the Universe;Knopf;{New York} 
  
\bibitem{Gu09}
\rf\nn Gu Z\dualand\nn Wen X;2012;Nucl.Phys. B;863;90
 
 \bibitem{Wen03}
 \rf\nn Wen X;2003;PRD;68;065003
 
 \bibitem{Levin05}
 \rf\nnn Levin M A\dualand\nn Wen X;2005;RMP;77;871 
 
\bibitem{Levin06}
\rf\nnn Levin M A\dualand\nn Wen X;2006;PRB;73;035122

\bibitem{steady}
\rf\nn Tegmark M\dualand\nn Yeh L;1994;Physica A;202;342


\end{thebibliography}
\end{document}